\documentclass[aps,prd,onecolumn,eqsecnum, nofootinbib]{revtex4-2}
\usepackage{amssymb}
\usepackage{amsmath}
\usepackage{amsfonts}
\usepackage{upgreek}
\usepackage{bm}
\usepackage{graphicx}
\usepackage{hyperref}
\DeclareSymbolFont{EulerScript}{U}{eus}{m}{n}
\SetSymbolFont{EulerScript}{bold}{U}{eus}{b}{n}
\DeclareSymbolFontAlphabet\scrpt{EulerScript}
\newcommand{\KK}{{k}}
\newcommand{\D}{{\mathfrak{D}}}
\newcommand{\C}{{\scrpt C}}
\newcommand{\LL}{{\scrpt L}}
\newcommand{\SSS}{{\scrpt S}}
\newcommand{\UU}{{\scrpt U}}
\newcommand{\VV}{{\scrpt V}} 
\newcommand{\Lie}{{\pounds}} 
\allowdisplaybreaks
\begin{document}
\title{Self-gravitating anisotropic fluid. II: Newtonian theory}  
\author{Tom Cadogan and Eric Poisson}  
\affiliation{Department of Physics, University of Guelph, Guelph,
  Ontario, N1G 2W1, Canada} 
\date{May 30, 2024} 
\begin{abstract}
This paper is the second in a sequence of three devoted to the formulation of a theory of self-gravitating anisotropic fluids in both Newtonian gravity and general relativity. In the first paper we set the stage, placed our work in context, and provided an overview of the results obtained in this paper and the next. In this second paper we develop the Newtonian theory, inspired by a real-life example of an anisotropic fluid, the (nematic) liquid crystal. We apply the theory to the construction of static and spherical stellar models. In the third paper we port the theory to general relativity, and exploit it to build relativistic stellar models. In addition to the usual fluid variables (mass density, velocity field), the Newtonian theory features a director vector field $\bm{c}(t,\bm{x})$, whose length provides a local measure of the size of the anisotropy, and whose direction gives the local direction of anisotropy. The theory is defined in terms of a Lagrangian which implicates all the relevant forms of energy: kinetic energy (with contributions from the velocity field and the time derivative of the director vector), internal energy (with isotropic and anisotropic contributions), gravitational interaction energy, and gravitational-field energy. This Lagrangian is easy to motivate, and it provides an excellent starting point for a relativistic generalization in the third paper. The equations of motion for the fluid, and Poisson's equation for the gravitational potential, follow from a variation of the action functional, given by the time integral of the Lagrangian. Because our stellar models feature a transition from an anisotropic phase at high density to an isotropic phase at low density, a substantial part of the paper is devoted to the development of a mechanics for the interface fluid, which mediates the phase transition.
\end{abstract} 
\maketitle

\section{Introduction} 
\label{sec:intro} 

This sequence of three papers aims to develop Newtonian and relativistic theories of self-gravitating anisotropic fluids, and to apply them to the construction of anisotropic stellar structures. The motivation for this work was discussed at great length in paper I \cite{cadogan-poisson:24a}; it comes largely from a desire to provide more satisfying models of stellar anisotropy in the context of general relativity. The formulation of the relativistic theory will be undertaken in paper III \cite{cadogan-poisson:24c}. In this paper we take the necessary first step of constructing a Newtonian theory, which will serve as a direct inspiration in the generalization to a relativistic setting. We also apply the Newtonian theory to the elaboration of anisotropic stellar models. 

To formulate our Newtonian theory of a self-gravitating anisotropic fluid, we take guidance in a real-life instance of an anisotropic fluid: the (nematic) liquid crystal \cite{degennes:74, outram:18, khoo:22}, in which long organic molecules are preferentially aligned in a common direction to create the anisotropy. Following Ericksen \cite{ericksen:60}, we motivate the model in Sec.~\ref{sec:dumbbell} in terms of a simple microscopic picture, in which the fluid consists of diatomic molecules. The position of the center of mass of a given molecule is denoted $\bm{x}$, and this center of mass moves with a velocity $\bm{v}$. The relative separation between the atoms, proportional to the director vector $\bm{c}$, defines the direction and magnitude of the anisotropy; the relative velocity is proportional to the director velocity $\bm{w} = d\bm{c}/dt$. The fluid picture emerges by going to a continuum limit, in which $\bm{v}$, $\bm{c}$, and $\bm{w}$ become vector fields, functions of time and position $\bm{x}$. The state of the fluid is further determined by its mass density, a density of internal energy that contains both isotropic and anisotropic contributions, and other variables that can be derived from these. 

We specify the theory in Sec.~\ref{sec:lagrangian} in terms of Lagrangian, which is then integrated over time to form an action functional. In addition to the fluid's kinetic and internal energies, the Lagrangian includes an interaction energy with the gravitational field, and the field's own energy. The Lagrangian formulation implies that our theory is restricted to conservative interactions, and it forbids the incorporation of dissipative effects created, for example, by viscosity. In spite of this limitation, we find that the formulation in terms of a Lagrangian provides a compelling entry point into the theory; the Lagrangian is easy to motivate, as it is built from all the forms of energy that are relevant to an anisotropic fluid coupled to gravity. More importantly, we find that the Lagrangian provides a straightforward path to generalize the theory to a relativistic setting. As was discussed at the end of Sec.~VI in paper I \cite{cadogan-poisson:24a}, the relativistic version of the Lagrangian is a fairly obvious extension of the Newtonian one, but there is nothing obvious about the generalization of the fluid's equations of motion. And as was mentioned at the end of Sec.~III in paper I, the resulting theory can always be modified after the fact to include dissipative effects, so that the limitation is not too severe after all. In Sec.~\ref{sec:variation} we undertake a variation of the action functional. The equations of motion for the fluid, and the field equation for gravity, are written down in Sec.~\ref{sec:fluid}, and shown to give rise to conservation statements for the total linear momentum, angular momentum, and energy.  

The need to invoke a phase transition in our anisotropic stellar models was explained in Sec.~V of paper I \cite{cadogan-poisson:24a}: The equations of stellar structure are generically singular at the surface, and we cure the problem by postulating the existence of a phase transition from an anisotropic phase at high density to an isotropic phase at low density. We wish to describe this phase transition in all generality, well beyond the immediate context of our static and spherically symmetric stellar models. This desire launches us on a rather long digression in Secs.~\ref{sec:interface} and \ref{sec:twophase}, in which we develop the relevant mechanics. In this general context, the phase transition is idealized as to occur on a time-dependent, two-dimensional surface, and it is mediated by an interface fluid that is itself anisotropic. 

In the first stage of our digression (Sec.~\ref{sec:interface}) we formulate a complete theory for the interface fluid. Because it lives on a moving surface, a number of the standard ingredients of the bulk theory require a substantial modification. First, the mathematical description of the two-dimensional surface is somewhat involved, and the relevant elements of a differential geometry of moving surfaces\footnote{This differential geometry is surprisingly cumbersome compared with that of a three-dimensional hypersurface embedded in a curved spacetime. The source of complexity has to do with the absolute nature of time in Newtonian physics. In the spacetime setting, the choice of intrinsic coordinates placed on the hypersurface is unrestricted, and there is no need to distinguish the time coordinate from the spatial coordinates. In the Newtonian setting, time is absolute, and the time coordinate is necessarily separated from the spatial coordinates; this segregation yields an awkward formulation of the differential geometry. This observation suggests that a parametrized description of the moving surface, with Newtonian time expressed as a function of the intrinsic parameters, would be far more elegant and perhaps more practical.} are reviewed in Appendix \ref{sec:moving_surface}, following an approach initiated by Grinfeld \cite{grinfeld:13}. Second, the variation of the interface fluid is delicate to describe, because the notion of an Eulerian change of a fluid variable does not exist in this context: the surface is displaced during a variation, and a comparison at the same spatial position cannot be undertaken. Fortunately, however, the Lagrangian change (a comparison at the same fluid element) is perfectly well defined, and the variation of the interface action can be formulated entirely in terms of such changes. Third, the interface fluid comes with an areal density of mass that is finite, but in view of its idealized localization on a two-dimensional surface, it possesses a volume density that is formally infinite. This implies that the gravitational field is discontinuous on the surface, which complicates the variation of the gravitational action --- variables are usually required to be differentiable. The difficulty can be dealt with --- the gravitational field is averaged across the surface --- but this prescription requires a thorough justification that we provide in Appendix~\ref{sec:interface_gravity}.

In the second stage of our digression (Sec.~\ref{sec:twophase}) we join the interface fluid to the anisotropic and isotropic phases of the bulk fluid, and thereby form a combined system. The complete Lagrangian gives rise to equations of motion for each phase of the fluid, and dynamical equations for the interface fluid. In addition, it produces junction conditions that permit the bulk variables to be connected across each side of the interface. This complete set of equations permits the exploration of the physics of a two-phase fluid in any context. In particular, the equations can be specialized to describe a static and spherically symmetric configuration, as we shall describe next, but the equations are not initially restricted to such situations. The theory, for example, provides the required foundation to study the dynamical stability of stellar models, to describe their deformation under an applied tidal field, and to compute their entire spectrum of normal modes of vibration.

In Sec.~\ref{sec:stellar_model} we do specialize the fluid equations to static and spherically symmetric configurations, and build models of anisotropic stars. We make specific choices of equations of state for the two-phase fluid --- our stars are polytropes --- and integrate the structure equations numerically. Some of our results were previously summarized in Sec.~IV of paper I \cite{cadogan-poisson:24a}. In this last section we explore the parameter space more broadly. We construct sequences of equilibrium configurations parametrized by the central density $\rho(r=0)$, and show that these terminate at a maximum value of the central density. This is quite unlike what occurs in the case of isotropic polytropes, for which the sequences continue indefinitely. 

\section{Diatomic molecule and anisotropic fluid}
\label{sec:dumbbell}

To provide a motivation for what is to come, we follow Ericksen \cite{ericksen:60} and develop a simple molecular model for the anisotropic fluid. The model features a molecule that consists of two atoms. The first has a mass $m_1$ and is situated at position $\bm{r}_1$ in some reference frame. The second has a mass $m_2$ and position $\bm{r}_2$. We let $m := m_1 + m_2$ denote the molecule's total mass. 

We introduce the position $\bm{x}$ of the molecule's center of mass, which is determined by $m \bm{x} = m_1 \bm{r}_1 + m_2 \bm{r}_2$. We introduce also the separation vector $\bm{\eta} := \bm{r}_2 - \bm{r}_1$, as well as a rescaled version $\bm{c} := (m_1 m_2/m^2)^{1/2}\, \bm{\eta}$, which we call the {\it director vector}. In terms of these we have that the individual positions are
\begin{equation}
\bm{r}_1 = \bm{x} - \sqrt{m_2/m_1}\, \bm{c}, \qquad
\bm{r}_2 = \bm{x} + \sqrt{m_1/m_2}\, \bm{c}.
\end{equation}
The individual velocities are then
\begin{equation}
\bm{v}_1 = \bm{v} - \sqrt{m_2/m_1}\, \bm{w}, \qquad
\bm{v}_2 = \bm{v} + \sqrt{m_1/m_2}\, \bm{w}, 
\end{equation}
where $\bm{v} := d\bm{x}/dt$ is the center-of-mass velocity, and $\bm{w} := d\bm{c}/dt$ is a rescaled version of the relative velocity. It follows from these results that the molecule's total kinetic energy is
\begin{equation}
\frac{1}{2} m_1 v_1^2 + \frac{1}{2} m_2 v_2^2 = \frac{1}{2} m (v^2 + w^2),
\end{equation}
that its total momentum is
\begin{equation}
m_1 \bm{v}_1 + m_2 \bm{v}_2 = m \bm{v},
\end{equation}
and that its total angular momentum is
\begin{equation}
m_1 \bm{r}_1 \times \bm{v}_1 + m_2 \bm{r}_2 \times \bm{v}_2
= m (\bm{x} \times \bm{v} + \bm{c} \times \bm{w}).
\end{equation}

We next consider a very large collection of these molecules, and describe it in terms of a continuous distribution of matter. In this description we have that $\bm{c}$, $\bm{v}$, and $\bm{w}$ become functions of time $t$ and position $\bm{x}$, and that $\bm{w}$ is now related to $\bm{c}$ according to
\begin{equation}
w^a = \frac{D c^a}{dt} := \partial_t c^a + v^b \nabla_b c^a,
\label{w_def}
\end{equation}
where $D/dt$ denotes a covariant material derivative. The mass density at $t$ and $\bm{x}$ is denoted $\rho$, and according to the previous results, we have that $\frac{1}{2} \rho (v^2 + w^2)$ is the fluid's density of kinetic energy, that $\rho \bm{v}$ is the density of linear momentum, and that $\rho ( \bm{x} \times \bm{v} + \bm{c} \times \bm{w})$ is the density of angular momentum.

The director field $\bm{c}(t,\bm{x})$ specifies a preferred direction at every point in the fluid, making it anisotropic. In his presentation of the theory of nematic liquid crystals, de Gennes \cite{degennes:74} takes  $\bm{c}$ to be a unit vector. We do not follow this practice here. In our description, the magnitude $|\bm{c}|$ of the director vector provides a measure of the size of the anisotropy, and the unit vector $\bm{c}/|\bm{c}|$ describes the direction of the anisotropy.  

\section{Fluid Lagrangian}
\label{sec:lagrangian}

We take the anisotropic fluid and its associated gravitational field to be governed by the Lagrangian
\begin{equation}
L = \int_V \bigl[ \tfrac{1}{2} \rho(v^2 + w^2) - \varepsilon - \omega + \rho U \bigr]\, dV
- \frac{1}{8\pi G} \int \nabla_a U \nabla^a U\, dV,
\label{Lag1}
\end{equation}
in which the first integral is over the region $V$ occupied by the fluid, while the second integral is over all space. We employ an arbitrary coordinate system $x^a$ with metric $g_{ab}$, and $dV := \sqrt{g}\, d^3x$ is the invariant volume element; $g := \mbox{det}[g_{ab}]$. We denote by $\partial V$ the closed, two-dimensional surface that bounds the region $V$; this is the boundary of the fluid. 

The first term inside the first integral is recognized as the density of kinetic energy, written in terms of the fluid's velocity $v^a$ and the director velocity $w^a$; we use the notation $v^2 := g_{ab} v^a v^b$ and $w^2 := g_{ab} w^a w^b$. In the second term we have $\varepsilon$, the isotropic contribution to the density of internal energy. In the third term we have the anisotropic contribution\footnote{The anisotropic contribution to the density of internal energy could be generalized from this simple expression. The irreducible pieces of $\nabla_a c_b$ are given by the trace part $\frac{1}{3} g_{ab} \nabla_c c^c$, the symmetric-tracefree part $\nabla_{\langle a} c_{b\rangle} := \nabla_{(a} c_{b)} - \frac{1}{3} g_{ab} \nabla_c c^c$, and the antisymmetric part $\nabla_{[a} c_{b]}$; here the round brackets indicate symmetrization of the indices, while the square brackets indicate antisymmetrization. The most general expression for $\omega$ would include three separate terms involving the square of each irreducible piece, and three separate coupling constants.}   
\begin{equation}
\omega := \frac{1}{2} \kappa \Xi, \qquad
\Xi := \nabla_a c_b \nabla^a c^b,
\end{equation}
with $\kappa > 0$ playing the role of a coupling constant; this is minimized when $c^a$ is uniform. The director field has the dimension of a length, and it follows that $\Xi$ is dimensionless, so that $\kappa$ must have the dimension of energy density, or $(\mbox{mass density}) (\mbox{velocity})^2$. Finally, the fourth term in the first integral gives us the interaction energy between the fluid and the gravitational field, while the second integral represents the field's own energy; the Newtonian potential $U$ is defined so that $\rho \nabla_a U$ is the density of gravitational force acting on the fluid. 

The isotropic internal energy $\varepsilon$ and the coupling constant $\kappa$ are assumed to be functions of $\rho$ only. While these quantities could also depend on the specific entropy $s$, we choose to rule out such a dependence for the sake of simplicity; the fluid is barotropic. We note that since the fluid is governed by a Lagrangian, there is no dissipation of energy, and therefore no production of entropy. 

It is sometimes useful to rescale $\varepsilon$, $\kappa$, and $\omega$ by the density, so we introduce
\begin{equation}
\hat{\varepsilon} := \varepsilon/\rho, \qquad
\hat{\kappa} := \kappa/\rho, \qquad
\hat{\omega} := \omega/\rho.
\end{equation}
Derivatives of these quantities with respect to $\rho$ will be needed. We introduce 
\begin{equation}
p := \rho^2 \frac{d\hat{\varepsilon}}{d\rho}
= \rho^2 \frac{d}{d\rho} (\varepsilon/\rho) 
\label{d_epsilon}
\end{equation}
as an isotropic thermodynamic pressure, and similarly, 
\begin{equation}
\lambda := \rho^2 \frac{d\hat{\kappa}}{d\rho} = \rho^2 \frac{d}{d\rho} (\kappa/\rho).
\label{lambda_def}
\end{equation} 
It is understood that $p$ and $\lambda$ have the same dimension, and are functions of $\rho$ only. 

An alternative expression for the Lagrangian is 
\begin{equation}
L = \int_V \rho \LL\, dV - \frac{1}{8\pi G} \int \nabla_a U \nabla^a U\, dV,
\label{Lag2}
\end{equation}
where
\begin{equation}
\LL := \frac{1}{2} v^2 + \frac{1}{2} w^2 - \hat{\varepsilon} - \hat{\omega} + U.
\label{LL_def}
\end{equation}
The action functional is
\begin{equation}
S = \int_{t_1}^{t_2} L\, dt,
\end{equation}
where the integration is carried out between two arbitrary reference times, $t_1$ and $t_2$. 

\section{Variation of the action}
\label{sec:variation}

In this section we carry out a variation of the action $S$. We begin in Sec.~\ref{subsec:var_U} with a variation with respect to the gravitational potential $U$, and in Sec.~\ref{subsec:var_fluid} we move on to a variation with respect to the fluid configuration.

It is a subtle matter to vary an action functional in fluid mechanics, because the variation must be constrained by mass and entropy conservation: the mass and entropy of a fluid element are to be held constant during the variation. An additional complication is that the variation must also preserve the identity of each fluid element. In an Eulerian approach to the variation, all these constraints can be implemented with Lagrange multipliers (see Ref.~\cite{seliger-whitham:68} for a discussion in the Newtonian context, and Ref.~\cite{schutz:70} for a relativistic generalization). We prefer to impose them implicitly through a Lagrangian approach, which takes a point of view that centers on the fluid element. Throughout this section we rely heavily on the variational techniques introduced by Schutz and Sorkin \cite{schutz-sorkin:77}; these are based on the Lagrangian perturbation theory of perfect fluids developed by Friedman and Schutz \cite{friedman-schutz:78a}.   

\subsection{Gravitational potential}
\label{subsec:var_U}

We first carry out a variation of the action with respect to the gravitational potential $U$, keeping the fluid variables fixed. We have that $\delta S = \int \delta L\, dt$ with
\begin{equation}
\delta L = \int_V \rho\, \delta U\, dV - \frac{1}{4\pi G} \int \nabla^a U \nabla_a \delta U\, dV.
\end{equation}
We express the integrand of the second integral as $\nabla_a (\delta U\, \nabla^a U) - \delta U\, \nabla^2 U$, where $\nabla^2 := g^{ab} \nabla_a \nabla_b$ is the Laplacian operator. Next we use Gauss's theorem to write the volume integral of the divergence term as a surface integral. This produces
\begin{equation}
\delta L = -\frac{1}{4\pi G} \oint_\infty \delta U\, \nabla^a U\, dS_a
+ \frac{1}{4\pi G} \int (\nabla^2 U + 4\pi G \rho )\delta U\, dV.
\end{equation}
We indicate that the surface integral is evaluated at infinity, because the domain of the volume integral is all space.

The variational principle requires $\delta U$ to be arbitrary everywhere, but to vanish at infinity. Under these conditions, we have that $\delta S = 0$ yields Poisson's equation
\begin{equation}
\nabla^2 U = -4\pi G \rho
\label{poisson} 
\end{equation}
for the gravitational potential. We assume that the fluid is the sole source of gravity. 

\subsection{Fluid variables}
\label{subsec:var_fluid}

The task before us now is to vary the action with respect to the fluid variables, keeping $U$ fixed. In this context the second integral of Eq.~(\ref{Lag2}) is no longer required, and this allows us to simplify the Lagrangian to 
\begin{equation}
L = \int_V \rho \LL\, dV,
\label{Lag3}
\end{equation}
with $\LL$ still defined by Eq.~(\ref{LL_def}). 

\subsubsection{Formalism} 

The variation is described in terms of two independent variables. The first is $\xi^a$, the Lagrangian displacement vector, which takes a fluid element at position $x^a$ in the reference configuration and places it at a new position $x^a + \xi^a$. The second is $\delta c^a$, the Eulerian change\footnote{It is possible to adopt instead the Lagrangian change $\Delta c^a$, but this produces computations that are more cumbersome; the end results are identical.} in the director field. 

We adopt the formalism of Friedman and Schutz \cite{friedman-schutz:78a}, in which Lagrangian and Eulerian changes are related by
\begin{equation}
\Delta Q^{a\cdots}_{b\cdots} = \delta Q^{a\cdots}_{b\cdots}  + \Lie_\xi Q^{a\cdots}_{b\cdots}
\label{Delta_vs_delta} 
\end{equation}
for any tensor $Q^{a\cdots}_{b\cdots}$; $\Lie_\xi$ denotes Lie differentiation in the direction of $\xi^a$. In terms of the Lagrangian displacement we have that
\begin{equation}
\Delta \rho = -\rho \nabla_a \xi^a, \qquad
\Delta v^a = \partial_t \xi^a, \qquad
\Delta g_{ab} = \nabla_a \xi_b + \nabla_b \xi_a.
\label{Delta} 
\end{equation}
The expression for $\Delta \rho$ reflects the fact that the variation preserves the mass of a fluid element. The expression for $\Delta g_{ab}$ follows from the fact that $\delta g_{ab} = 0$.

The integral identity $\delta \int_V \rho j\, dV = \int_V \rho \Delta j\, dV$, valid for any scalar $j$, implies that the variation of the Lagrangian can be computed as
\begin{equation}
\delta L = \int_V \rho \Delta \LL\, dV,
\label{delta_L} 
\end{equation}
where, according to Eq.~(\ref{LL_def}),
\begin{equation}
\Delta \LL = \Delta (\tfrac{1}{2} v^2) + \Delta(\tfrac{1}{2} w^2) - \Delta \hat{\varepsilon}
- \Delta \hat{\omega} + \Delta U.
\label{delta_LL} 
\end{equation}
From this we readily obtain $\delta S = \int \delta L\, dt$.

The variation is carried out according to the familiar rules. The fields $\xi^a$ and $\delta c^a$ are arbitrary and independent within $V$, but they are required to vanish on the boundary,  
\begin{equation}
\xi^a(\partial V) = 0 = \delta c^a(\partial V).
\label{onS}
\end{equation}
They are also required to vanish at the reference times $t_1$ and $t_2$ featured in the action integral, 
\begin{equation}
\xi^a(t_{1,2}) = 0 = \delta c^a(t_{1,2});
\label{at_t12} 
\end{equation}
these equations apply in the entire domain $V$. 

\subsubsection{Computations}

We set out to compute each term that appears in Eq.~(\ref{delta_LL}). We begin with $\Delta (\frac{1}{2} v^2)$, which we calculate as
\begin{equation}
\Delta (\tfrac{1}{2} v^2) = \Delta (\tfrac{1}{2} g_{ab} v^a v^b) = \frac{1}{2} v^a v^b \Delta g_{ab} + v_a \Delta v^a
= v^a v^b \nabla_a \xi_b + v_a \partial_t \xi^a,
\end{equation}
where we made use of Eq.~(\ref{Delta}).

Next we proceed with $\Delta (\frac{1}{2} w^2)$, which we calculate as
\begin{equation}
\Delta (\tfrac{1}{2} w^2) = \delta (\tfrac{1}{2} w^2) + \xi^c \nabla_c (\tfrac{1}{2} w^2)    
= w_a \delta w^a + w_a \xi^c \nabla_c w^a.
\end{equation}
To obtain $\delta w^a$ we begin with the definition of Eq.~(\ref{w_def}), which implies that $\delta w^a = \partial_t \delta c^a + \delta v^b \nabla_b c^a + v^b \nabla_b \delta c^a$. In this we replace $\delta v^b$ with $\Delta v^b = \partial_t \xi^b$ using Eq.~(\ref{Delta_vs_delta}), and get
\begin{equation}
\delta w^a = \partial_t \delta c^a + v^b\nabla_b \delta c^a
+ \nabla_b c^a \bigl( \partial_t \xi^b - \xi^c \nabla_c v^b + v^c \nabla_c \xi^b \bigr).
\end{equation}
Making the substitution in $\Delta (\frac{1}{2} w^2)$, we arrive at
\begin{equation}
\Delta (\tfrac{1}{2} w^2) = w_b \nabla_a c^b\, \partial_t \xi^a
+ w_b \nabla_a c^b v^c \nabla_c \xi^a
+ \bigl( w_b \nabla_a w^b - w_b \nabla_c c^b \nabla_a v^c \bigr) \xi^a
+ w_a\, \partial_t \delta c^a + w_a v^b \nabla_b \delta c^a.
\end{equation}

For $\Delta \hat{\varepsilon}$ we make use of Eqs.~(\ref{d_epsilon}) and (\ref{Delta}), and obtain
\begin{equation}
\Delta \hat{\varepsilon} = -\frac{p}{\rho}\, \nabla_a \xi^a.
\label{Del_epsilon} 
\end{equation}
In a more complete treatment in which $\hat{\varepsilon}$ would depend on the specific entropy $s$ in addition to the mass density $\rho$, the isotropic pressure of Eq.~(\ref{d_epsilon}) would be defined by a partial derivative at constant $s$, and Eq.~(\ref{Del_epsilon}) would reflect the fact that the variation preserves the entropy of a fluid element. In our simplified treatment in which the fluid is homentropic, this constraint is imposed automatically. 

The computation of $\Delta \hat{\omega}$ is carried out as
\begin{equation}
\Delta \hat{\omega} = \Delta (\tfrac{1}{2} \hat{\kappa} \Xi)
= \tfrac{1}{2} \Xi\, \Delta \hat{\kappa} + \hat{\kappa}\, \Delta (\tfrac{1}{2} \Xi).
\end{equation}
For $\Delta \hat{\kappa}$ we import Eqs.~(\ref{lambda_def}) and (\ref{Delta}), and get $\Delta \hat{\kappa} = -(\lambda/\rho) \nabla_a \xi^a$. For $\Delta (\tfrac{1}{2} \Xi)$ we write
\begin{equation}
\Delta (\tfrac{1}{2} \Xi) = \delta (\tfrac{1}{2} \Xi) + \xi^a \nabla_a (\tfrac{1}{2} \Xi)
= \nabla^b c^c \bigl( \nabla_b \delta c_c + \nabla_{ab} c_c\, \xi^a \bigr).
\end{equation}
Collecting results, we arrive at
\begin{equation}
\Delta \hat{\omega} = -\frac{1}{2} \frac{\lambda}{\rho}\, \Xi \nabla_a \xi^a
+ \hat{\kappa} \nabla^b c_a \nabla_b \delta c^a
+ \hat{\kappa} \nabla^b c^c \nabla_{ab} c_c\, \xi^a.
\end{equation}

Finally, we have that $\Delta U = \delta U + \xi^a \nabla_a U$. We note, however, that $\delta U$ is independent of $\xi^a$ and $\delta c^a$, and we recall that the variation with respect to the gravitational potential was already carried out in Sec.~\ref{subsec:var_U}. In the context of this section we must therefore write $\Delta U = \xi^a \nabla_a U$.

\subsubsection{Variation of the action}

With all the results collected previously, we find that after some organization, $\rho \Delta \LL$ becomes 
\begin{align}
\rho \Delta \LL &= \bigl( \rho v_a + \rho w_b \nabla_a c^b \bigr) \partial_t \xi^a
+ \bigl( T_a^{\ b} + \nabla_a c^c J^b_{\ c} \bigr) \nabla_b \xi^a
+ \Bigl\{ \rho w_b \bigl[ \nabla_a w^b - \nabla_a (v^c \nabla_c c^b) \bigr]
+ J^{bc} \nabla_{ab} c_c + \rho \nabla_a U \Bigr\} \xi^a
\nonumber \\ & \quad \mbox{} 
+ \rho w_a\, \partial_t \delta c^a + J^b_{\ a} \nabla_b \delta c^a,  
\label{rho_DLL1}
\end{align}
where
\begin{equation}
T_{ab} := \rho v_a v_b + \bigl( p + \tfrac{1}{2} \lambda \Xi \bigr) g_{ab}
+ \kappa \nabla_a c_c \nabla_b c^c
\label{T_def}
\end{equation}
and
\begin{equation}
J_{ab} := \rho v_a w_b - \kappa \nabla_a c_b. 
\label{J_def}
\end{equation} 
Notice that $T_{ab}$ is a symmetric tensor, but that $J_{ab}$ possesses no such symmetry. 

We recall that the variation of the Lagrangian is given by $\delta L = \int \rho \Delta \LL\, dV$. In the expression of Eq.~(\ref{rho_DLL1}) we see terms involving $\nabla_b \xi^a$ and $\nabla_b \delta c^a$, and we deal with these by integrating by parts, sending all divergences to surface integrals. We also see terms involving $\partial_t \xi^a$ and $\partial_t \delta c^a$, and we deal with these by incorporating them within a total time derivative.  After these manipulations and some simplification, we arrive at 
\begin{align}
\delta L &= \frac{d}{dt} \int_V \bigl[ (\rho v_a + \rho w_b \nabla_a c^b) \xi^a
+ \rho w_a\, \delta c^a \bigr] dV
+ \oint_{\partial V} \bigl[ (T_a^{\ b} + \nabla_a c^c J^b_{\ c}) \xi^a + J^b_{\ a}\, \delta c^a \bigr]\, dS_b
\nonumber \\ & \quad \mbox{} 
- \int_V \Bigl\{ \partial_t(\rho v_a) + \nabla_b T_a^{\ b} - \rho \nabla_a U
+ \bigl[ \partial_t(\rho w_b) + \nabla_c J^c_{\ b} \bigr] \nabla_a c^b \Bigr\} \xi^a\, dV
\nonumber \\ & \quad \mbox{} 
- \int_V \bigl[ \partial_t(\rho w_a) + \nabla_b J^b_{\ a} \bigr]\, \delta c^a\, dV.
\label{delta_L_final}
\end{align}
The variation of the action is $\delta S = \int \delta L\, dt$. With the variation rules spelled out previously, we eliminate the total time derivative and the surface integral. What remains are the two volume integrals involving $\xi^a$ and $\delta c^a$. 

\section{Fluid equations and conservation laws} 
\label{sec:fluid}

In this section we deduce the dynamical equations for the fluid variables (Sec.~\ref{subsec:fluid_equations}), obtain a wave equation for the director vector (Sec.~\ref{subsec:wave}), and derive conservation laws for the fluid's total energy, momentum, and angular momentum (Sec.~\ref{subsec:global}). 

\subsection{Fluid equations}
\label{subsec:fluid_equations} 

We return to $\delta S = \int \delta L\, dt$ and the variation of Eq.~(\ref{delta_L_final}), and demand that $\delta S = 0$ for arbitrary and independent variations $\xi^a$ and $\delta c^a$. We obtain two sets of dynamical equations for the fluid variables. The first is
\begin{equation}
\partial_t(\rho v_a) + \nabla_b T_a^{\ b} -\rho \nabla_a U = 0,
\label{fluid1}
\end{equation}
which governs the fluid velocity. The second is
\begin{equation}
\partial_t (\rho w_a) + \nabla_b J^b_{\ a} = 0,
\label{fluid2}
\end{equation}
which governs the director velocity. We recall that $w^a$ is related to the director field $c^a$ by
\begin{equation}
w^a := \partial_t c^a + v^b \nabla_b c^a. 
\end{equation}
We also recall that
\begin{equation}
T_{ab} := \rho v_a v_b + \bigl( p + \tfrac{1}{2} \lambda \Xi \bigr) g_{ab}
+ \kappa \nabla_a c_c \nabla_b c^c, 
\label{T_def_repeat}
\end{equation}
and note that Eq.~(\ref{fluid1}) gives it an interpretation as a flux tensor for the momentum density $\rho v_a$; $\lambda$ is defined by Eq.~(\ref{lambda_def}) and $\Xi := \nabla_a c_b \nabla^a c^b$. Finally, we recall that 
\begin{equation}
J_{ab} := \rho v_a w_b - \kappa \nabla_a c_b,
\label{J_def_repeat}
\end{equation} 
and observe that Eq.~(\ref{fluid2}) provides it with an interpretation as a flux tensor for the director momentum density $\rho w_a$.

The fluid equations (\ref{fluid1}) and (\ref{fluid2}) are written in the form of conservation laws. They can be presented in mechanical form by appealing to the continuity equation
\begin{equation}
\partial_t \rho + \nabla_a (\rho v^a) = 0. 
\label{continuity} 
\end{equation}
The equation gives rise to the identity
\begin{equation}
\partial_t (\rho u^a) = \rho \frac{Du^a}{dt} - \nabla_b (\rho u^a v^b), 
\end{equation}
valid for any vector field $u^a$; here $D u^a/dt := \partial_t u^a + v^b \nabla_b u^a$ is the covariant material derivative.

When we apply the identity to $v^a$ and combine it with Eq.~(\ref{fluid1}), we obtain
\begin{equation}
\rho \frac{Dv^a}{dt} + \nabla_b S^{ab} - \rho \nabla^a U = 0,
\label{fluid3}
\end{equation}
where
\begin{equation}
S^{ab} := T^{ab} - \rho v^a v^b = \bigl( p + \tfrac{1}{2} \lambda \Xi \bigr) g^{ab}
+ \kappa \nabla^a c^c \nabla^b c_c
\label{S_def}
\end{equation}
is the fluid's stress tensor. Equation (\ref{fluid3}) is a generalized version of Euler's equation.

When we apply the identity to $w^a$ and combine it with Eq.~(\ref{fluid2}), we obtain
\begin{equation}
\rho \frac{Dw^a}{dt} + \nabla_b K^{ba} = 0,
\label{fluid4}
\end{equation}
where
\begin{equation}
K^{ba} := J^{ba} - \rho v^b w^a = -\kappa \nabla^b c^a. 
\label{K_def}
\end{equation}

The fluid equations come with equations of state that relate $\varepsilon$, $\kappa$, $p$, and $\lambda$ to the mass density $\rho$. They are also supplemented with Poisson's equation (\ref{poisson}) for the Newtonian potential $U$. 

\subsection{Director wave} 
\label{subsec:wave} 

As an elementary application of the fluid equations, we consider a wave of director field traveling in a homogeneous medium at rest. We set $\rho = \mbox{constant}$, $\kappa = \mbox{constant}$, and $v^a = 0$. We take the amplitude of the wave to be sufficiently small that terms quadratic in $c^a$ can be neglected in Eq.~(\ref{fluid1}). The only relevant equation for our purposes is then Eq.~(\ref{fluid2}), which becomes $\rho \partial_t w_a + \nabla_b J^b_{\ a} = 0$, or $\rho \partial_{tt} c_a - \kappa \nabla^2 c_a = 0$, or
\begin{equation}
\biggl( -\frac{1}{c_{\rm dir}^2} \frac{\partial^2}{\partial t^2} + \nabla^2 \biggr) c_a = 0.
\end{equation}
This is a wave equation for the director field, with the wave's traveling speed given by
\begin{equation}
c_{\rm dir} := \sqrt{\kappa/\rho}.
\end{equation}
This assignment confirms that $\kappa$ must be positive. For reasonable equations of state we can expect $c_{\rm dir}$ and the speed of sound $c_{\rm s} := \sqrt{dp/d\rho}$ to be of the same order of magnitude. 
 
\subsection{Global conservation laws}
\label{subsec:global} 

The fluid equations listed in Sec.~\ref{subsec:fluid_equations} give rise to global conservation laws for momentum, angular momentum, and energy. We establish these laws in this section. For our purposes here it is helpful to consider a region of space $\VV$ that is fixed in time and surrounds the fluid distribution; $\VV$ is strictly larger than $V$, and the fluid stays confined within $\VV$ as it evolves in time. The region $\VV$ is bounded by the closed surface $\partial\VV$; because all fluid variables are confined to $V \subset \VV$, they vanish on $\partial\VV$.

In our developments below, the equations pertaining to the conservation of momentum and angular momentum are all formulated in Cartesian coordinates, with the covariant derivative $\nabla_a$ reducing to the partial derivative $\partial_a$. The equations pertaining to the conservation of energy are formulated in any coordinate system. 

\subsubsection{Momentum and director momentum}

Recalling the diatomic molecule of Sec.~\ref{sec:dumbbell} and the passage to a continuum description, we define
\begin{equation}
P^a := \int_\VV \rho v^a\, dV
\label{mom}
\end{equation}
to be the fluid's total momentum. In addition, we define
\begin{equation}
P_{\rm dir}^a := \int_\VV \rho w^a\, dV
\label{dir_mom}
\end{equation}
to be the momentum associated with the motion of the director field. We shall show that both quantities are conserved by the fluid's dynamics:
\begin{equation}
\frac{d}{dt} P^a = 0, \qquad
\frac{d}{dt} P^a_{\rm dir}= 0.
\end{equation}

We begin with Eq.~(\ref{mom}), which we differentiate with respect to time. Because $\VV$ is fixed in time, this is
\begin{equation}
\frac{d}{dt} P^a = \int_\VV \partial_t (\rho v^a)\, dV.
\end{equation}
We insert Eq.~(\ref{fluid1}) and send the volume integral of $\nabla_b T^{ab}$ to a surface integral. This gives
\begin{equation}
\frac{d}{dt} P^a = -\oint_{\partial\VV} T^{ab}\, dS_b + \int_\VV \rho \nabla^a U\, dV.
\end{equation}
The surface integral evaluates to zero by virtue of the fact that all fluid variables vanish on $\partial\VV$. The volume integral can be computed by inserting the solution to Poisson's equation for $U$; this is done in Sec.~1.4.3 of Poisson and Will \cite{poisson-will:14}, and the integral is shown to vanish. We have arrived at $dP^a/dt = 0$.

Similar manipulations involving Eqs.~(\ref{dir_mom}) and (\ref{fluid2}) produce
\begin{equation}
\frac{d}{dt} P^a_{\rm dir} = -\oint_{\partial\VV} J^{ba}\, dS_b,
\end{equation}
and the surface integral vanishes. We have established that $dP_{\rm dir}^a/dt = 0$.

\subsubsection{Angular momentum}

Taking again our inspiration from Sec.~\ref{sec:dumbbell}, we introduce
\begin{equation}
{\cal J}^{ab} := {\cal L}^{ab} + {\cal S}^{ab}
\end{equation}
as the total angular-momentum tensor, with
\begin{equation}
{\cal L}^{ab} := \int_\VV \rho (x^a v^b - v^a x^b)\, dV
\end{equation}
denoting the contribution from the fluid's motion, and
\begin{equation}
{\cal S}^{ab} := \int_\VV \rho (c^a w^b - w^a c^b)\, dV
\end{equation}
denoting the contribution from the director field. We shall show that
\begin{equation}
\frac{d}{dt} {\cal L}^{ab} = 0, \qquad
\frac{d}{dt} {\cal S}^{ab} = 0,
\end{equation}
so that $d {\cal J}^{ab}/dt = 0$. While we expected the total angular momentum to be conserved, we actually find that ${\cal L}^{ab}$ and ${\cal S}^{ab}$ are separately conserved by the fluid's dynamics.

We calculate the rate of change of ${\cal L}^{ab}$ as 
\begin{align}
\frac{d}{dt} {\cal L}^{ab} &= \int_\VV \bigl[ x^a \partial_t(\rho v^b) - \partial_t (\rho v^a) x^b \bigr]\, dV
\nonumber \\
&= \int_\VV \bigl( -x^a \nabla_c T^{cb} + \rho x^a \nabla^b U + x^b \nabla_c T^{cb} - \rho x^b \nabla^a U \bigr)\, dV
\nonumber \\
&= \oint_{\partial\VV} \bigl( -x^a T^{cb} + x^b T^{ca} \bigr)\, dS_c
+ \int_\VV \bigl( T^{ab} - T^{ba} \bigr)\, dV
+ \int_\VV \rho \bigl( x^a \nabla^b U - x^b \nabla^a U \bigr)\, dV.
\end{align}
As usual the surface integral vanishes, and the symmetry of the momentum flux tensor implies that the first volume integral evaluates to zero. The second volume integral is shown to vanish in Sec.~1.4.3 of Poisson and Will \cite{poisson-will:14}. We arrive at $d {\cal L}^{ab}/dt = 0$.

For ${\cal S}^{ab}$ it is useful to rely on the integral identity
\begin{equation}
\frac{d}{dt} \int_\VV \rho j^{a\cdots}_{b\cdots}\, dV = \int_\VV \rho \frac{d}{dt} j^{a\cdots}_{b\cdots}\, dV,
\label{identity} 
\end{equation}
valid for any Cartesian tensor $j^{a\cdots}_{b\cdots}$ by virtue of the continuity equation (\ref{continuity}). We apply it to $c^a w^b - w^a c^b$ and get
\begin{align}
\frac{d}{dt} {\cal S}^{ab} &= \int_\VV \rho \biggl( c^a \frac{dw^b}{dt} - c^b \frac{dw^a}{dt} \biggr)\, dV
\nonumber \\
&= \int_\VV \bigl( -c^a \nabla_c K^{cb} + c^b \nabla_c K^{ca} \bigr)\, dV
\nonumber \\
&= \oint_{\partial\VV} \bigl( -c^a K^{cb} + c^b K^{ca} \bigr)\, dS_c
+ \int_\VV \bigl( \nabla_c c^a K^{cb} - \nabla_c c^b K^{ca} \bigr)\, dV.
\end{align}
We inserted Eq.~(\ref{fluid4}) in the second step, and integrated by parts in the third step. As usual the surface integral is zero, and the volume integral vanishes as well, by virtue of the explicit expression for $K^{ab}$ --- refer back to Eq.~(\ref{K_def}). We have established that $d {\cal S}^{ab}/dt = 0$.

\subsubsection{Energy}

The Lagrangian of Eq.~(\ref{Lag1}) motivates the following expression for the fluid's total energy:
\begin{equation}
E = {\cal T}_{\rm cm} + {\cal T}_{\rm dir} + E_{\rm iso} + E_{\rm aniso} + E_{\rm inter} + E_{\rm field},
\label{E_total} 
\end{equation}
where
\begin{subequations}
\label{E_pieces} 
\begin{align}
{\cal T}_{\rm cm} &:= \int_\VV \tfrac{1}{2} \rho v^2\, dV, \\
{\cal T}_{\rm dir} &:= \int_\VV \tfrac{1}{2} \rho w^2\, dV, \\
E_{\rm iso} &:=\int_\VV \varepsilon\, dV = \int_\VV \rho\, \hat{\varepsilon}\, dV, \\
E_{\rm aniso} &:=\int_\VV \omega\, dV = \int_\VV \rho\, \hat{\omega}\, dV, \\
E_{\rm inter} &:= -\int_\VV \rho U\, dV, \\
E_{\rm field} &:= \frac{1}{8\pi G} \int \nabla_a U \nabla^a U\, dV, \\
\end{align}
\end{subequations}
are respectively the kinetic energy in center-of-mass motion (${\cal T}_{\rm cm}$), the kinetic energy in director motion (${\cal T}_{\rm dir}$), the isotropic contribution to the internal energy ($E_{\rm iso}$), the anisotropic contribution ($E_{\rm aniso}$), the interaction energy between the fluid and the gravitational field ($E_{\rm inter}$), and the gravitational-field energy ($E_{\rm field}$). 

Familiar manipulations allow us to compute the rate of change of these quantities. We obtain
\begin{subequations}
\label{Edot_pieces} 
\begin{align}
\frac{d}{dt} {\cal T}_{\rm cm} &= \int_\VV \bigl( S^{ab} \nabla_a v_b + \rho v^a\nabla_a U \bigr)\, dV, \\
\frac{d}{dt} {\cal T}_{\rm dir} &= -\int_\VV \kappa \nabla_a c_b \nabla^a w^b\, dV, \\
\frac{d}{dt} E_{\rm iso} &= -\int_\VV p \nabla_a v^a\, dV, \\
\frac{d}{dt} E_{\rm aniso} &= \int_\VV \bigl( -\tfrac{1}{2} \lambda \Xi\, \nabla_a v^a
- \kappa \nabla_a c_c \nabla_b c^c\, \nabla^a v^b + \kappa \nabla_a c_b \nabla^a w^b \bigr)\, dV, \\
\frac{d}{dt} (E_{\rm inter} + E_{\rm field}) &= -\int_\VV \rho v^a \nabla_a U\, dV.
\end{align}
\end{subequations}
All these results are based on the integral identity of Eq.~(\ref{identity}), in which the tensor $j^{a\cdots}_{b\cdots}$ is taken to be a scalar $j$. For ${\cal T}_{\rm cm}$ we apply the identity to $j = \frac{1}{2} v^2$, insert Eq.~(\ref{fluid3}), integrate by parts, and eliminate the surface term. For ${\cal T}_{\rm dir}$ we let $j = \frac{1}{2} w^2$ and use Eq.~(\ref{fluid4}). For $E_{\rm iso}$ we make use of Eq.~(\ref{d_epsilon}) to calculate $d\hat{\varepsilon}/dt = (d\hat{\varepsilon}/d\rho)(d\rho/dt)$, and rewrite Eq.~(\ref{continuity}) in the form $d\rho/dt = -\rho \nabla_a v^a$. For $E_{\rm aniso}$ we let $j = \hat{\omega} = \frac{1}{2} \hat{\kappa} \Xi$, and use Eq.~(\ref{lambda_def}) to compute $d\hat{\kappa}/dt$. The calculation of $d\Xi/dt$ proceeds as
\begin{align}
\frac{1}{2} \frac{d\Xi}{dt} &= \frac{1}{2} \bigl( \partial_t \Xi + v^c \nabla_c \Xi \bigr)
\nonumber \\
&= \nabla_a c_b \bigl( \nabla^a \partial_t c^b + v^c \nabla_c^{\ a} c^b \bigr)
\nonumber \\
&= \nabla_a c_b \bigl[ \nabla^a (w^b - v^c \nabla_c c^b) + v^c \nabla_c^{\ a} c^b \bigr]
\nonumber \\
&= \nabla_a c_b \bigl( \nabla^a w^b - \nabla^a v^c \nabla_c c^b \bigr); 
\end{align}
in the last step we cancelled out the terms featuring second derivatives of $c^b$, taking advantage of the fact that the Riemann tensor vanishes in flat spacetime. Finally, for $E_{\rm inter} + E_{\rm field}$ we insert $dU/dt = \partial_t U  + v^a\nabla_a U$ inside the first integral, integrate the second integral by parts, eliminate the surface term at infinity, use Poisson's equation $\nabla^2 U = -4\pi G \rho$, and observe the cancellation of terms involving $\rho \partial_t U$.

With these results established, we return to Eq.~(\ref{E_total}), which we differentiate with respect to time. We insert Eqs.~(\ref{Edot_pieces}) and substitute the expression of Eq.~(\ref{S_def}) for the stress tensor. We arrive at
\begin{equation}
\frac{dE}{dt} = 0,
\end{equation}
the statement of energy conservation. 

\section{Surface fluid} 
\label{sec:interface} 

In Sec.~IV of paper I \cite{cadogan-poisson:24a} we explained the need for a phase transition in our anisotropic stallar models, from an anisotropic phase at high densities to an isotropic phase at low densities. The phase transition takes place on a two-dimensional surface, and it is mediated by an interface fluid that is itself anisotropic. Here we prepare the way for a discussion of this phase transition in Sec.~\ref{sec:twophase} by formulating mechanical laws for a surface fluid in isolation. We begin in Sec.~\ref{subsec:kinematics} with a description of the kinematics of a surface fluid, we derive a local statement of mass conservation in Sec.~\ref{subsec:mass}, and introduce a Lagrangian formalism of fluid perturbations in Sec.~\ref{subsec:perturbation}. We proceed in Sec.~\ref{subsec:dynamics1} with the specification of an action functional for the surface fluid, and vary this action to obtain the fluid's equations of motion. 

The approach adopted here follows Grinfeld \cite{grinfeld:09} fairly closely. The mathematical developments rely on a differential geometry of moving surfaces, which summarized in Appendix~\ref{sec:moving_surface}. 

\subsection{Kinematics of a surface fluid}
\label{subsec:kinematics}

We examine a fluid distribution confined to a two-dimensional surface $\SSS(t)$ embedded in a three-dimensional, Euclidean space. The surface is described in arbitrary coordinates $x^a$ by the parametric equations $x^a = X^a(t,\theta^A)$, in which $\theta^A$ are coordinates intrinsic to $\SSS(t)$. We let $e^a_A := \partial_A X^a$ be tangent vectors on $\SSS(t)$, and $n^a$ is the unit normal --- refer to Sec.~\ref{subsec:description} for additional details. 

A given fluid element on the surface moves with $\theta^A = \uptheta^A(t)$, and its motion in the ambiant space is described by
\begin{equation}
x^a = X^a\bigl(t, \theta^A = \uptheta^A(t) \bigr).
\end{equation}
The velocity of the fluid element is then $v^a = \partial_t X^a(t,\theta) + \dot{\uptheta}^A \partial_A X^a$, or
\begin{equation}
v^a = W^a + \dot{\uptheta}^A\, e^a_A
= W_n\, n^a + \bigl( W^A + \dot{\uptheta}^A \bigr) e^a_A,
\label{velocity} 
\end{equation}
where $W^a := \partial_t X^a$ is the grid velocity [see Eq.~(\ref{grid_velocity})], which is decomposed into normal and tangential components [see Eq.~(\ref{Wa_decomp})]; an overdot indicates differentiation with respect to $t$. From Eq.~(\ref{velocity}) we find that the normal component of the fluid velocity is
\begin{equation}
v_n = W_n,
\end{equation} 
and that the tangential components are
\begin{equation} 
v^A = W^A + \dot{\uptheta}^A.
\end{equation} 

Under a reparametrization of the surface (Sec.~\ref{subsec:reparametrization}), $\psi^M = \Psi^M(t,\theta^A)$, the description of the fluid's motion changes to $\psi^M := \uppsi^M(t)$, with 
\begin{equation} 
\dot{\uppsi}^M = \partial_t \Psi^M + \dot{\uptheta}^A \partial_A \Psi^M.
\label{psi_dot}
\end{equation}
Conversely, the inverse transformation $\theta^A = \Theta^A(t,\psi^M)$ turns a motion described by $\psi^M = \uppsi^M(t)$ to $\theta^A = \uptheta^A(t)$, with 
\begin{equation}
\dot{\uptheta}^A = \partial_t \Theta^A + \dot{\uppsi}^M \partial_M \Theta^A.
\label{theta_dot} 
\end{equation}
The fluid velocity is
\begin{equation}
v^a = W_n n^a + \bigl( W^M + \dot{\uppsi}^M \bigr) e^a_M
\end{equation}
when it is expressed in terms of the new surface coordinates $\psi^M$. The normal component is still $v_n = W_n$, and the new tangential components are $v^M = W^M + \dot{\uppsi}^M$. 

The normal component of the fluid velocity, equal to $W_n$, transforms as a surface scalar [see Eq.~(\ref{Wn_transf})],
\begin{equation}
v_n(t,\psi) = v_n(t,\theta).
\end{equation} 
The tangential components of the fluid velocity transform as a surface vector,
\begin{equation}
v^M(t,\psi) = v^A(t,\theta)\, \partial_A \Psi^M, \qquad
v^A(t,\theta) = v^M(t,\psi)\, \partial_M \Theta^A.
\end{equation}
To arrive at this conclusion we begin with the expression for $v^M$, insert Eqs.~(\ref{psi_dot}) and (\ref{WA_transf}), and use Eq.~(\ref{diff_identities}) to simplify the result. It follows from this that the fluid velocity $v^a$ forms a set of surface scalars:
\begin{equation}
v^a(t,\psi) = v^a(t,\theta).
\end{equation}
The situation is very different for the grid velocity (Sec.~\ref{subsec:reparametrization}): its tangential components $W^A$ {\it do not} transform as a surface vector, and $W^a$ {\it are not} surface scalars. 

We may always choose $\theta^A$ to be a system of {\it Lagrangian coordinates}, such that the motion of a fluid element on $\SSS(t)$ is described by $\theta^A = \mbox{constant}$. With this choice the fluid velocity $v^a$ coincides with the grid velocity $W^a$. We shall adopt Lagrangian coordinates in most of the following developments.

\subsection{Conservation of mass}
\label{subsec:mass} 

Let $\UU(t)$ be a {\it material patch} of the surface $\SSS(t)$. By this we mean a portion of the surface that moves with the fluid, in the sense that fluid elements within $\UU(t)$ stay within the patch as the fluid moves, and fluid elements outside the patch stay outside; fluid elements do not cross the boundary. The patch $\UU(t)$ corresponds to a time-independent domain $D$ of a system $\theta^A$ of Lagrangian coordinates.

The fluid within $\UU(t)$ has a mass
\begin{equation}
M = \int_{\UU(t)} \sigma\, dS = \int_D \sigma(t,\theta) \sqrt{\Omega}\, d^2\theta,
\end{equation}
where $\sigma$ is the mass density and $\sqrt{\Omega}\, d^2\theta$ is the element of surface area in Lagrangian coordinates (with $\Omega := \mbox{det}[\Omega_{AB}]$, $\Omega_{AB}$ denoting the induced metric on the surface). Differentiating this with respect to $t$ and making use of Eq.~(\ref{dFdt}), we obtain
\begin{equation}
\frac{dM}{dt} = \int_D \Bigl[ \partial_t \sigma + \sigma \bigl( v_n K + D_A v^A \bigr) \Bigr] \sqrt{\Omega}\, d^2\theta,
\end{equation}
where $K := \Omega^{AB} K_{AB}$ is the trace of the extrinsic curvature on $\SSS(t)$, and $D_A$ the covariant-derivative operator compatible with the induced metric (refer to Sec.~\ref{subsec:description}). 

Because $M$ is conserved for any material patch $\UU(t)$, the surface density must satisfy
\begin{equation}
\partial_t \sigma + \sigma \bigl( v_n K + D_A v^A \bigr) = 0.
\label{continuity1}
\end{equation}
The equation is formulated in Lagrangian coordinates, and the time derivative in $\partial_t \sigma(t,\theta)$ is carried out at constant $\theta^A$, that is, along the trajectory of a given fluid element; this is the familiar {\it material derivative} of fluid mechanics.

The quantity $\partial_t \sigma$ does {\it not} transform as a surface scalar [see Eq.~(\ref{partial_scalar})], and Eq.~(\ref{continuity1}) is therefore not invariant under a change of surface coordinates. To remedy this we appeal to the Hadamard time derivative (Sec.~\ref{subsec:hadamard}), $\D_t \sigma := \partial_t \sigma - v^A \partial_A \sigma$, which does transform as a surface scalar. (The Hadamard derivative is actually defined in terms of the grid velocity $W^A$. But we recall that the grid velocity coincides with the fluid velocity in Lagrangian coordinates). Equation (\ref{continuity1}) becomes
\begin{equation}
\D_t \sigma + D_A(\sigma v^A) + \sigma v_n K = 0.
\label{continuity2}
\end{equation}
This is now a scalar equation that can be formulated in any system $\psi^M$ of surface coordinates. In this generic formulation, $\D_t := \partial_t - W^M \partial_M$ would now involve the grid velocity $W^M$, and $v^M = W^M + \dot{\uppsi}^M$ is decomposed into grid and intrinsic velocities; we still have that $v_n = W_n$.

\subsection{Fluid perturbation}
\label{subsec:perturbation}

In this section we extend the kinematics of a surface fluid to a perturbed configuration. We introduce a surface version of the Lagrangian displacement vector, as well as define and compute the Lagrangian variation of various fluid quantities. These constructions will allow us, in Sec.~\ref{subsec:dynamics1}, to formulate a variational principle for the surface fluid.  

\subsubsection{Lagrangian displacement}

We examine a one-parameter family of fluid configurations on a one-parameter family of surfaces $\SSS(t;\epsilon)$, with $\epsilon$ standing for the parameter; the unperturbed configuration corresponds to $\epsilon = 0$. We continue to work with Lagrangian coordinates $\theta^A$, and imagine that a given fluid element keeps its label $\theta^A$ as $\epsilon$ increases away from zero. For each value of $\epsilon$ the motion of the fluid element is given by
\begin{equation}
x^a = X^a(t,\theta^A;\epsilon),
\end{equation}
with $\theta^A = \mbox{constant}$. 

The perturbed surface $\SSS(t;\epsilon)$ comes with a full set of geometric quantities. We have the tangent vectors $e^a_A(\epsilon)$, the normal vector $n^a(\epsilon)$, the induced metric $\Omega_{AB}(\epsilon)$, the extrinsic curvature $K_{AB}(\epsilon)$, and so on. These are defined in the usual way from the embedding relations $X^a(t,\theta;\epsilon)$.

The surface {\it Lagrangian displacement} vector $\xi^a$ is defined by
\begin{equation}
\xi^a(t,\theta) := X^a(t,\theta;\epsilon) - X^a(t,\theta;0)
= \epsilon \partial_\epsilon X^a; 
\label{xi_def} 
\end{equation}
the derivative with respect to $\epsilon$ is evaluated at $\epsilon = 0$. The displacement vector takes a fluid element at position $\theta^A$ on the unperturbed interface $\SSS(t;0)$ and places it at the same position $\theta^A$ on the perturbed surface $\SSS(t;\epsilon)$. The vector can be decomposed according to
\begin{equation}
\xi^a(t,\theta) = \xi_n n^a(0) + \xi^A e^a_A(0),
\label{xi_decomp}
\end{equation}
in terms of basis vectors defined on the unperturbed surface $\SSS(t;0)$.

\subsubsection{Lagrangian and Eulerian variations}

In the standard perturbation theory of a three-dimensional fluid, as formulated by Friedman and Schutz \cite{friedman-schutz:78a}, one defines the Eulerian variation of a tensor $Q^{a\cdots}_{b\cdots}$ by comparing the tensor at the same spatial position in the perturbed and unperturbed configurations. The definition is
\begin{equation}
\delta Q^{a\cdots}_{b\cdots} := Q^{a\cdots}_{b\cdots}(t,x;\epsilon)
- Q^{a\cdots}_{b\cdots}(t,x;0).
\label{euler_bulk}
\end{equation}
One also defines the Lagrangian variation of the tensor by comparing it at the same fluid element in the perturbed and unperturbed configurations. The tensor components are given in a frame attached to Lagrangian coordinates $y^j$, and the definition is
\begin{equation}
\Delta Q^{j\cdots}_{k\cdots} := Q^{j\cdots}_{k\cdots}(t,y;\epsilon)
- Q^{j\cdots}_{k\cdots}(t,y;0).
\label{lagrange_bulk} 
\end{equation}
The Lagrangian and Eulerian variations are related by
\begin{equation} 
\Delta Q^{a\cdots}_{b\cdots} = \delta Q^{a\cdots}_{b\cdots} + \Lie_\xi Q^{a\cdots}_{b\cdots},
\label{EvsL_bulk} 
\end{equation}
where $\Lie_\xi$ denotes the Lie derivative in the direction of the Lagrangian displacement vector. 

The definitions of Eqs.~(\ref{euler_bulk}) and (\ref{lagrange_bulk}) cannot be applied directly (or at all) to a fluid confined to a two-dimensional surface. In the case of the Eulerian variation, Eq.~(\ref{euler_bulk}) cannot be applied when the tensor $Q^{a\cdots}_{b\cdots}$ is constructed from fluid variables, because it is then defined on the surface $\SSS(t;\epsilon)$ only. It is not defined beyond the surface, and because the surface alters its position as $\epsilon$ increases, it is impossible to provide a comparison at the same spatial position. There is, therefore, no meaningful notion of an Eulerian variation in such circumstances. A comparison becomes possible, however, when the tensor is defined globally in the ambiant space, and in such cases we can readily apply Eq.~(\ref{euler_bulk}). Examples of such tensors are the gravitational field $g_a = \nabla_a U$ and the metric $g_{ab}$.  

In the case of the Lagrangian variation, Eq.~(\ref{lagrange_bulk}) can be applied directly when the tensor is intrinsic to the surface and of the form $Q^{A\cdots}_{B\cdots}$. We then have, by definition
\begin{equation}
\Delta Q^{A\cdots}_{B\cdots}(t,\theta) := Q^{A\cdots}_{B\cdots}(t,\theta;\epsilon)
- Q^{A\cdots}_{B\cdots}(t,\theta;0) = \epsilon \partial_\epsilon Q^{A\cdots}_{B\cdots},
\label{lagrange_inter1}
\end{equation}
where $\theta^A$ are Lagrangian coordinates on $\SSS(t;\epsilon)$. In the case of an ambiant tensor of the form $Q^{a\cdots}_{b\cdots}$, however, Eq.~(\ref{lagrange_bulk}) cannot be applied, because the system $\theta^A$ is two-dimensional and the transformation $x^a = X^a(t,\theta;\epsilon)$ cannot be inverted uniquely. We must therefore alter the definition for such tensors. 

We first examine the case of a vector $p^a$. To define its Lagrangian variation we construct the geometric vector $\bm{p} = p^a \bm{e}_a$, where $\bm{e}_a$ are the basis vectors associated with the ambiant coordinates $x^a$. We then let
\begin{equation}
\Delta \bm{p} := \bm{p}(t,\theta;\epsilon) - \bm{p}(t,\theta;0),
\end{equation}
where $\bm{p}(t\,\theta;\epsilon)$ is the geometric vector evaluated at position $\theta^A$ on $\SSS(t;\epsilon)$ for the perturbed fluid configuration, while $\bm{p}(t,\theta;0)$ is the vector evaluated at the same position $\theta^A$ on $\SSS(t;0)$ for the unperturbed configuration. Because $\theta^A$ are Lagrangian coordinates, the comparison is carried out at the same fluid element. 

We calculate this as
\begin{equation}
\Delta \bm{p} = p^a(t,\theta;\epsilon)\, \bm{e}_a \bigr|_{x = X(t,\theta;\epsilon)}
- p^a(t,\theta;0)\, \bm{e}_a \bigr|_{x = X(t,\theta;0)},
\end{equation}
and to account for the change in the basis vectors as they are moved from $X(t,\theta;\epsilon)$ to $X(t,\theta;0)$, we write
\begin{equation}
\bm{e}_a \bigr|_{x = X(t,\theta;\epsilon)}
= \bm{e}_a \bigr|_{x = X(t,\theta;0) + \xi} 
= \bm{e}_a \bigr|_{x = X(t,\theta;0)} + \xi^b \partial_b \bm{e}_a
= \bm{e}_a \bigr|_{x = X(t,\theta;0)} + \Gamma^c_{ba} \xi^b\, \bm{e}_c, 
\end{equation}
where $\Gamma^c_{ba}$ are Christoffel symbols associated with the metric $g_{ab}$. We arrive at
\begin{equation} 
\Delta \bm{p} = \Delta p^a\, \bm{e}_a \bigr|_{x = X(t,\theta;0)}
\end{equation}
with
\begin{equation}
\Delta p^a := p^a(t,\theta;\epsilon) - p^a(t,\theta;0) + \Gamma^a_{bc} \xi^b p^c
= \epsilon \partial_\epsilon p^a + \Gamma^a_{bc} \xi^b p^c.
\label{Delta_vector}
\end{equation}
Because $\Delta \bm{p}$ is a geometric vector, its components $\Delta p^a$ transform as they should under a change of ambiant coordinates. 

When $\bm{p}$ is defined globally in the ambiant space, we have that
\begin{equation}
p^a(t,\theta;\epsilon) = p^a\bigl(x=X(t,\theta;\epsilon);\epsilon \bigr)
= p^a\bigl(x=X(t,\theta;0) + \xi;\epsilon \bigr)
= p^a(x;\epsilon) + \xi^b \partial_b p^a,
\end{equation}
while $p^a(t,\theta;\epsilon) = p^a(x;0)$. It follows that
\begin{equation}
p^a(t,\theta;\epsilon) - p^a(t,\theta;0) = \delta p^a + \xi^b \partial_b p^a,
\end{equation}
where $\delta p^a := p^a(x;\epsilon) - p^a(x;0)$ is the Eulerian variation of Eq.~(\ref{euler_bulk}), with $x$ standing for the point on $\SSS(t;0)$ with coordinates $x^a = X^a(t,\theta;0)$. Combining this with Eq.~(\ref{Delta_vector}), we find that the relation between Lagrangian and Eulerian variations is given by
\begin{equation}
\Delta p^a = \delta p^a + \xi^b \nabla_b p^a.
\label{LvsE_inter}
\end{equation}
We emphasize that this relation is meaningful only when $p^a$ is defined globally, both on and off the surface. Equation (\ref{LvsE_inter}) is to be contrasted with Eq.~(\ref{EvsL_bulk}), which is obtained under a different definition of the Lagrangian variation.  

Similar considerations can be applied to a covector $q_a$. We begin with the geometric object $\bm{q} = q_a \bm{\omega}^a$, where $\bm{\omega}^a$ are basis covectors, and we define its Lagrangian variation to be
\begin{equation}
\Delta \bm{q} := \bm{q}(t,\theta;\epsilon) - \bm{q}(t,\theta;0),
\end{equation}
where $\bm{q}(t\,\theta;\epsilon)$ is the covector evaluated at position $\theta^A$ on $\SSS(t;\epsilon)$ for the perturbed fluid configuration, while $\bm{q}(t,\theta;0)$ is the covector evaluated at the same position $\theta^A$ on $\SSS(t;0)$ for the unperturbed configuration. A calculation similar to the one carried out previously allows us to express this as
\begin{equation}
\Delta \bm{q} = \Delta q_a\, \bm{\omega}^a \bigr|_{x = X(t,\theta;0)}
\end{equation}
with
\begin{equation}
\Delta q_a := q_a(t,\theta;\epsilon) -q_a(t,\theta;0) - \Gamma^c_{ba} \xi^b q_c
= \epsilon \partial_\epsilon q_a - \Gamma^c_{ba} \xi^b q_c.
\label{Delta_covector}
\end{equation}
When $q_a$ is defined globally in the ambiant space, we have that
\begin{equation}
\Delta q_a = \delta q_a + \xi^b \nabla_b q_a,
\end{equation}
where $\delta q_a := q_a(x;\epsilon) - q_a(x;0)$ is the Eulerian variation.

The rules generalize to an arbitrary tensor $Q^{a\cdots}_{b\cdots}$; we have that
\begin{equation}
\Delta Q^{a\cdots}_{b\cdots} = Q^{a\cdots}_{b\cdots}(t,\theta;\epsilon)
- Q^{a\cdots}_{b\cdots}(t,\theta;0) + \Gamma^a_{dc} \xi^d Q^{c\cdots}_{b\cdots} + \cdots
- \Gamma^c_{db} \xi^d Q^{a\cdots}_{c\cdots} - \cdots,
\end{equation}
with a Christoffel-symbol term inserted for each tensor index. When the tensor is defined globally, we also have that
\begin{equation}
\Delta Q^{a\cdots}_{b\cdots} = \delta Q^{a\cdots}_{b\cdots} + \xi^c \nabla_c Q^{a\cdots}_{b\cdots}.
\end{equation}
This is again to be contrasted with Eq.~(\ref{EvsL_bulk}), which is obtained under a different definition of the Lagrangian variation.  

An example of a tensor defined globally is the metric $g_{ab}$. According to our definition, its Lagrangian variation on $\SSS(t;0)$ is
\begin{equation}
\Delta g_{ab} = 0.
\label{Dg}
\end{equation}
This follows because the metric is independent of $\epsilon$, so that $\delta g_{ab} = 0$, and because its covariant derivative vanishes. 

\subsubsection{Velocity perturbation}

The perturbed velocity field is
\begin{equation}
v^a(t,\theta;\epsilon) = \partial_t X^a(t,\theta;\epsilon)
= \partial_t \bigl[ X^a(t,\theta;0) + \xi^a(t,\theta) \bigr]
= v^a(t,\theta;0) + \partial_t \xi^a(t,\theta),
\end{equation}
and according to Eq.~(\ref{Delta_vector}), its Lagrangian variation is
\begin{equation}
\Delta v^a(t,\theta) = \partial_t \xi^a(t,\theta) + \Gamma^a_{bc} \xi^b v^c.
\label{Dv_def}
\end{equation}
This can be decomposed according to
\begin{equation}
\Delta v^a = \Delta v_n\, n^a(0) + \Delta v^A\, e^a_A(0),
\label{Dv_decomp}
\end{equation}
in terms of the basis vectors defined on $\SSS(t;0)$. It is useful to note that $\Delta v_n$ is {\it not equal} to $v_n(\epsilon) - v_n(0)$, and that $\Delta v^A$ is {\it not equal} to $v^A(\epsilon) - v^A(0)$. The reason is that the decomposition of Eq.~(\ref{Dv_decomp}) features the unperturbed basis vectors, while a decomposition of $v^a(\epsilon)$ into components $v_n(\epsilon)$ and $v^A(\epsilon)$ would involve the vectors $n^a(\epsilon)$ and $e^a_A(\epsilon)$.

To compute $\Delta v^a$ we begin with Eq.~(\ref{xi_decomp}), which we differentiate with respect to $t$ at constant $\theta^A$. For $\partial_t n^a$ we substitute Eq.~(\ref{partial_n1}), and for $\partial_t e^a_A$ we invoke Eq.~(\ref{partial_eaA}); in these equations we insert $W^a = v^a$, because the fluid velocity coincides with the grid velocity when we work with Lagrangian coordinates. We arrive at
\begin{equation}
\Delta v^a = \bigl( \partial_t \xi_n + \xi^A \partial_A v_n - K_{AB} \xi^A v^B \bigr) n^a
+ \bigl[ \partial_t \xi^A - \xi_n D^A v_n + \xi^B D_B v^A
+ (\xi_n v^B + v_n \xi^B) K^A_{\ B} \bigr] e^a_A
\label{Dv_result}
\end{equation}
after simplification. The projections are
\begin{subequations}
\label{Dv_components} 
\begin{align} 
\Delta v_n &= \partial_t \xi_n + \xi^A \partial_A v_n - K_{AB} \xi^A v^B,  
\label{Dv_normal} \\ 
\Delta v^A &= \partial_t \xi^A - \xi_n D^A v_n + \xi^B D_B v^A
+ (\xi_n v^B + v_n \xi^B) K^A_{\ B}.
\label{Dv_tangent}
\end{align} 
\end{subequations}
On the right-hand side of Eqs.~(\ref{Dv_result}) and (\ref{Dv_components}), all geometric quantities refer to the unperturbed surface $\SSS(t;0)$. 

\subsubsection{Perturbation of the induced metric}

We wish to compute
\begin{equation}
\Delta \Omega_{AB} := \Omega_{AB}(t,\theta;\epsilon) - \Omega_{AB}(t,\theta;0)
= \epsilon \partial_\epsilon \Omega_{AB},
\end{equation}
the Lagrangian variation of the induced metric on the interface. To simplify the manipulations we take the ambiant coordinates $x^a$ to be Cartesian; there is no loss of generality implied in this choice.

The induced metric on $\SSS(t;\epsilon)$ is
\begin{equation}
\Omega_{AB}(t,\theta;\epsilon)
= \delta_{ab}\, \partial_A X^a(t,\theta;\epsilon)\, \partial_B X^b(t,\theta;\epsilon),
\end{equation}
and differentiation with respect to $\epsilon$ produces
\begin{equation}
\Delta \Omega_{AB} = \delta_{ab} \bigl( \partial_A \xi^a\, e^b_B
+ e^a_A\, \partial_B \xi^b \bigr);
\end{equation}
we made use of Eq.~(\ref{xi_def}), and it is understood that here, $e^a_A \equiv e^a_A(\epsilon = 0)$. To calculate $\partial_A \xi^a$ we begin with Eq.~(\ref{xi_decomp}), which we differentiate with respect to $\theta^A$. For $\partial_A n^a$ we insert Eq.~(\ref{partialA_na1}), and for $D_A e^a_C$ we use Eq.~(\ref{GW2}); in these equations we set $\Gamma^a_{bc} = 0$ by virtue of our choice of ambiant coordinates. We arrive at
\begin{equation}
\Delta \Omega_{AB} = 2\xi_n K_{AB} + D_A \xi_B + D_B \xi_A,
\label{DOmega1}
\end{equation} 
in which all geometric quantities refer to $\SSS(t;0)$. It follows from Eq.~(\ref{DOmega1}) that
\begin{equation}
\frac{1}{\sqrt{\Omega}} \Delta \sqrt{\Omega} = \frac{1}{2} \Omega^{AB} \Delta \Omega_{AB}
= \xi_n K + D_A \xi^A,
\label{DOmega2}
\end{equation}
where $K := \Omega^{AB} K_{AB}$.

\subsubsection{Perturbation of an integrated quantity}

Let $\UU(t;\epsilon)$ be a material patch on the family $\SSS(t;\epsilon)$ of surfaces. Still working in Lagrangian coordinates $\theta^A$, the patch corresponds to a domain $D$ that is independent of both $t$ and $\epsilon$. The integral of a scalar $j$ over the patch is
\begin{equation}
J(t;\epsilon) := \int_{\UU(t;\epsilon)} j\, dS
= \int_D j(t,\theta;\epsilon) \sqrt{\Omega(\epsilon)}\, d^2\theta.
\end{equation}
The perturbation of this quantity is $\delta J := J(t;\epsilon) - J(t;0)$, and it can be expressed as
\begin{equation}
\delta J = \int_D \Delta \bigl( j \sqrt{\Omega} \bigr)\, d^2\theta.
\end{equation}
Inserting Eq.~(\ref{DOmega2}), we arrive at
\begin{equation}
\delta J = \int_{\UU(t;0)} \Bigl[ \Delta j + j \bigl( \xi_n K + D_A \xi^A \bigr) \Bigr]\, dS.
\label{delta_J1}
\end{equation}
The result applies to any scalar field, and to any material patch on the surface. 

The mass of any material patch is necessarily preserved during the perturbation. This implies that
\begin{equation}
\Delta \sigma + \sigma \bigl( \xi_n K + D_A \xi^A \bigr) = 0, 
\label{Dsigma}
\end{equation}
and this equation allows us to express $\Delta \sigma$ in terms of the Lagrangian displacement.

When the integrated quantity is of the form
\begin{equation}
J(t;\epsilon) := \int_{\UU(t;\epsilon)} \sigma n\, dS
\end{equation}
for some scalar $n$, the perturbation is given by
\begin{equation} 
\delta J = \int_{\UU(t;0)} \sigma \Delta n\, dS. 
\label{delta_J2}
\end{equation}
This follows by virtue of Eq.~(\ref{delta_J1}) with $j = \sigma n$, as well as Eq.~(\ref{Dsigma}). 

\subsection{Dynamics of a surface fluid} 
\label{subsec:dynamics1} 

In this subsection we apply the kinematics developed previously to formulate a variational principle for an anisotropic fluid confined to a two-dimensional surface $\SSS(t)$. Here we take the surface fluid to be isolated; it is surrounded by vacuum. We shall consider the coupled dynamics of bulk and interface fluids in Sec.~\ref{sec:twophase}. In our developments below we denote the surface's boundary by the closed curve $\C(t)$. 

Adapting Eq.~(\ref{Lag1}) to a two-dimensional setting, we assign the Lagrangian
\begin{equation}
L_{\rm surface} = \int_{\SSS(t)} \sigma \LL_{\rm surface}\, dS, \qquad
\LL_{\rm surface} := \tfrac{1}{2} v^2 - \hat{e} - \hat{\chi} + U
\label{Lag_surface} 
\end{equation}
to the surface fluid, where $\sigma$ is the mass density, $v^2 := g_{ab} v^a v^b$ with $v^a$ denoting the fluid velocity, $\hat{e}$ is the isotropic contribution to the internal energy per unit mass, and 
\begin{equation}
\hat{\chi} := \tfrac{1}{2} \hat{\mu} c^2, \qquad c^2 := g_{ab} c^a c^b
\end{equation}
is the anisotropic contribution, assumed to be proportional to $c_a c^a$, with $c^a$ denoting the director vector and $\frac{1}{2} \hat{\mu}$ standing for the factor of proportionality. For simplicity we take $\hat{e}$ and $\hat{\mu}$ to depend on $\sigma$ only (they could also depend on specific entropy), and define the thermodynamic derivatives 
\begin{equation}
\eta := -\sigma^2 \frac{d \hat{e}}{d \sigma}, \qquad
\tau := -\sigma^2 \frac{d \hat{\mu}}{d \sigma}. 
\label{eta_def}
\end{equation}
In the case of an isotropic fluid, $\eta$ would be recognized as the surface tension; $\tau$ is a new quantity associated with the anisotropic contribution to the internal energy. The action functional for the surface fluid is $S_{\rm surface} = \int L_{\rm surface}\, dt$, with a domain of integration bounded by the arbitrary times $t_1$ and $t_2$.

In addition to its two-dimensional nature, the surface Lagrangian of Eq.~(\ref{Lag_surface}) has two features that distinguishes it from the bulk Lagrangian of Eq.~(\ref{Lag1}). The first is that it does not include a kinetic term $\frac{1}{2} w^2$ associated with the director velocity $w^a$. The second is that $\hat{\chi}$, the anisotropic contribution to the internal energy, is quadratic in $c^a$ instead of $\nabla_a c^b$. These choices are motivated by economy and simplicity, and reflect a notion that the surface Lagrangian describes an effective theory expressed as a derivative expansion; the expansion is truncated at zeroth order, and it neglects all derivatives of the director vector. These choices will serve us well in Sec.~\ref{sec:twophase}, when we join the surface fluid with two phases of an anisotropic bulk fluid. 

We wish to calculate the change in $L_{\rm surface}$ that results from a change of fluid configuration. This is described in part by $\xi^a$, the Lagrangian displacement vector, and in part by $\Delta c^a$, the Lagrangian variation of the director vector. The variation is constrained by the statement of mass conservation, so that $\Delta \sigma$ shall be given by Eq.~(\ref{Dsigma}). We shall also put $\delta U = 0$; the independent variation of the surface action with respect to the gravitational potential is carried out in Appendix~\ref{sec:interface_gravity}. All calculations below are performed in Lagrangian coordinates; the final statement of the fluid equations, however, will be given in an arbitrary system of surface coordinates.  

According to Eq.~(\ref{delta_J2}) the variation of the Lagrangian is given by
\begin{equation}
\delta L_{\rm surface} = \int_{\SSS(t)} \sigma \Delta \LL_{\rm surface}\, dS,
\end{equation}
where 
\begin{equation}
\Delta \LL_{\rm surface} = \Delta (\tfrac{1}{2} v^2) - \Delta \hat{e} - \Delta \hat{\chi} + \Delta U.
\end{equation}
To compute $\Delta (\tfrac{1}{2} v^2)$ we combine Eqs.~(\ref{Dv_result}) and (\ref{Dg}), and obtain
\begin{equation}
\Delta (\tfrac{1}{2} v^2) = v_n \partial_t \xi_n + v_A \partial_t \xi^A
+ \xi_n \bigl( -v^A D_A v_n + K_{AB} v^A v^B \bigr)
+ \xi^A \bigl( v_n D_A v_n + v_B D_A v^B \bigr).
\end{equation}
To compute $\Delta \hat{e}$ we invoke Eq.~(\ref{eta_def}) and insert Eq.~(\ref{Dsigma}); we get 
\begin{equation}
\Delta \hat{e} = \frac{\eta}{\sigma}\bigl( \xi_n K + D_A \xi^A \bigr).
\end{equation}
A similar computation returns $\Delta \hat{\mu}$, and this gives us
\begin{equation}
\Delta \hat{\chi} = \frac{\tau c^2}{2\sigma} \bigl( \xi_n K + D_A \xi^A \bigr) + \hat{\mu} c_a \Delta c^a. 
\end{equation}
With $\delta U$ set to zero, the discussion of Appendix~\ref{sec:interface_gravity} --- refer to Eq.~(\ref{DeltaU_interface}) --- justifies the relation 
\begin{equation}
\Delta U = \xi_n \langle n^a \nabla_a U \rangle + \xi^A D_A U,
\end{equation}
where $\langle n^a \nabla_a U \rangle$ is the arithmetic average of the (discontinuous) normal derivative of the gravitational potential on each side of the surface.  

We make the substitutions in $\Delta \LL_{\rm surface}$, rearrange, and arrive at
\begin{align}
\sigma \Delta \LL_{\rm surface} &=
\sigma \partial_t \bigl( v_n \xi_n + v_A \xi^A \bigr)
- D_A\bigl[ (\eta + \tfrac{1}{2} \tau c^2) \xi^A \bigr]
- \sigma \hat{\mu} c_a\, \Delta c^a 
\nonumber \\ & \quad \mbox{}
- \xi_n \Bigl[ \sigma\bigl( \partial_t v_n + v^A D_A v_n - K_{AB} v^A v^B - \langle n^a \partial_a U \rangle \bigr)
+ (\eta + \tfrac{1}{2} \tau c^2) K \Bigr]
\nonumber \\ & \quad \mbox{}
- \xi^A \Bigl[ \sigma (\partial_t v_A - v_n D_A v_n - v_B D_A v^B -D_A U)
- D_A(\eta + \tfrac{1}{2} \tau c^2) \Bigr].
\label{dL1}
\end{align}
Integration over $\SSS(t)$ produces
\begin{align}
\delta L_{\rm surface} &= 
\frac{d}{dt} \int_{\SSS(t)} \sigma \partial_t \bigl( v_n \xi_n + v_A \xi^A \bigr)\, dS 
- \oint_{\C(t)} (\eta + \tfrac{1}{2} \tau c^2) \xi^A\, d\ell_A 
- \int_{\SSS(t)} \sigma \hat{\mu} c_a\, \Delta c^a\, dS  
\nonumber \\ & \quad \mbox{}
- \int_{\SSS(t)} \xi_n \Bigl[ \sigma\bigl( \partial_t v_n + v^A D_A v_n - K_{AB} v^A v^B
- \langle n^a \partial_a U \rangle \bigr) + (\eta + \tfrac{1}{2} \tau c^2) K \Bigr]\, dS
\nonumber \\ & \quad \mbox{}
- \int_{\SSS(t)} \xi^A \Bigl[ \sigma (\partial_t v_A - v_n D_A v_n - v_B D_A v^B -D_A U)
- D_A(\eta + \tfrac{1}{2} \tau c^2) \Bigr]\, dS. 
\label{dL2}
\end{align}
The total time derivative originates from the identity  
\begin{equation}
\int_{\SSS(t)} \sigma \partial_t n\, dS = \frac{d}{dt} \int_{\SSS(t)} \sigma n\, dS, 
\end{equation}
where $n$ is any scalar field on the surface. This is a direct consequence of $\partial_t (\sigma \sqrt{\Omega}) = 0$, which follows from Eqs.~(\ref{continuity1}) and (\ref{partial_Omega2}), with the grid velocity $W^a$ identified with the fluid velocity $v^a$; we also used the fact that in Lagrangian coordinates, the surface $\SSS(t)$ corresponds to a time-independent domain of integration. The integral over $\C(t)$ originates from an application of the two-dimensional version of Gauss's theorem (see Sec.~\ref{subsec:gauss}); $d\ell_A$ is a normal-directed element of length on $\C(t)$.

The variation of the action is $\delta S_{\rm surface} = \int \delta L_{\rm surface}\, dt$, and it is subjected to the conditions that $\xi^a$ and $\Delta c^a$ should vanish everywhere on $\SSS(t)$ at the boundary times $t = t_1$ and $t = t_2$, and that they should vanish on the boundary curve $\C(t)$ at any time $t$; otherwise $\xi^a$ and $\Delta c^a$ are arbitrary. Demanding that $\delta S_{\rm surface} = 0$ for arbitrary and independent $\xi_n$, $\xi_A$, and $\Delta c^a$ returns the dynamical equations
\begin{subequations}
\label{fluid_eqns1} 
\begin{align}
0 &= \sigma \bigl( \partial_t v_n + v^A D_A v_n - K_{AB} v^A v^B
- \langle n^a \partial_a U \rangle \bigr) + (\eta + \tfrac{1}{2} \tau c^2) K, \\
0 &= \sigma \bigl( \partial_t v_A - v_n D_A v_n - v_B D_A v^B -D_A U \bigr) - D_A (\eta + \tfrac{1}{2} \tau c^2),
\label{fluid_eqns1b} \\ 
0 &= \sigma \hat{\mu} c_a.
\label{fluid_eqns1c} 
\end{align}
\end{subequations}
We may use Eq.~(\ref{partial_Omega1}) --- again with $W^a = v^a$ --- to raise the vector index in Eq.~(\ref{fluid_eqns1b}). This gives
\begin{equation}
0 = \sigma \bigl( \partial_t v^A - v_n D^A v_n + v^B D_B v^A + 2 v_n K^A_{\ B} v^B - D^A U \bigr)
- D^A (\eta + \tfrac{1}{2} \tau c^2).
\label{fluid_eqns2}
\end{equation}

Equations (\ref{fluid_eqns1}), together with the mass-conservation equation (\ref{continuity1}) and equations of state $\eta = \eta(\sigma)$, $\tau = \tau(\sigma)$ constitute a complete set of dynamical equations for the surface fluid. These equations are formulated in Lagrangian coordinates, and they are not covariant under a change of surface coordinates. To remedy this we replace the partial time-derivative operator $\partial_t$ (which does not return a tensor when acting on a surface tensor; see Sec.~\ref{subsec:reparametrization}) with the Hadamard derivative $\D_t$ (which does return a tensor; see Sec.~\ref{subsec:hadamard}). The final set of equations shall be
\begin{subequations}
\label{fluid_eqns3} 
\begin{align}
0 &= \sigma \bigl( \D_t v_n + 2 v^A D_A v_n - K_{AB} v^A v^B
- \langle n^a \partial_a U \rangle \bigr) + (\eta + \tfrac{1}{2} \tau c^2) K, \\
0 &= \sigma \bigl( \D_t v^A - v_n D^A v_n + v^B D_B v^A + 2 v_n K^A_{\ B} v^B - D^A U \bigr)
- D^A (\eta + \tfrac{1}{2} \tau c^2), \\
0 &= \sigma \hat{\mu} c_a.  
\end{align}
\end{subequations}
These equations are joined with
\begin{equation}
\D_t \sigma + D_A(\sigma v^A) + \sigma v_n K = 0, 
\label{continuity3}
\end{equation}
the statement of mass conservation. When acting on a scalar $f$, the Hadamard derivative is defined by $\D_t f := \partial_t f - W^A \partial_A f$; when acting on a vector $p^A$, it is defined by $\D_t p^A := \partial_t p^A - W^B \partial_B p^A + p^B \partial_B W^A = \partial_t p^A - W^B D_B p^A + p^B D_B W^A$, where $W^A$ are the tangential components of the grid velocity.

Equations (\ref{fluid_eqns3}) and (\ref{continuity3}) can be formulated in any system $\psi^M$ of surface coordinates. In generic coordinates, the grid and fluid velocities are related by $v_n = W_n$ and $v^M = W^M + \dot{\uppsi}^M$, with $\psi^M = \uppsi^M(t)$ describing the trajectory of a fluid element on $\SSS(t)$. In Lagrangian coordinates $\theta^A$, we have that $\uptheta^A(t) = \mbox{constant}$, and the grid and fluid velocities coincide. Equations (\ref{fluid_eqns3}) and (\ref{continuity3}) agree with Eqs.~(50) of Grinfeld's paper \cite{grinfeld:09}; the author, however, provides no detail of derivation.

The dynamical equation for the director vector, $c^a = 0$, implies that in the theory formulated here, an isolated surface fluid must be isotropic. The conclusion will change when the surface is placed between two distinct phases of an anisotropic fluid.  

\section{Two-phase fluid}
\label{sec:twophase} 

In this section we examine a situation in which a bulk fluid presents two distinct phases, an anisotropic phase occupying a region $V_-$ of three-dimensional space, and an isotropic phase occupying a region $V_+$; the phases are joined at a common boundary $\SSS(t)$, on which there is an interface fluid. (A variation on this theme would have a second anisotropic phase in $V_+$, distinct from the one in $V_-$.) This situation brings the developments of Secs.~\ref{sec:lagrangian}, \ref{sec:variation}, \ref{sec:fluid}, and \ref{sec:interface} under a common setting, and provides a complete foundation for the anisotropic stellar models constructed below in Sec.~\ref{sec:stellar_model}.

We begin in Sec.~\ref{subsec:Lagrangian} with the specification of the Lagrangian for the combined system (bulk and intersurface fluids coupled to a gravitational field). We vary the action in Sec.~\ref{subsec:variation} and extract the fluid equations in Sec.~\ref{subsec:fluid_eqns}, together with the junction conditions satisfied by the bulk variables. Poisson's equation for the gravitational potential is derived in Sec.~\ref{subsec:poisson}, and in Sec.~\ref{subsec:static} the complete system of equations is specialized to a static configuration. 

\subsection{Lagrangian} 
\label{subsec:Lagrangian} 

The complete Lagrangian for this system is
\begin{equation}
L = L_{\rm aniso} + L_{\rm iso} + L_{\rm interface} + L_{\rm field},
\end{equation}
where
\begin{equation}
L_{\rm aniso} = \int_{V_-} \rho \bigl[ \tfrac{1}{2} (v^2 + w^2) - \hat{\varepsilon} - \hat{\omega} + U \bigr]\, dV
\end{equation}
is the Lagrangian of an anisotropic fluid, as previously given by Eq.~(\ref{Lag1}),
\begin{equation}
L_{\rm iso} = \int_{V_+} \rho \bigl( \tfrac{1}{2} v^2 - \hat{\varepsilon} + U \bigr)\, dV
\end{equation}
is the Lagrangian of an isotropic fluid (a special case with $c^a = 0 = w^a$), 
\begin{equation}
L_{\rm interface} = \int_{\SSS(t)} \sigma \bigl( \tfrac{1}{2} v^2 - \hat{e} - \hat{\chi} + U \bigr)\, dS
\end{equation}
is the surface Lagrangian of Eq.~(\ref{Lag_surface}), and
\begin{equation}
L_{\rm field} = -\frac{1}{8\pi G} \int \nabla_a U \nabla^a U\, dV
\end{equation}
is the contribution from the gravitational field. We recall that $\rho$ is the volume density of mass of the bulk fluid, $v^2 := g_{ab} v^a v^b$ with $v^a$ denoting the fluid velocity, $w^2 := g_{ab} w^a w^b$ with $w^a := \partial_t c^a + v^b \nabla_b c^a$ denoting the director velocity, $\hat{\varepsilon}$ is the isotropic contribution to the internal energy per unit mass, and $\hat{\omega} := \frac{1}{2} \hat{\kappa} \Xi$ with $\Xi := \nabla_a c_b \nabla^a c^b$ is the anisotropic contribution. We recall also that $\sigma$ is the surface density of mass of the interface fluid, $\hat{e}$ is the isotropic contribution to the internal energy per unit mass, and $\hat{\chi} := \frac{1}{2} \hat{\mu} c_a c^a$ is the anisotropic contribution. Finally, we recall that $U$ stands for the gravitational potential. It is understood that $\hat{\varepsilon}$ and $\hat{\kappa}$ are functions of $\rho$ only, while $\hat{e}$ and $\hat{\mu}$ are functions of $\sigma$ only. 

We assume that the fluid velocity $v^a$ is shared between the bulk and interface fluids, and is continuous across $\SSS(t)$, so that the velocity of a fluid element in $V_\pm$ agrees with the velocity of an element on $\SSS(t)$ when the element of bulk fluid is made to approach the surface. Continuity of the velocity implies that fluid elements do not cross the interface. We also assume that the director vector $c^a$ is shared between the anisotropic fluid in $V_-$ and the interface fluid, and is continuous on the $V_-$ side of $\SSS(t)$; the vector vanishes everywhere in $V_+$.  

\subsection{Variation of the action}
\label{subsec:variation} 

The complete action for the system is $S = \int L\, dt$, and we wish to vary this action with respect to the fluid configuration. The fluid variation is described by $\xi^a$ and $\delta c^a$ within the anisotropic phase, by
$\xi^a$ within the isotropic phase, and by $\xi^a$ and $\Delta c^a$ on the interface. We assume that $\xi^a$ is continuous across $\SSS(t)$, and that $\Delta c^a$ is continuous on the $V_-$ side of $\SSS(t)$. 
The variation is subjected to the usual constraints that the variation fields vanish everywhere at the reference times $t = t_{1,2}$, and that they vanish at all times on the boundary $\partial V$ of the combined region $V = V_+ + V_-$. This boundary {\it does not} include $\SSS(t)$, on which $\xi^a$ and $\Delta c^a$ shall also be arbitrary.

The variation of $L_{\rm aniso}$ was already carried out in Sec.~\ref{subsec:var_fluid}, and the outcome was displayed in Eq.~(\ref{delta_L_final}). The fluid equations in $V_-$ are then given by Eqs.~(\ref{fluid1}), (\ref{fluid2}), (\ref{T_def_repeat}), and (\ref{J_def_repeat}). To obtain the fluid equations in $V_+$ we could subject $L_{\rm iso}$ to a similar variation, but it is simpler to set $c^a = 0 = w^a$ in the preceding equations; the outcome is the standard Euler equation for a perfect fluid.

To identify the appropriate junction conditions on $\SSS(t)$ we focus our attention on the contributions to $\delta L$ that originate from the interface. When the surface is viewed from $V_-$ we write the directed surface element as $dS_a = n_a\, dS$, where $n_a$ is the surface's unit normal; when it is viewed from $V_+$ we write $dS_a = -n_a\, dS$ instead. The choices of sign reflect the fact that $n_a$ is taken to point from the negative to the positive side of the interface. For the bulk terms we obtain
\begin{subequations}
\begin{align}
\delta L_{\rm aniso}[\SSS(t)] &= \int_{\SSS(t)} \bigl( T^-_{ab} n^b\, \xi^a + J^b_{\ a} n_b\, \Delta c^a \bigr) dS, \\
\delta L_{\rm iso}[\SSS(t)] &= \int_{\SSS(t)} \bigl( -T^+_{ab} n^b\, \xi^a \bigr) dS,
\label{delta_Lbulk}
\end{align}
\end{subequations}
in which $T^\pm_{ab}$ is the momentum-flux tensor defined in $V_\pm$. In our expression for $\delta L_{\rm aniso}$ we expressed $\delta c^a$ in terms of $\Delta c^a$ by making use of Eq.~(\ref{LvsE_inter}), $\Delta c^a = \delta c^a + \xi^b \nabla_b c^a$. This is allowed, because the director vector is defined globally in $V_-$. 

The variation of $L_{\rm interface}$ was carried out in Sec.~\ref{subsec:dynamics1}, with a final result displayed in Eq.~(\ref{dL2}). When we eliminate terms that vanish according to the variations rules, the expression becomes 
\begin{align}
\delta L_{\rm interface} &= 
- \int_{\SSS(t)} \xi_n \Bigl[ \sigma\bigl( \partial_t v_n + v^A D_A v_n - K_{AB} v^A v^B
- \langle n^a \partial_a U \rangle \bigr) + (\eta + \tfrac{1}{2} \tau c^2) K \Bigr]\, dS
\nonumber \\ & \quad \mbox{}
- \int_{\SSS(t)} \xi^A \Bigl[ \sigma (\partial_t v_A - v_n D_A v_n - v_B D_A v^B -D_A U)
- D_A(\eta + \tfrac{1}{2} \tau c^2) \Bigr]\, dS
\nonumber \\ & \quad \mbox{}
- \int_{\SSS(t)} \sigma \hat{\mu} c_a\, \Delta c^a\, dS. 
\label{delta_Linter}
\end{align}
Because $L_{\rm field}$ does not depend on the fluid variables, it does not contribute to $\delta L[\SSS(t)]$.  

\subsection{Fluid equations}
\label{subsec:fluid_eqns} 

Demanding that $\delta S = 0$ for arbitrary and independent variations $\xi^a$ and $\delta c^a$ within $V_-$ returns the same dynamical equations as in Sec.~\ref{subsec:fluid_equations}, 
\begin{subequations}
\label{dyn1_aniso}
\begin{align}
0 &= \partial_t(\rho v_a) + \nabla_b T_{\ a}^{-b} -\rho \nabla_a U, \\
0 &= \partial_t (\rho w_a) + \nabla_b J^b_{\ a},
\end{align} 
\end{subequations}
where
\begin{subequations}
\begin{align} 
T^-_{ab} &:= \rho v_a v_b + \bigl( p + \tfrac{1}{2} \lambda \Xi \bigr) g_{ab}
+ \rho \hat{\kappa} \nabla_a c_c \nabla_b c^c, \\ 
J_{ab} &:= \rho v_a w_b - \rho \hat{\kappa} \nabla_a c_b, 
\end{align}
\end{subequations}
with $p := \rho^2 d\hat{\varepsilon}/d\rho$ and $\lambda := \rho^2 d \hat{\kappa}/d\rho$. The equations can also be presented in the mechanical form of Eq.~(\ref{fluid3}) and (\ref{fluid4}).

Demanding that $\delta S = 0$ for an arbitrary $\xi^a$ within $V_+$ returns
\begin{equation}
0 = \partial_t(\rho v_a) + \nabla_b T_{\ a}^{+b} -\rho \nabla_a U, \\
\label{dyn1_iso} 
\end{equation}
where
\begin{equation}
T^+_{ab} := \rho v_a v_b + p g_{ab}. 
\end{equation}
In both cases the dynamical equations come with 
\begin{equation}
0 = \partial_t \rho + \nabla_a (\rho v^a), 
\end{equation}
the statement of mass conservation. 

Demanding that $\delta S = 0$ for arbitrary and independent $\xi_n$, $\xi^A$, and $\Delta c^a$ on $\SSS(t)$ returns
\begin{subequations}
\label{dyn2} 
\begin{align}
0 &= \sigma \bigl( \partial_t v_n + v^A D_A v_n - K_{AB} v^A v^B
- \langle n^a \partial_a U \rangle \bigr) + (\eta + \tfrac{1}{2} \tau c^2) K + \bigl[ T_{ab} \bigr] n^a n^b, \\
0 &= \sigma \bigl( \partial_t v_A - v_n D_A v_n - v_B D_A v^B -D_A U \bigr)
- D_A (\eta + \tfrac{1}{2} \tau c^2) + \bigl[ T_{ab} \bigr] e^a_A n^b, \\
0 &= \sigma \hat{\mu} c_a - J^b_{\ a} n_b, 
\end{align}
\end{subequations}
together with
\begin{equation} 
0 = \partial_t \sigma + \sigma \bigl( v_n K + D_A v^A \bigr).
\end{equation}
Here, $v_n$ and $v_A$ are respectively the normal and tangential components of the fluid velocity ($v^a = v_n\, n^a + v^A\, e^a_A$), $K_{AB}$ is the extrinsic curvature on $\SSS(t)$, $K := K^A_{\ A}$ is its trace, $D_A$ indicates covariant differentiation with respect to the intrinsic coordinates $\theta^A$, $\eta := -\sigma^2 d\hat{e}/d\sigma$, $\tau := -\sigma^2 d\hat{\mu}/d\sigma$, and
\begin{equation}
\langle \psi \rangle := \frac{1}{2} \psi_+(\SSS) + \frac{1}{2} \psi_-(\SSS),\qquad
\bigl[ \psi \bigr] := \psi_+(\SSS) - \psi_-(\SSS)
\end{equation}
are respectively the average and jump of a quantity $\psi$ across the interface. Equations (\ref{dyn2}) are modified versions of Eqs.~(\ref{fluid_eqns1}), and they are formulated in Lagrangian coordinates $\theta^A$. The equations can be re-expressed in terms of $v^A$ as in Eq.~(\ref{fluid_eqns2}), and in terms of the Hadamard time derivative as in Eq.~(\ref{fluid_eqns3}); this final form is covariant under a change of surface coordinates. 

\subsection{Poisson's equation}
\label{subsec:poisson} 

Variation of the action with respect to the gravitational field involves manipulations described in Sec.~\ref{subsec:var_U} and Appendix \ref{sec:interface_gravity}. It produces the Poisson equation
\begin{equation}
\nabla^2 U = -4\pi G \bigl\{ \rho + \sigma \delta(\ell) \bigr\}
\label{dyn3} 
\end{equation}
for the gravitational potential, with $\ell$ denoting the distance away from the interface, as measured along geodesics (straight lines) that cross $\SSS(t)$ orthogonally.

\subsection{Static configuration} 
\label{subsec:static} 

In the special case of a static fluid configuration, Eqs.~(\ref{dyn1_aniso}) and (\ref{dyn1_iso}) reduce to
\begin{subequations}
\label{fluid_static1} 
\begin{align}
0 &= \nabla_a \bigl( p + \tfrac{1}{2} \lambda \Xi \bigr)
+ \rho \hat{\kappa} \nabla^b c^c \nabla_{ab} c_c - \rho \nabla_a U, \\
0 &= \nabla_b \bigl( \rho \hat{\kappa} \nabla^b c_a \bigr) 
\end{align}
\end{subequations}
in the anisotropic phase, and
\begin{equation}
0 = \nabla_a p - \rho \nabla_a U
\label{fluid_static2}
\end{equation}
in the isotropic phase. The junction conditions of Eq.~(\ref{dyn2}) become 
\begin{subequations}
\label{fluid_static3} 
\begin{align}
0 &= \bigl[ p \bigr] + \bigl( \eta + \tfrac{1}{2} \tau c^2 \bigr) K - \tfrac{1}{2} \lambda \Xi 
- \rho \hat{\kappa} \bigl( n^a \nabla_a c_c \bigr) \bigl( n^b \nabla_b c^c \bigr) 
- \sigma \langle n^a \partial_a U \rangle, \\
0 &= D_A\bigl( \eta + \tfrac{1}{2} \tau c^2 \bigr)
+ \rho \hat{\kappa} \bigl( e^a_A \nabla_a c_c \bigr) \bigl( n^b \nabla_b c^c \bigr) + \sigma D_A U, \\
0 &= \sigma \hat{\mu} c_a + \rho \hat{\kappa} \bigl( n^b \nabla_b c_a \bigr). 
\end{align}
\end{subequations} 
Poisson's equation (\ref{dyn3}) stays the same. 

It is instructive to note that when the fluid is everywhere isotropic, in $V_\pm$ and on $\SSS(t)$, the junction conditions reduce to $[p] + \eta K = \sigma \langle n^a \nabla_a U \rangle$ and $D_A \eta = -\sigma D_A U$. When gravity can be neglected we end up with $[p] + \eta K = 0$, the statement of the Young-Laplace law; in this case the surface tension must be uniform on the interface: $D_A \eta = 0$. 

\section{Anisotropic stellar models}
\label{sec:stellar_model} 

In this section we apply the fluid mechanics developed in Sec.~\ref{sec:twophase} to the formulation of anisotropic stellar models. We take the fluid configuration to be static and spherically symmetric, and make specific choices of equations of state for the fluid. We recall from Sec.~IV of paper I \cite{cadogan-poisson:24a}, and from the developments of Sec.~\ref{sec:twophase}, that the stellar models must feature a phase transition at some critical density $\rho_{\rm crit}$; the fluid is anisotropic at higher densities, and isotropic at lower densities. In Sec.~\ref{subsec:inner_outer} we write down the structure equations that apply in the star's  anisotropic inner core and isotropic outer shell, and in Sec.~\ref{subsec:junction} we obtain the appropriate junction conditions. We specialize the equations to a polytropic equation of state in Sec.~\ref{subsec:polytrope}, and present our numerical results in Sec.~\ref{subsec:results}. In Sec.~\ref{subsec:analytical} we provide an approximate, analytical solution to the structure equations for the specific case of an $n=1$ polytrope, for which $\kappa = \varepsilon \propto \rho^2$.  

\subsection{Inner core and outer shell}
\label{subsec:inner_outer} 

We take the star to consist of an inner core $V_-$ surrounded by an outer shell $V_+$. In the inner core the fluid is anisotropic and characterized by
\begin{equation}
\kappa = \varepsilon,
\end{equation}
from which it follows that $\lambda = p$. The assumed spherical symmetry implies that all fluid variables depend on the radial coordinate $r$ only, and that $c := c^r$ is the only nonvanishing component of the director vector. The fluid equations (\ref{dyn3}) and (\ref{fluid_static1}) become
\begin{subequations}
\label{inner_core}
\begin{align}
0 &= \biggl( 1 + \frac{1}{2} c^{\prime 2} + \frac{c^2}{r^2} \biggr) r p' + \frac{Gm\rho}{r}
+ (\varepsilon + p) \biggl[ -\biggl(2 + \frac{r\varepsilon'}{\varepsilon} \biggr) c^{\prime 2}
+ \frac{4}{r} c c' - \frac{2}{r^2} c^2 \biggr], \\
0 &= r^2 c'' + \biggl( 2 + \frac{r \varepsilon'}{\varepsilon} \biggr) r c' - 2 c, \\
0 &= r m' - 4\pi r^3 \rho,
\end{align}
\end{subequations}
in which a prime indicates differentiation with respect to $r$. The mass function $m(r)$ is related to the gravitational potential by $U' = -Gm/r^2$. It is noteworthy that the equations are invariant under a reflection $c \to -c$ of the director vector.

In the outer shell the fluid is isotropic. The fluid equations reduce to 
\begin{subequations}
\label{outer_shell}
\begin{align}
0 &= rp' + \frac{Gm\rho}{r}, \\
0 &= rm' - 4\pi r^3 \rho; 
\end{align}
\end{subequations}
these equations follow from Eqs.~(\ref{dyn3}) and (\ref{fluid_static2}). 

\subsection{Junction conditions}
\label{subsec:junction}

The phase transition between the anisotropic core and the isotropic shell occurs on a sphere of radius $r = r_{\rm crit}$, and the junction conditions derived from Eqs.~(\ref{dyn3}) and (\ref{fluid_static3}) are
\begin{subequations}
\label{junction1}
\begin{align}
0 &= \bigl[ p \bigr] - (\kappa + \tfrac{1}{2} \lambda) c^{\prime 2} - \frac{\lambda c^2}{r_{\rm crit}^2} 
+ \bigl( \eta + \tfrac{1}{2} \tau c^2 \bigr) K + \frac{G \langle m \rangle \sigma}{r_{\rm crit}^2},\\
0 &= \kappa c' + \sigma \hat{\mu} c, \\
0 &= [m] - 4\pi r_{\rm crit}^2 \sigma.
\end{align}
\end{subequations} 
We recall that $[\psi] := \psi_+ - \psi_-$ is the jump of a quantity $\psi$ across the interface, and $\langle \psi \rangle := \frac{1}{2}(\psi_+ + \psi_-)$ is its average. We recall also that $\sigma$, $\eta$, $\tau$, and $\hat{\mu}$ are quantities that characterize the interface fluid ($\sigma$ is the surface mass density), and that $K = 2/r_{\rm crit}$ is the sphere's extrinsic curvature.

We assume that the equation of state relating $\varepsilon$ to $\rho$ is the same in both phases of the fluid. This implies continuity of the internal energy across the interface, $[\varepsilon] = 0$, and continuity of the pressure, $[p] = 0$. We also assume, for simplicity, that the gravitational influence of the interface can be neglected, so that the mass function is continuous to a good approximation, $[m] \simeq 0$. With these choices Eq.~(\ref{junction1}) becomes
\begin{equation}
\bigl( \eta + \tfrac{1}{2} \tau c^2 \bigr) K = -\frac{Gm\sigma}{r_{\rm crit}^2} 
+ (\varepsilon_{\rm crit} + \tfrac{1}{2} p_{\rm crit}) c^{\prime 2}
+ \frac{p_{\rm crit} c^2}{r_{\rm crit}^2} 
\label{junction2}
\end{equation}
and
\begin{equation}
\sigma \hat{\mu} c = - \varepsilon_{\rm crit} c',
\label{junction3}
\end{equation}
where $\varepsilon_{\rm crit}$ and $p_{\rm crit}$ are respectively the internal energy density and pressure corresponding to the critical density $\rho_{\rm crit}$, while $c$ and $c'$ are respectively the director vector and its derivative evaluated at $r = r_{\rm crit}$ from the anisotropic side of the fluid. Provided that the right-hand side of Eq.~(\ref{junction2}) is dominated by the anisotropy terms proportional to $c^{\prime 2}$ and $c^2$, we have that the effective surface tension $\eta + \frac{1}{2} \tau c^2$ is positive on the interface. On the other hand, Eq.~(\ref{junction3}) reveals that $\hat{\mu}$ is negative when $c'/c$ is positive, which is the behavior revealed by the numerical integrations presented below. 

\subsection{Polytrope}
\label{subsec:polytrope} 

We now specialize the stellar model further and choose the polytropic equation of state
\begin{equation}
\varepsilon = n \KK \rho^{1 + 1/n}
\end{equation}
for both phases of the anisotropic fluid; here $\KK$ and $n$ are constants. The equation of state implies that $p := \rho^2 d(\varepsilon/\rho)/d\rho = \KK \rho^{1+1/n} = \varepsilon/n$.

To integrate the structure equations we introduce a Lane-Emden variable $\vartheta$, defined by
\begin{equation}
\rho = \rho_{\rm crit} \vartheta^n, 
\end{equation}
where $\rho_{\rm crit}$ is the previously introduced critical density. From this it follows that
\begin{equation}
p = p_{\rm crit} \vartheta^{n+1}, \qquad
\varepsilon = n p_{\rm crit} \vartheta^{n+1},
\end{equation}
where $p_{\rm crit} := \KK \rho_{\rm crit}^{1+1/n}$. The new variable lies in the interval $\vartheta_c \geq \vartheta \geq 0$, with
\begin{equation} 
\vartheta_c := (\rho_c/\rho_{\rm crit})^{1/n}, \qquad
\rho_c := \rho(r=0) 
\end{equation}
standing for the central value $\vartheta(r=0)$, and with $\vartheta = 0$ describing the stellar surface. The phase transition occurs at $\vartheta = 1$, and to produce an anisotropic polytrope we shall choose a $\vartheta_c$ that is larger than unity.

We also introduce the dimensionless variables $\chi$, $u$, and $v$, defined by
\begin{equation}
m = \frac{4\pi}{3} \rho_c r^3 \chi, \qquad
c = \beta r u, \qquad
c' = \beta v,
\end{equation}
where $\beta$ is a dimensionless parameter that measures the degree of anisotropy. The equation $m' = 4\pi r^2 \rho$ for the mass function implies that $\chi(r=0) = 1$. We scale $u$, and thereby provide a precise meaning for $\beta$, by demanding that $u(r=0) = 1$. The definition of $v$ then implies that $v(r=0) = 1$, and we have that
\begin{equation}
\beta := c'(r=0).
\end{equation} 
Finally, we introduce a dimensionless radial variable $\zeta$ with
\begin{equation}
r^2 = r_0^2\, \zeta, \qquad
r_0^2 := \frac{3}{2\pi} (n+1) \frac{p_{\rm crit}}{G \rho_{\rm crit}^2}.
\end{equation}
In terms of the new variable we have that $r d/dr = 2\zeta d/d\zeta$.

The structure equations become
\begin{subequations}
\label{inner_core_dimless}
\begin{align}
0 &= \Bigl\{ 1 + \beta^2 \bigl[ u^2 - (n+\tfrac{1}{2}) v^2 \bigr] \Bigr\} \frac{d\vartheta}{d\zeta}
+ \vartheta_c^n\, \chi - \beta^2 (u-v)^2 \frac{\vartheta}{\zeta},
\label{dvartheta_dzeta_inner} \\
0 &= \frac{d\chi}{d\zeta} + \frac{3}{2\zeta} \bigl[ \chi - (\vartheta/\vartheta_c)^n \bigr], \\
0 &= \frac{du}{d\zeta} + \frac{1}{2\zeta} (u-v), \\
0 &= \frac{dv}{d\zeta} + (n+1) \frac{v}{\vartheta}\, \frac{d\vartheta}{d\zeta}
- \frac{1}{\zeta} (u-v)
\end{align}
\end{subequations}
in the inner core, and
\begin{subequations}
\label{outer_shell_dimless}
\begin{align}
0 &= \frac{d\vartheta}{d\zeta} + \vartheta_c^n\, \chi, \\
0 &= \frac{d\chi}{d\zeta} + \frac{3}{2\zeta} \bigl[ \chi - (\vartheta/\vartheta_c)^n \bigr]
\end{align}
\end{subequations}
in the outer shell. A clever implementation of these equations selects $\vartheta$ as the independent variable, and integrates for $\zeta(\vartheta)$, $\chi(\vartheta)$, $u(\vartheta)$, and $v(\vartheta)$. The system of Eqs.~(\ref{inner_core_dimless}) is integrated in the interval $\vartheta_c \leq \vartheta \leq 1$, and Eqs.~(\ref{outer_shell_dimless}) are then integrated in $1 \geq \vartheta \geq 0$. We impose continuity of $\zeta$ and $\chi$ at $\vartheta = 1$, and allow $u$ and $v$ to jump to zero. 

Integration of the structure equations returns the surface values $\zeta_s := \zeta(\vartheta=0)$ and $\chi_s := \chi(\vartheta=0)$. From these we calculate the star's mass
\begin{equation}
M = M_{\rm unit}\, \vartheta_c^n \zeta_s^{3/2} \chi_s, \qquad
M_{\rm unit} := \frac{4\pi}{3} \rho_{\rm crit} r_0^3,
\label{star_mass}
\end{equation}
and its radius
\begin{equation}
R = R_{\rm unit}\, \zeta_s^{1/2}, \qquad
R_{\rm unit} := r_0.
\label{star_radius} 
\end{equation}
We note that the mass unit $M_{\rm unit}$ and radius unit $R_{\rm unit}$ are independent of the central density $\theta_c^n := \rho_c/\rho_{\rm crit}$. 

\subsection{Numerical results}
\label{subsec:results} 

For a given polytropic index $n$, the stellar models obtained by integrating Eqs.~(\ref{inner_core_dimless}) and (\ref{outer_shell_dimless}) form a two-parameter family labelled by the scaled central density $\rho_c/\rho_{\rm crit}$ and the anisotropy parameter $\beta$. A small sample of our numerical results (for $n=1.0$) were previously presented in paper I \cite{cadogan-poisson:24a}. Here we broaden our exploration of the parameter space.  

\begin{figure}
\includegraphics[width=0.32\linewidth]{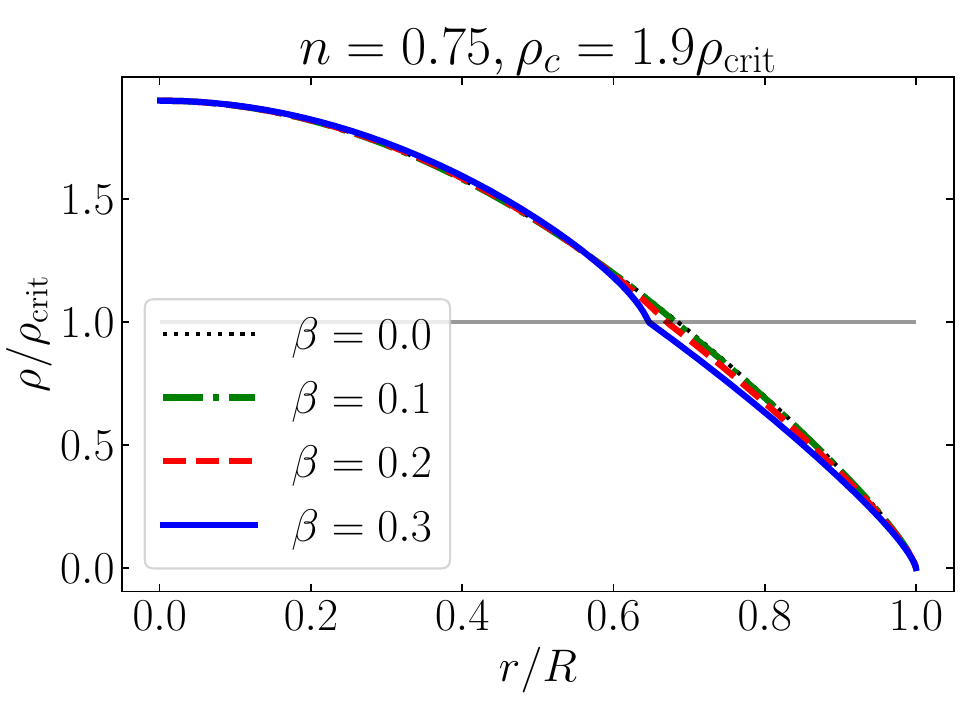}
\includegraphics[width=0.32\linewidth]{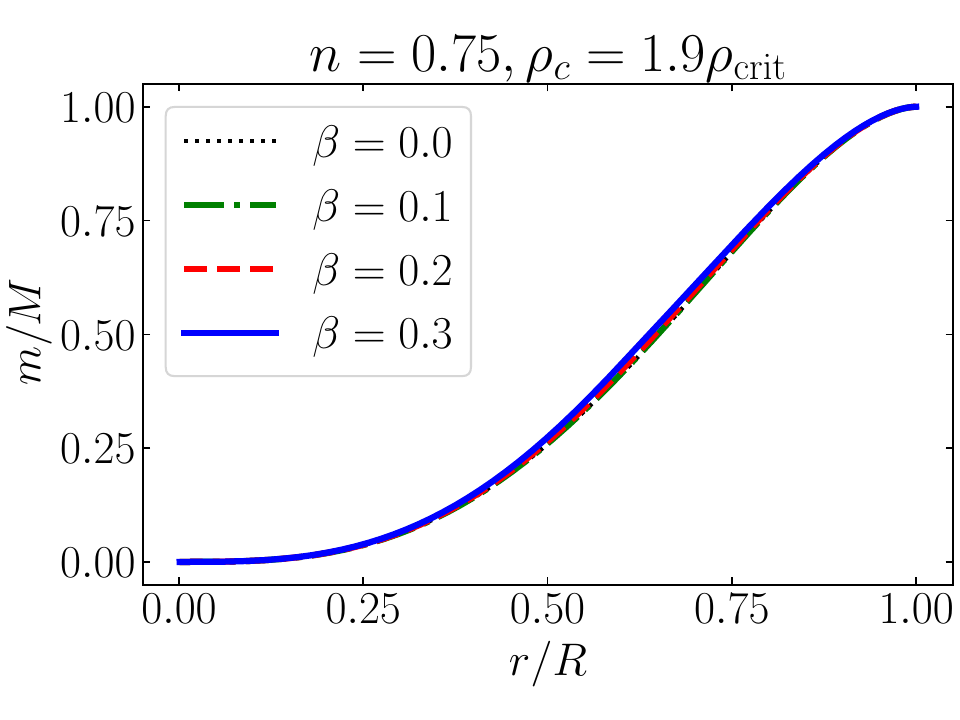}
\includegraphics[width=0.32\linewidth]{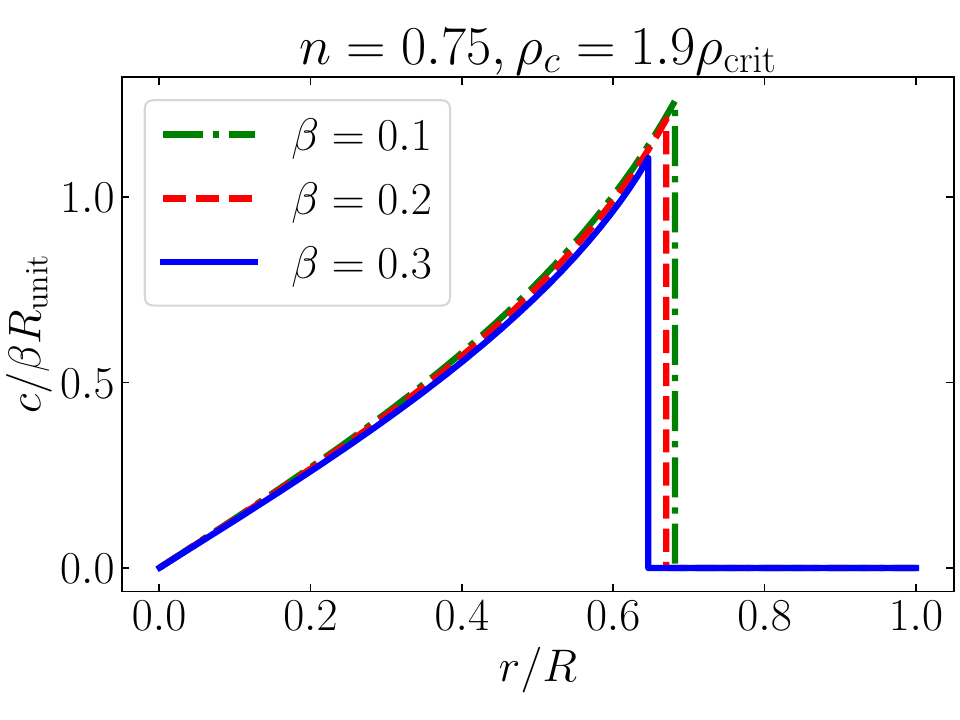}
\caption{Radial profiles of the density, mass, and director vector for stellar models with $n=0.75$ and $\rho_c/\rho_{\rm crit} = 1.9$. Left: $\rho/\rho_{\rm crit}$ as a function of $r/R$. Middle: $m/M$ as a function of $r/R$. Right: $c/(\beta R_{\rm unit})$ as a function of $r/R$. Dotted black curves: isotropic polytrope with $\beta = 0$. Dash-dotted green curves: $\beta = 0.1$. Dashed red curves: $\beta = 0.2$. Solid blue curves: $\beta = 0.3$.}  
\label{fig:fig1} 
\end{figure}

\begin{figure}
\includegraphics[width=0.32\linewidth]{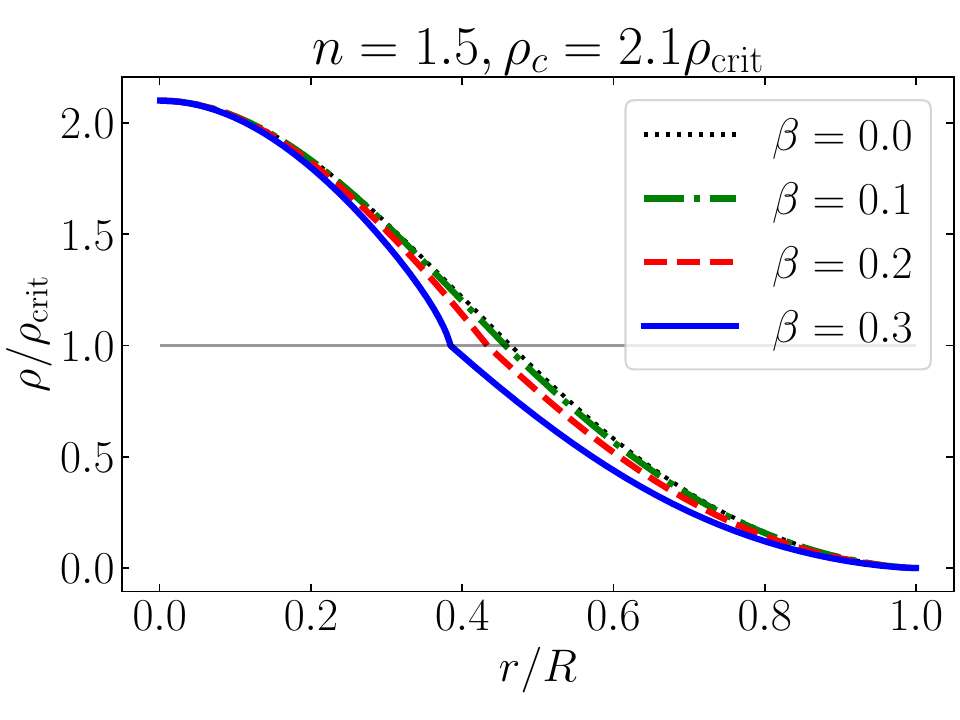}
\includegraphics[width=0.32\linewidth]{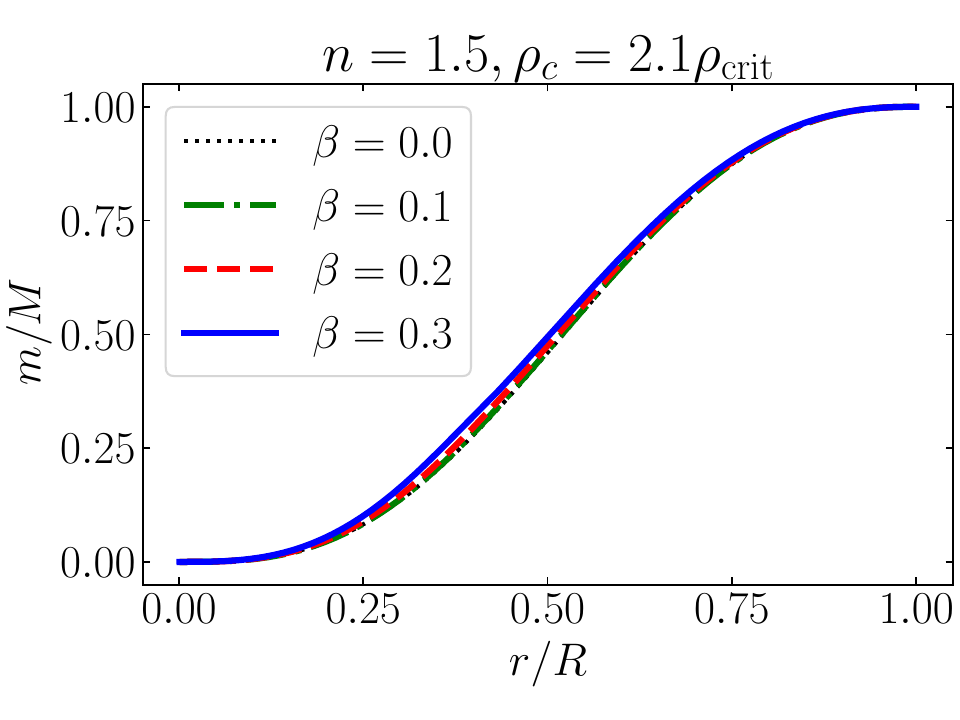}
\includegraphics[width=0.32\linewidth]{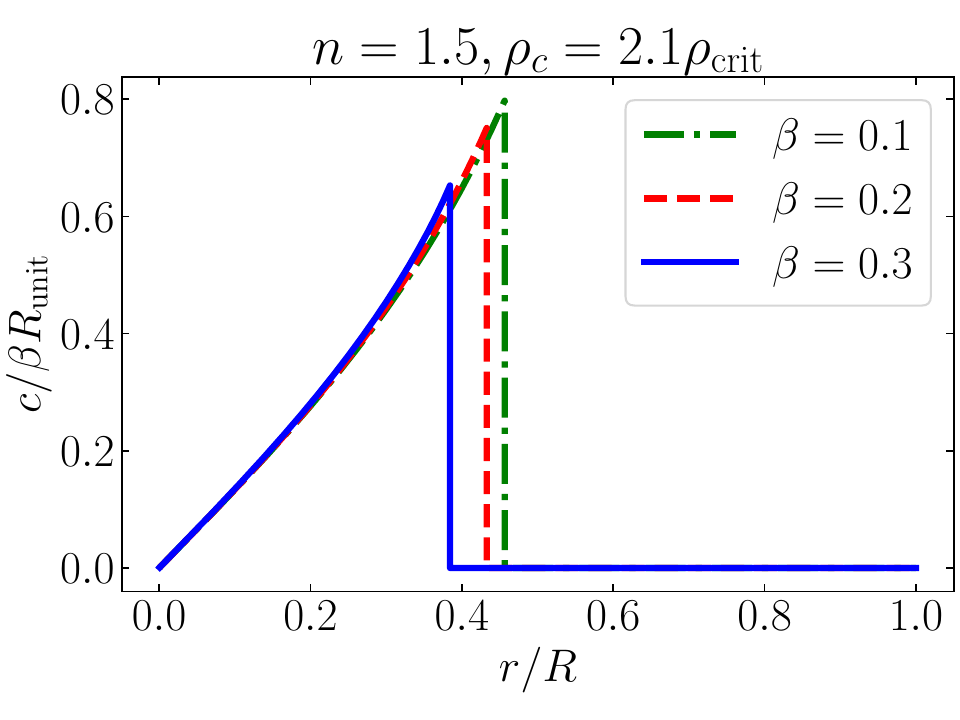}
\caption{Radial profiles of the density, mass, and director vector for stellar models with $n=1.5$ and $\rho_c/\rho_{\rm crit} = 2.1$.} 
\label{fig:fig2} 
\end{figure}

\begin{figure}
\includegraphics[width=0.32\linewidth]{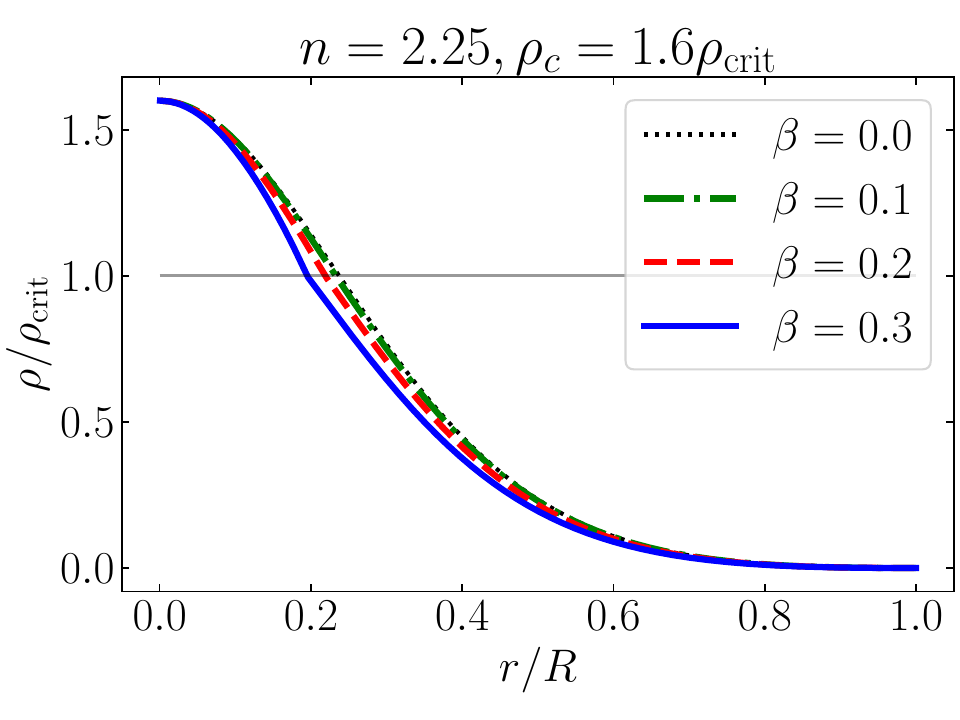}
\includegraphics[width=0.32\linewidth]{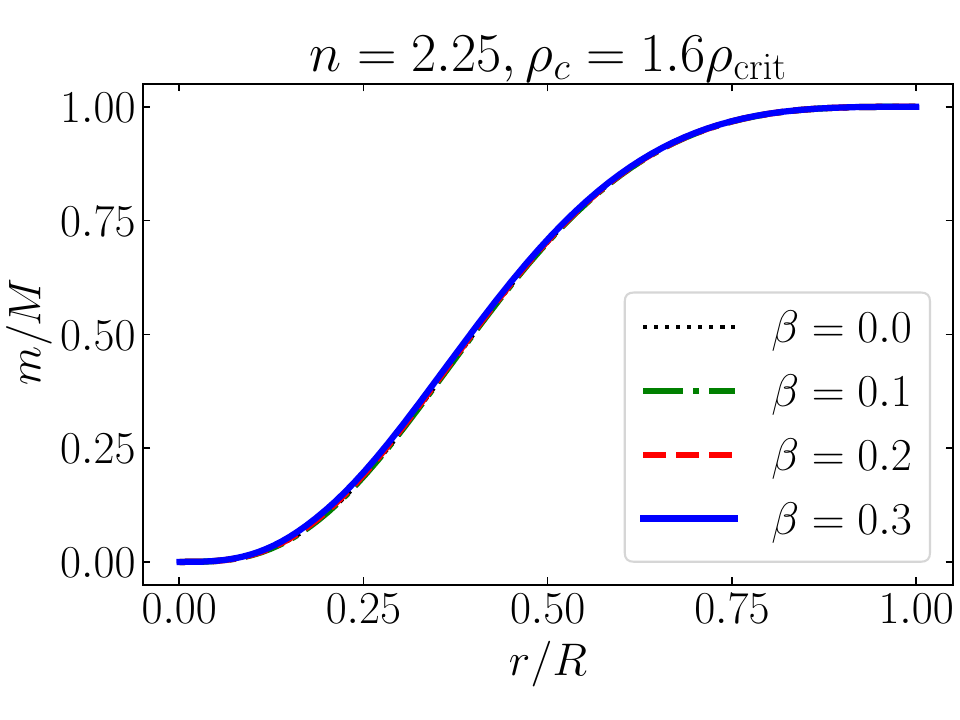}
\includegraphics[width=0.32\linewidth]{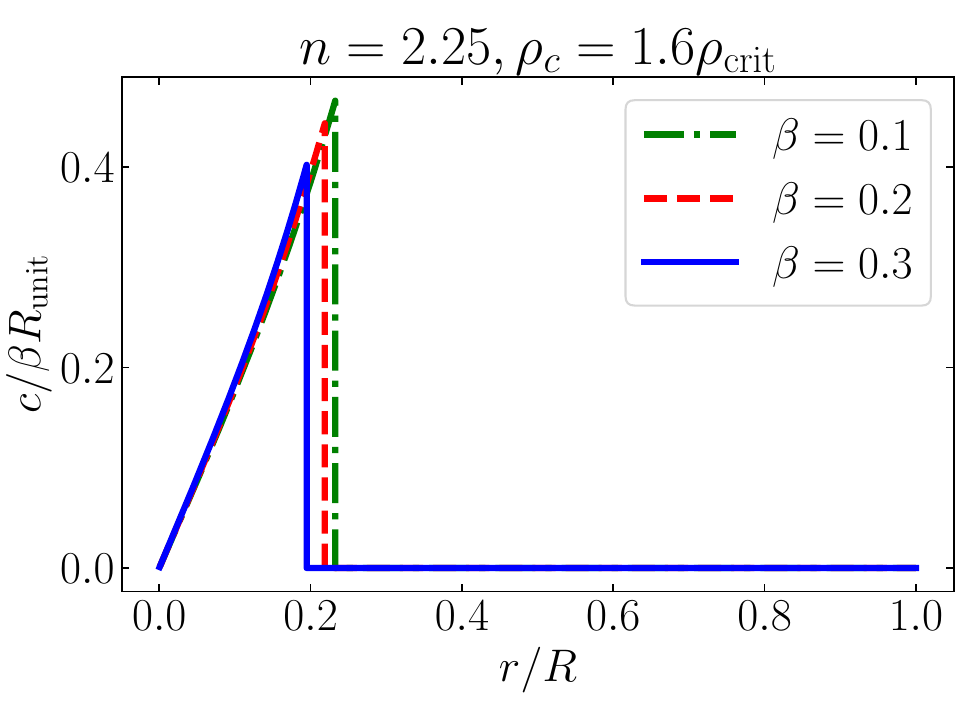}
\caption{Radial profiles of the density, mass, and director vector for stellar models with $n=2.25$ and $\rho_c/\rho_{\rm crit} = 1.6$.} 
\label{fig:fig3} 
\end{figure}

We begin with Fig.~\ref{fig:fig1}, which shows density, mass, and director profiles for stellar models with $n=0.75$ and $\rho_c/\rho_{\rm crit} = 1.9$. The qualitative features are the same as those observed in paper I. For a given $r/R$, the scaled density $\rho/\rho_{\rm crit}$ decreases with increasing $\beta$, while the scaled mass $m/M$ increases with increasing $\beta$. The density displays a discontinuity in its first derivative when it becomes equal to the critical density; this is where the phase transition occurs. The director vector increases monotonically with $r/R$, until it jumps to zero at the phase transition. The scaled critical radius $r_{\rm crit}/R$ decreases with increasing $\beta$. The same features are observed in Figs.~\ref{fig:fig2} and \ref{fig:fig3}, which display radial profiles for stellar models with $n=1.5$ and $\rho_c/\rho_{\rm crit} = 2.1$, and $n=2.25$ and $\rho_c/\rho_{\rm crit} = 1.6$, respectively.  

\begin{figure}
\includegraphics[width=0.32\linewidth]{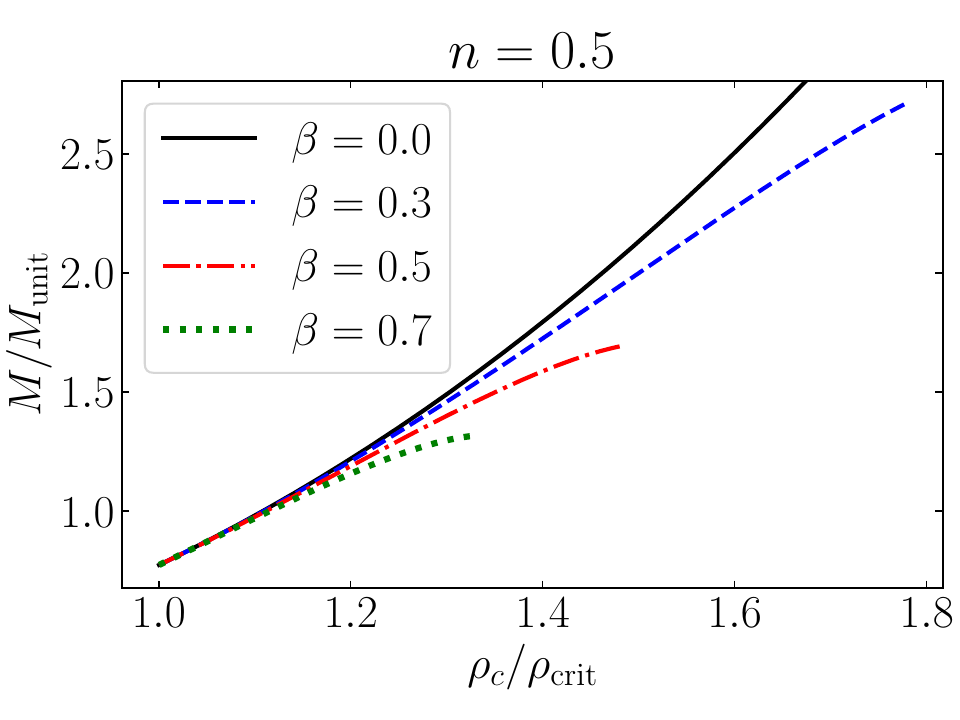}
\includegraphics[width=0.32\linewidth]{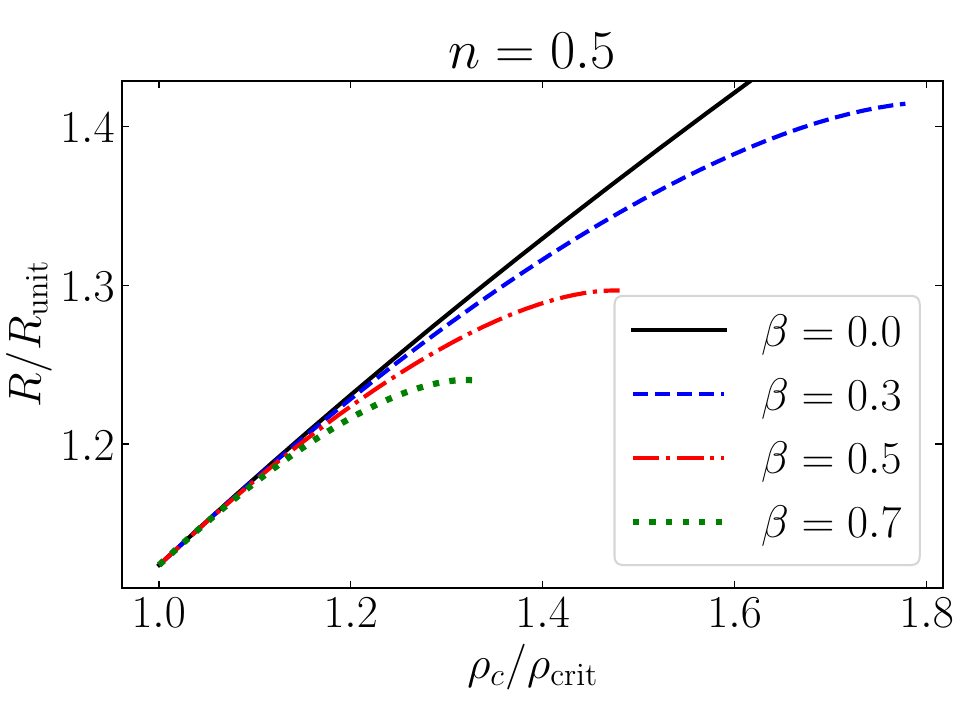}
\includegraphics[width=0.32\linewidth]{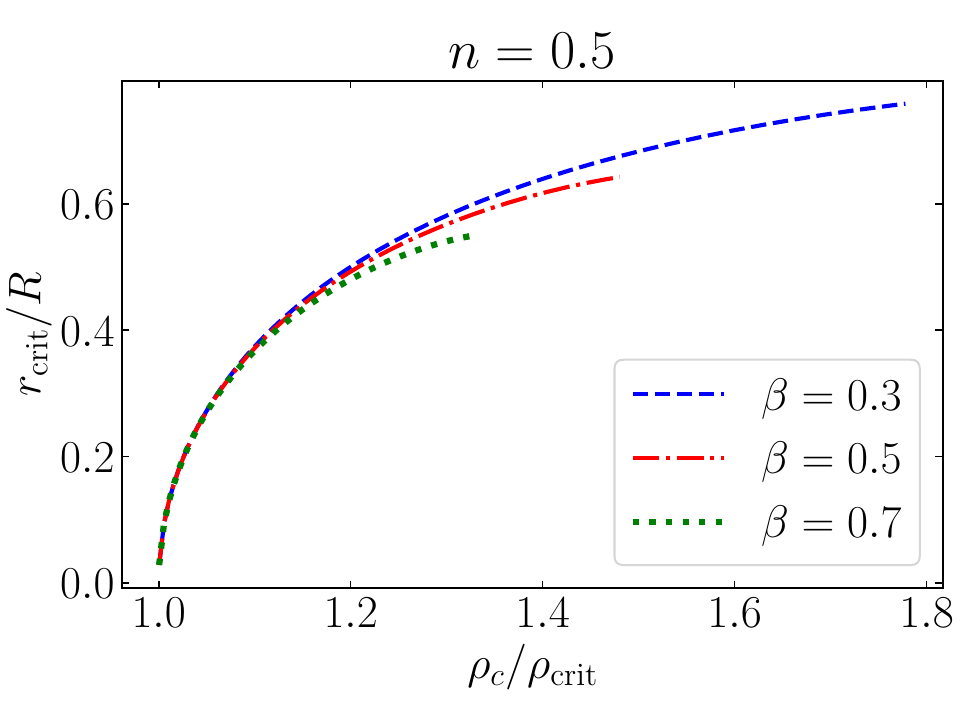}
\caption{Stellar mass $M$, stellar radius $R$, and critical radius $r_{\rm crit}$ as functions of the central density for stellar models with $n=0.5$. Left: $M/M_{\rm unit}$ as a function of $\rho_c/\rho_{\rm crit}$. Middle: $R/R_{\rm unit}$ as a function of $\rho_c/\rho_{\rm crit}$. Right: $r_{\rm crit}/R$ as a function of $\rho_c/\rho_{\rm crit}$. Solid black curves: $\beta = 0$ (isotropic model). Dashed blue curves: $\beta = 0.3$. Dash-dotted red curves: $\beta = 0.5$. Dotted green curves: $\beta = 0.7$. Below $\rho_c/\rho_{\rm crit} = 1$ all stellar models are isotropic.}  
\label{fig:fig4} 
\end{figure} 

\begin{figure}
\includegraphics[width=0.32\linewidth]{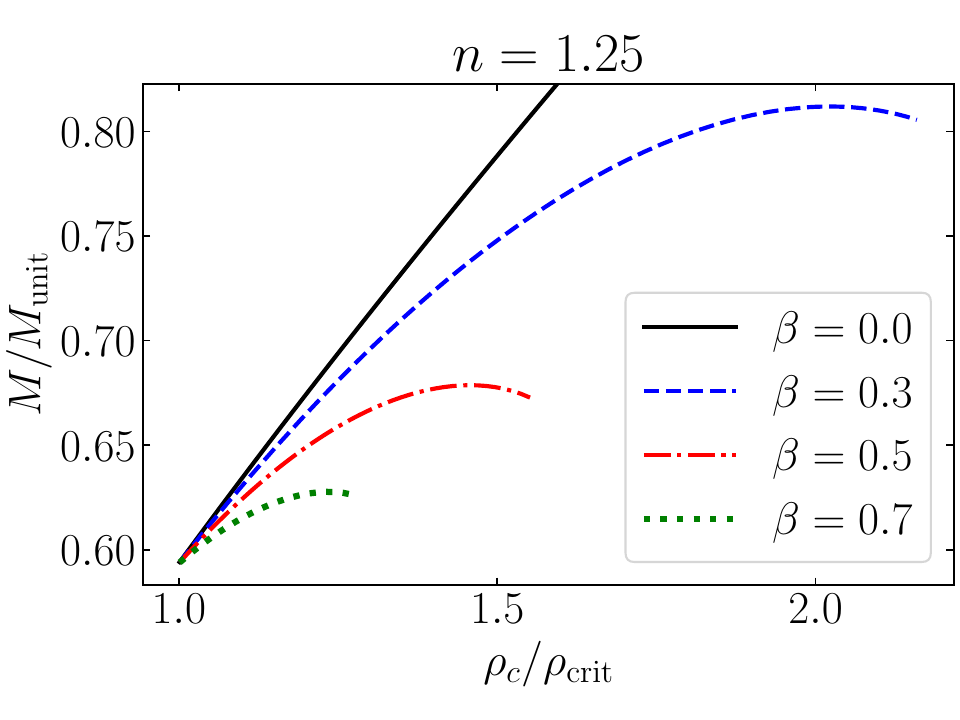}
\includegraphics[width=0.32\linewidth]{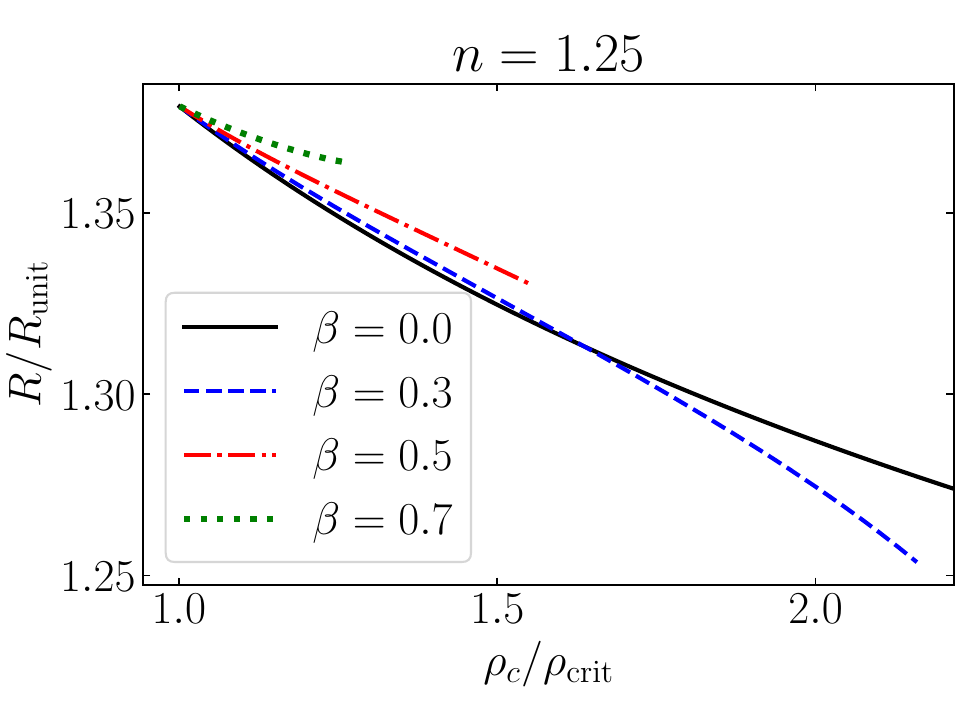}
\includegraphics[width=0.32\linewidth]{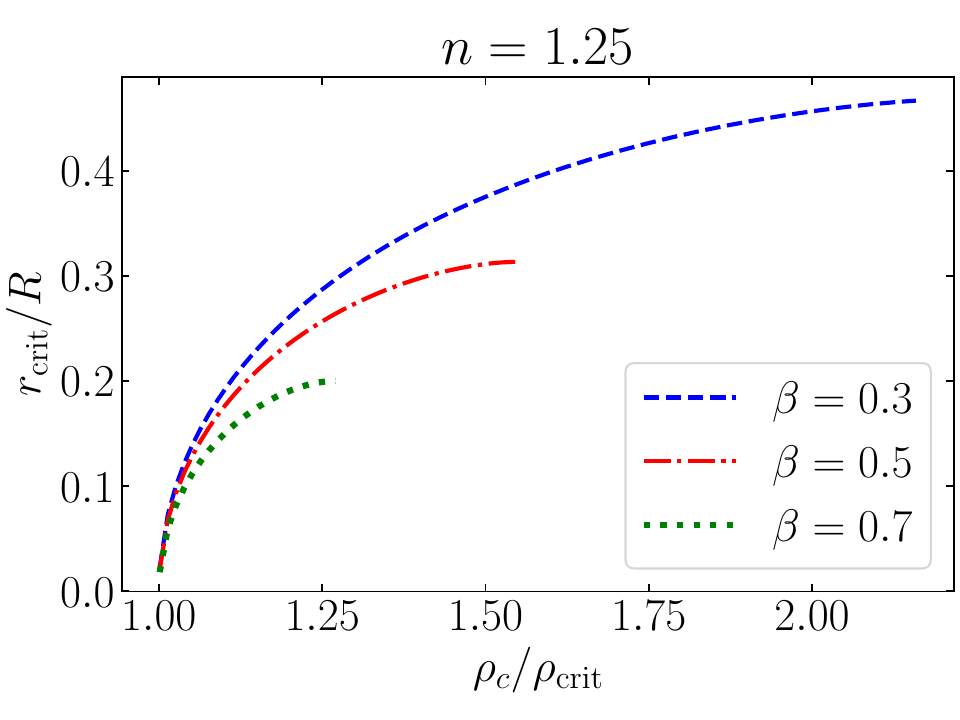}
\caption{Stellar mass, stellar radius, and critical radius as functions of the central density for stellar models with $n=1.25$.}  
\label{fig:fig5} 
\end{figure} 

\begin{figure}
\includegraphics[width=0.32\linewidth]{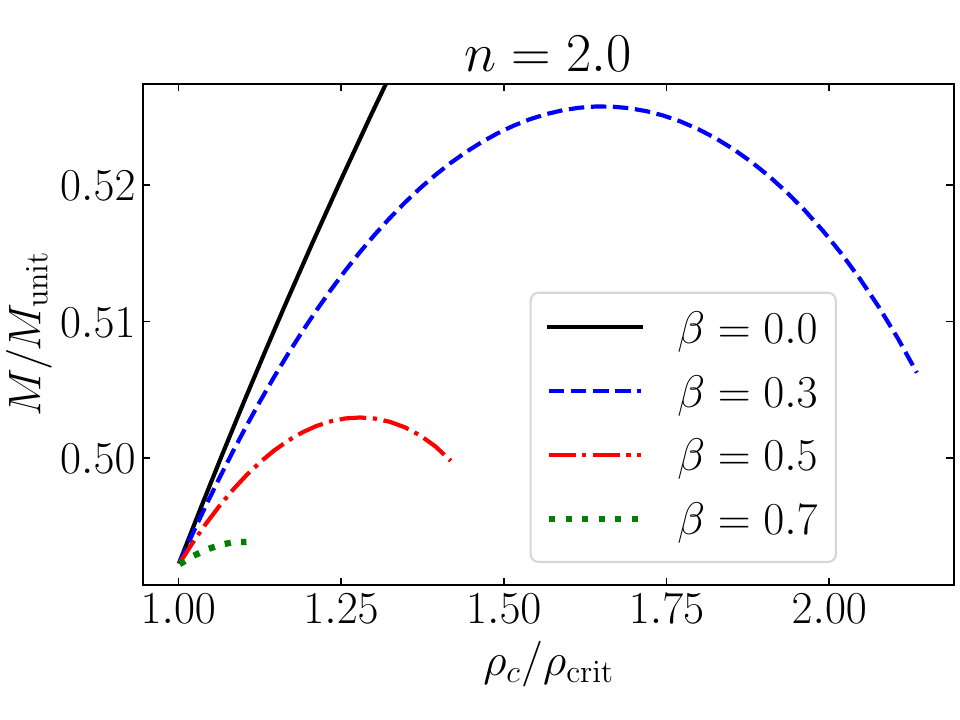}
\includegraphics[width=0.32\linewidth]{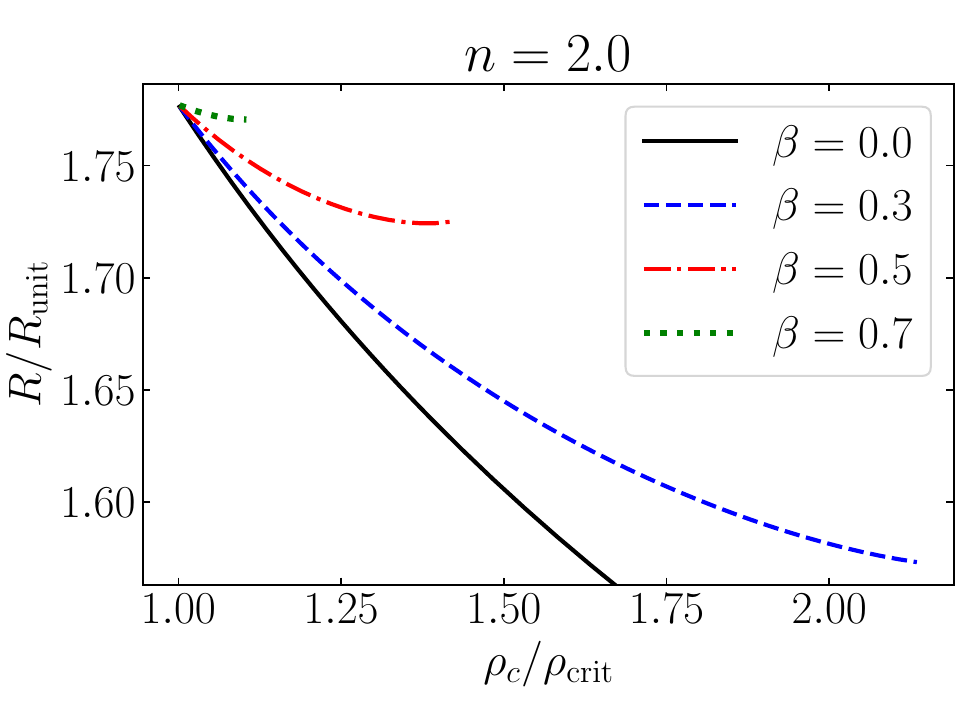}
\includegraphics[width=0.32\linewidth]{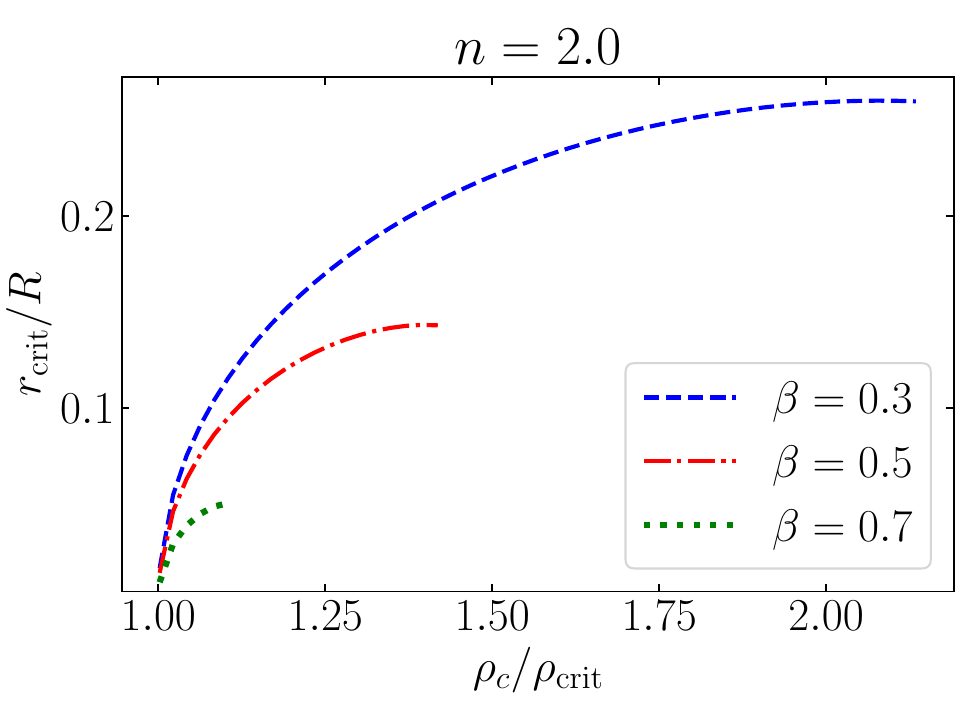}
\caption{Stellar mass, stellar radius, and critical radius as functions of the central density for stellar models with $n=2.0$.}  
\label{fig:fig6} 
\end{figure} 

\begin{figure}
\includegraphics[width=0.5\linewidth]{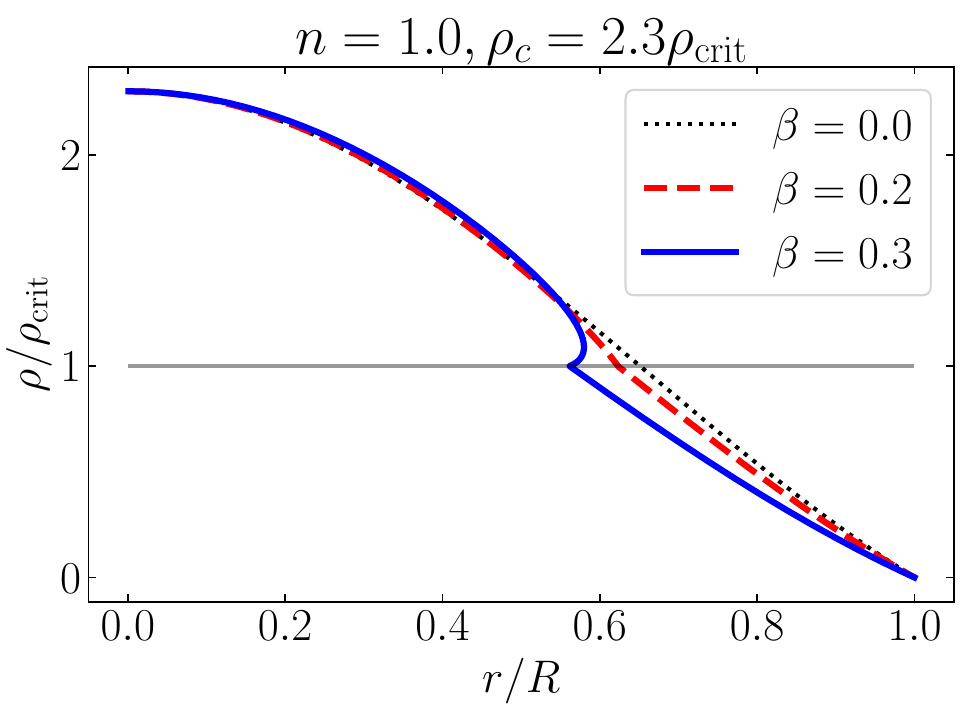}
\caption{Unphysical density profile. Plots of $\rho/\rho_{\rm crit}$ as a function of $r/R$ for stellar models with $n=1.0$ and $\rho_c/\rho_{\rm crit} = 2.3$. When $\beta$ exceeds a value $\beta_{\rm max}$ (here between 0.2 and 0.3), the density becomes multivalued near the phase transition. Dotted black curve: $\beta =0$. Dashed red curve: $\beta = 0.2$. Solid blue curve: $\beta = 0.3$.} 
\label{fig:fig7} 
\end{figure} 

In Fig.~\ref{fig:fig4} we plot the stellar mass $M$, stellar radius $R$, and critical radius $r_{\rm crit}$ as functions of $\rho_c/\rho_{\rm crit}$, for models with $n = 0.5$. We do the same in Fig.~\ref{fig:fig5} for models with $n = 1.25$, and in Fig.~\ref{fig:fig6} for $n = 2.0$. In all cases we observe that for a given $\rho_c/\rho_{\rm crit}$, $M$ and $r_{\rm crit}$ both decrease with increasing $\beta$. The stellar radius, however, presents a richer spectrum of behavior: while $R$ decreases with increasing $\beta$ for low values of $n$, it increases with $\beta$ for high values of $n$, and these is no clear ordering for intermediate values of $n$.  

The figures reveal that the equilibrium sequence of an anisotropic polytrope terminates at a maximum value of the central density. This is unlike the isotropic sequence, which keeps going indefinitely. The reason for the termination has to do with the factor 
\begin{equation}
1 + \beta^2 \bigl[ u^2 - (n+\tfrac{1}{2}) v^2 \bigr]
\end{equation}
that appears in front of $d\vartheta/d\zeta$ in Eq.~(\ref{dvartheta_dzeta_inner}). When the central density exceeds the maximum value, we find that the factor changes sign somewhere in the interval $\rho_c < \rho < \rho_{\rm crit}$, causing the structure equations to become singular. Such behavior is displayed in Fig.~\ref{fig:fig7} for an anisotropic polytrope with $n=1.0$, $\rho_c/\rho_{\rm crit} = 2.3$, and $\beta = 0.3$; we see that the density is multivalued near the phase transition. In paper I we speculated that an anisotropic star might become dynamically unstable when (or before) the density is about to become multivalued; a stability analysis will be required to decide whether the conjecture is valid.  

Another feature seen in Figs.~\ref{fig:fig5} and \ref{fig:fig6} is that the anisotropic sequences achieve a maximum stellar mass at some central density, beyond which the mass begins to decrease with increasing density. This is unlike the isotropic sequences, which produce a mass that is always increasing with central density. In the case of isotropic stars it is known that the turning point marks the onset of a dynamical instability to radial perturbations. The same may well be true in the case of our anisotropic stars, so that the sequences should be properly terminated at the configuration of maximum mass (when it occurs before the multivalued density). Again, a stability analysis is required to decide whether this conjecture is true.  

\subsection{Analytical treatment for an $n=1$ polytrope}
\label{subsec:analytical} 

The numerical results presented previously reveal that anisotropic polytropes can be obtained only when the anisotropy parameter $\beta$ does not exceed a maximal value $\beta_{\rm max}(\rho_c)$, which is smaller than unity. When $\beta$ is much smaller than unity we can simplify the structure equations by expanding all variables in powers of $\beta$. And when these equations are formulated for an $n=1$ polytrope, their solutions can be obtained analytically and expressed in terms of simple functions.

For our purposes here we continue to use $\vartheta$ and $u$ as dimensionless variables associated with $\rho$ and $c$, respectively, but we now express the mass variable and radial coordinate as
\begin{equation}
m = \frac{4\pi}{3} \rho_{\rm crit} r_1^3\, \bar{m}, \qquad 
r = r_1 x,
\end{equation}
where 
\begin{equation} 
r_1^2 := \frac{1}{2\pi} \frac{p_{\rm crit}}{G \rho_{\rm crit}^2} = \frac{1}{6} r_0^2.
\end{equation} 
In the inner core we expand the variables as
\begin{equation}
\vartheta_{\rm inner} = \vartheta_0 + \beta^2 \vartheta_2 + O(\beta^4), \qquad
\bar{m}_{\rm inner} = \bar{m}_0 + \beta^2 \bar{m}_2 + O(\beta^4), \qquad
u_{\rm inner} = u_0 + O(\beta^2),
\end{equation}
make the substitutions in Eq.~(\ref{inner_core}), and expand the equations in powers of $\beta^2$. To help with the analytical work we convert the system of first-order differential equations for $(\vartheta_0, \bar{m}_0)$ and $(\vartheta_2, \bar{m}_2)$ to second-order differential equations for $\vartheta_0$ and $\vartheta_2$.

The equation satisfied by $\vartheta_0$ is
\begin{equation}
x^2 \frac{d^2 \vartheta_0}{dx^2} + 2x \frac{d\vartheta_0}{dx} + x^2 \vartheta_0 = 0,
\end{equation}
and the appropriate solution is
\begin{equation}
\vartheta_0 = \bar{\rho}_c \frac{\sin x}{x},
\end{equation}
where $\bar{\rho}_c := \rho_c/\rho_{\rm crit}$ is the scaled central density. The mass variable is then determined by $d\bar{m}_0/dx = 3x^2\vartheta_0$, with solution
\begin{equation}
\bar{m}_0 = 3\bar{\rho}_c ( \sin x - x \cos x).
\end{equation}
Near $x = 0$ we have that $\vartheta_0 = \bar{\rho}_c(1 - \frac{1}{6} x^2 + \cdots)$ and $\bar{m}_0 = \bar{\rho}_c x^3 (1 - \frac{1}{10} x^2 + \cdots)$. 

The equation satisfied by $u_0$ is
\begin{equation}
x^2 \frac{d^2 u_0}{d x^2}
+ 2x \biggl( 1 + \frac{x}{\vartheta_0} \frac{d\vartheta_0}{dx} \biggr) \frac{d u_0}{dx}
- 2 u_0 = 0,
\end{equation}
and the solution is
\begin{equation}
u_0 = 3\biggl( \frac{1}{x} - \frac{\cos x}{\sin x} \biggr).
\end{equation}
Near $x = 0$ we have that $u_0 = x(1 + \frac{1}{15} x^2 + \cdots)$.

The equation satisfied by $\vartheta_2$ is
\begin{equation}
x^2 \frac{d^2 \vartheta_2}{dx^2} + 2x \frac{d\vartheta_2}{dx} + x^2 \vartheta_2 = S,
\end{equation}
with a source term given by
\begin{align}
S &:= \biggl[ \frac{3}{2} x^2 \Bigl( \frac{d u_0}{dx} \Bigr)^2 - u_0^2 \biggr]
\frac{d^2 \vartheta_0}{dx^2}
+ \biggl[ 3x^2 \frac{d u_0}{dx} \frac{d^2 u_0}{dx^2} 
+ 5x \Bigl( \frac{d u_0}{dx} \Bigr)^2 - 6 u_0 \frac{d u_0}{dx}
+ \frac{2}{x} u_0^2 \biggr] \frac{d\theta_0}{dx}
\nonumber \\ & \quad \mbox{} 
+ \biggl[ 4\Bigl( x \frac{d u_0}{dx} - u_0 \Bigr) \frac{d^2 u_0}{dx^2}
- 2 \Bigl( \frac{d u_0}{dx} \Bigr)^2 + \frac{4}{x} u_0 \frac{d u_0}{dx}
- \frac{2}{x^2} u_0^2 \Biggr] \vartheta_0.
\end{align}
The solution is
\begin{equation}
\vartheta_2 = \bar{\rho}_c \bigg( -\frac{3}{4} \frac{4 + \cos^2x + \cos^4 x}{x \sin^3x}
- \frac{9}{2} \frac{\cos x}{x^2} + \frac{45}{4} \frac{\sin x}{x^3}
+ 18 \frac{\cos x}{x^4} - \frac{27}{2} \frac{\sin x}{x^5} \biggr).
\end{equation}
The mass variable is then determined by $d\bar{m}_2/dx = 3x^2\vartheta_2$, which yields 
\begin{equation}
\bar{m}_2 = \frac{9}{4} \bar{\rho}_c \biggl[ \frac{(4-\cos^2 x)\cos x}{\sin^2 x}
- \frac{4-7\cos^2 x}{\sin x} - 15 \frac{\cos x}{x} + 9\frac{\sin x}{x^2} \biggr]. 
\end{equation}
Near $x = 0$ we have that $\vartheta_2 = -\frac{1}{12} \bar{\rho}_c x^2 (1 + \frac{13}{25} x^2 + \cdots)$ and $\bar{m}_2 = -\frac{1}{20} \bar{\rho}_c x^5(1 + \frac{13}{70} x^2 + \cdots)$. 

The foregoing results apply to the inner core. In the outer shell the fluid is isotropic, and the solutions to the structure equations are
\begin{subequations}
\begin{align} 
\vartheta_{\rm outer} &= \bar{\rho}_c \biggl( A \frac{\sin x}{x} + B \frac{\cos x}{x} \biggr), \\
\bar{m}_{\rm outer} &= 3 \bar{\rho}_c \bigl[ A (\sin x - x \cos x)
+ B(\cos x + x\sin x) \bigr].
\end{align} 
\end{subequations}
The constants $A$ and $B$ are determined by ensuring the continuity of $\vartheta$ and $\bar{m}$ at $x = x_{\rm crit}$, where $\vartheta = 1$, and where the phase transition occurs.

A complete analytical solution is constructed by applying the following steps. First, we specify the numerical values of $\beta$ and $\bar{\rho}_c$. Second, we determine $x_{\rm crit}$ by solving the transcendental equation $\vartheta_0 + \beta^2 \vartheta_2 = 1$; this requires a numerical treatment. Third, we determine $A$ and $B$ by imposing continuity of $\vartheta$ and $\bar{m}$ at $x = x_{\rm crit}$; this can be done analytically. Fourth and finally, we determine $x_s$, the value of the radial coordinate at the stellar surface, by solving $\vartheta_{\rm outer} = 0$; this must also be done numerically. With all this, we can build plots of $\rho/\rho_{\rm crit}$ and $m/M$ as functions of $r/R$, and compare them with the numerical results obtained previously.

As an example we select $\beta = 0.2$ and $\bar{\rho}_c = 2.0$. For this model we find that $x_{\rm crit} \simeq 1.824405$, $A \simeq 0.9498547$, $B \simeq 0.02897347$, and $x_s \simeq 3.111099$. For this choice of parameters we find that the analytical results are barely distinguishable from the numerical results, and we declare the approximation excellent. As would be expected, however, the approximation degrades as $\beta$ is increased. For $\beta = 0.3$ we find a substantial departure from the numerical results.  

\begin{acknowledgments} 
This work was supported by the Natural Sciences and Engineering Research Council of Canada.  
\end{acknowledgments} 

\appendix

\section{Differential geometry of a moving surface}
\label{sec:moving_surface}  

We develop a differential geometry of a two-dimensional surface embedded in a three-dimensional, Euclidean space. The surface is taken to depend on time, so that it moves and alters its shape as time marches on. A helpful reference on this topic is the book by Grinfeld \cite{grinfeld:13}. Our description differs from his in some of the implementation details, but the spirit is very much the same. Our developments rely to a large extent on techniques introduced in Chapter 3 of Ref.~\cite{poisson:b04}, hereafter referred to as the {\it Toolkit}.

We begin in Sec.~\ref{subsec:description} with the mathematical description of the surface in terms of intrinsic coordinates and embedding relations; we define various geometric quantities, such as the induced metric and extrinsic curvature. We consider a change in intrinsic coordinates in Sec.~\ref{subsec:reparametrization}, and review Gauss's theorem in Sec.~\ref{subsec:gauss}. In Sec.~\ref{subsec:grid} we introduce the surface's grid velocity (the velocity of grid points in a given parametrization), in Sec.~\ref{subsec:partial_metric} we compute the time derivative of the induced metric, and other simular computations are presented in Sec.~\ref{subsec:computations}. In Sec.~\ref{subsec:hadamard} we introduce the important concept of Hadamard time derivative on the surface, and in Sec.~\ref{subsec:integral} we derive a useful identity for the time derivative of an integrated quantity.

The results obtained here are used extensively in Secs.~\ref{sec:interface} and \ref{sec:twophase} of the paper. 

\subsection{Description of a moving surface}
\label{subsec:description}

A two-dimensional surface $\SSS(t)$ is embedded in a three-dimensional ambient space, which we take to be Euclidean. We use arbitrary coordinates $x^a$ in the ambiant space, and in these the metric tensor is $g_{ab}$. We let $\nabla_a$ be the covariant-derivative operator compatible with $g_{ab}$; the associated connection is denoted $\Gamma^a_{bc}$.

The surface is described by the parametric equations $x^a = X^a(t, \theta^A)$, in which $\theta^A$ (with $A=2,3$) are intrinsic coordinates on $\SSS(t)$. The vectors
\begin{equation}
e^a_A := \frac{\partial X^a}{\partial \theta^A} = \partial_A X^a 
\end{equation}
are tangent to $\SSS(t)$, and $n^a$ is normal to the surface, so that $n_a e^a_A = 0$; we take it to be a unit vector, so that $n_a n^a = 1$. The tangent vectors satisfy $e^b_B \nabla_b e^a_A = e^b_A \nabla_b e^a_B$, the statement that they are Lie transported along each other. 

The induced metric on $\SSS(t)$ is
\begin{equation}
\Omega_{AB}(t,\theta) := g_{ab}\, e^a_A e^b_B,
\label{Omega_def} 
\end{equation}
in which $g_{ab}$ is evaluated at $x^a = X^a(t,\theta)$. We let $\Omega^{AB}$ denote its matrix inverse, $\Omega := \mbox{det}[\Omega_{AB}]$, and $D_A$ is the covariant-derivative operator compatible with the induced metric; the associated connection is denoted $\Gamma^A_{BC}$. The Levi-Civita tensor on $\SSS(t)$ is denoted $\varepsilon_{AB}$, with $\varepsilon_{23} = \Omega^{1/2}$. The three-dimensional metric evaluated on $\SSS(t)$ admits the completeness relation
\begin{equation}
g^{ab} = n^a n^b + \Omega^{AB} e^a_A e^b_B. 
\label{completeness}
\end{equation}
We shall use the notation $e^A_a := \Omega^{AB} g_{ab} e^b_B$. 

It is useful to note that the normal vector can be expressed as
\begin{equation}
n_a = \frac{1}{2} \varepsilon_{abc} \varepsilon^{BC} e^b_B e^c_C,
\label{normal} 
\end{equation} 
where $\varepsilon_{abc}$ is the Levi-Civita tensor of the ambiant space. A quick computation will indeed confirm that this expression is compatible with the defining properties $n_a e^a_A = 0$ and $n_a n^a = 1$ of the normal vector. It follows from this that the surface Levi-Civita tensor can be expressed as  
\begin{equation}
\varepsilon_{AB} =\varepsilon_{abc}\, e^a_A e^b_B n^c; 
\label{LC} 
\end{equation}
insertion of Eq.~(\ref{normal}) on the right-hand side of Eq.~(\ref{LC}) returns the left-hand side after straightforward manipulations. 

The surface's extrinsic curvature is defined by [{\it Toolkit} Eq.~(3.31)]
\begin{equation}
K_{AB} := e^a_A e^b_B \nabla_a n_b;
\label{extrinsic_def}
\end{equation}
it is symmetric under the exchange of $A$ and $B$. We let $K := \Omega^{AB} K_{AB}$ be its trace. Tangential derivatives of the basis vectors are given by the Gauss-Weingarten equation [{\it Toolkit} Eq.~(3.33)]
\begin{equation}
e^b_B \nabla_b e^c_A = \Gamma^C_{AB}\, e^c_C - K_{AB} n^c;
\label{GW1}
\end{equation}
both sides of the equation are symmetric under the exchange of $A$ and $B$. An alternative formulation of this equation is
\begin{equation}
D_B e^c_A = -\Gamma^c_{ab}\, e^a_A e^b_B - K_{AB} n^c = D_A e^c_B.
\label{GW2}
\end{equation}
This can be derived by developing the left-hand side, 
\begin{equation}
D_B e^c_A = \partial_B e^c_A - \Gamma^C_{AB} e^c_C
= e^b_B \partial_b e^c_A - \Gamma^C_{AB} e^c_C
= e^b_B \bigl( \nabla_b e^c_A - \Gamma^c_{ab} e^a_A \bigr) - \Gamma^C_{AB} e^c_C,
\end{equation}
and inserting Eq.~(\ref{GW1}). 

\subsection{Surface reparametrization}
\label{subsec:reparametrization}

A change of intrinsic coordinates on $\SSS(t)$ is described by
\begin{equation}
\theta^A = \Theta^A(t, \psi^M), \qquad
\psi^M = \Psi^M(t, \theta^A),
\label{reparam} 
\end{equation}
where $\psi^M$ are the new coordinates. The surface is described by $x^a = Y^a(t, \psi)$ in the new parametrization, where 
\begin{equation}
Y^a(t,\psi) := X^a\bigl(t, \Theta(t,\psi)\bigr). 
\label{Y_vs_X} 
\end{equation}
Conversely, we have that 
\begin{equation}
X^a(t,\theta) = Y^a\bigl(t, \Psi(t,\theta)\bigr).
\label{X_vs_Y} 
\end{equation}

The differential expression of Eqs.~(\ref{reparam}) is
\begin{equation}
d\theta^A = \partial_t \Theta^A\, dt + \partial_M\Theta^A\, d\psi^M, \qquad
d\psi^M = \partial_t \Psi^M\, dt + \partial_A \Psi^M\, d\theta^A.
\end{equation}
Compatibility of these equations implies the set of identities
\begin{subequations} 
\label{diff_identities} 
\begin{align} 
0 &= \partial_t \Theta^A + \partial_t \Psi^M\, \partial_M \Theta^A, \\
0 &= \partial_t \Psi^M + \partial_t \Theta^A\, \partial_A \Psi^M, \\
\delta^A_B &= \partial_M \Theta^A\, \partial_B \Psi^M, \\
\delta^M_N &= \partial_A \Psi^M\, \partial_N \Theta^A.
\end{align}
\end{subequations} 

The new parametrization produces a new set of geometric objects, including the tangent vectors $e^a_M := \partial_M Y^a$, the induced metric $\Omega_{MN}$, the extrinsic curvature $K_{MN}$, and so on. All such objects transform as surface tensors. For example,
\begin{equation}
e^a_M = \partial_M X^a\bigl(t, \Theta(t,\psi)\bigr) = \partial_A X^a(t,\theta)\, \partial_M \Theta^A
= e^a_A\, \partial_M \Theta^A; 
\end{equation}
this is the standard transformation rule for covectors. 

An important aspect of the differential geometry of a moving surface is that the time-derivative operator $\partial_t$ {\it does not} return a surface tensor when acting on a surface tensor. To investigate this in the simplest context, let $f$ be a scalar field on $\SSS(t)$, and let us work out the transformation of $\partial_t f$, by which we mean ``partial derivative with respect to $t$, keeping the intrinsic coordinates fixed''. Because $f$ is a scalar, the functions $f(t,\theta)$ and $f(t,\psi)$ are related by
\begin{equation}
f(t,\psi^M) := f\bigl(t, \theta^A=\Theta^A(t,\psi^M)\bigr).
\end{equation}
Differentiation with respect to $t$ at fixed $\psi^M$ produces
\begin{equation}
\partial_t f(t,\psi) = \partial_t f(t,\theta) + \partial_t \Theta^A\, \partial_A f(t,\theta),
\label{partial_scalar} 
\end{equation}
and we see that indeed, $\partial_t f$ does not transform as a scalar under the coordinate transformation of Eq.~(\ref{reparam}). The reason for this is that the transformation between $\theta^A$ and $\psi^M$ is time dependent, so that keeping $\theta^A$ fixed is necessarily different from keeping $\psi^M$ fixed. The basic lesson is that the partial time derivative of a surface tensor will not itself be a surface tensor.

\subsection{Two-dimensional Gauss theorem}
\label{subsec:gauss}

Let a portion $\UU(t)$ of $\SSS(t)$ be enclosed by a curve $\C(t)$, which we describe by the parametric equations $\theta^A = \vartheta^A(t,\chi)$, with $\chi$ a running parameter on $\C(t)$. The vector
\begin{equation}
e^A := \frac{\partial \vartheta^A}{\partial \chi}
\end{equation}
is tangent to $\C(t)$, and $r^A$ shall be the unit normal vector, which points out of $\UU(t)$. 

Let $v^A$ be a vector field in $\UU(t)$. The statement of Gauss's theorem is that
\begin{equation}
\int_{\UU(t)} D_A v^A\, dS = \oint_{\C(t)} v^A d\ell_A,
\label{gauss} 
\end{equation}
where $dS := \sqrt{\Omega}\, d^2\theta$ is the element of surface area on $\SSS(t)$, while
\begin{equation}
d\ell_A := \varepsilon_{AB} e^B\, d\chi = r_A\, ds
\label{dell1}
\end{equation}
is a normal-directed element of length on $\C(t)$; $\varepsilon_{AB}$ is the Levi-Civita tensor on $\SSS(t)$, and $ds$ is the element of arclength on $\C(t)$. In Eq.~(\ref{gauss}) it is assumed that the direction along which $\C(t)$ is traversed is compatible (in accordance with the right-hand rule) with the direction chosen for $n^a$, the surface's unit normal.

This version of Gauss's theorem is established by following the strategy adopted in Sec.~3.3.1 of the {\it Toolkit}; we shall not go into the details here. To establish the second part of Eq.~(\ref{dell1}) we remark first that $\varepsilon_{AB} e^B$ must be proportional to $r_A$, since the construction is necessarily orthogonal to $e^A$. We therefore have $\varepsilon_{AB} e^B = f r_A$ for some $f$, and it follows that $f = \varepsilon_{AB} r^A e^B$. To perform the calculation we adopt coordinates $\theta^A = (\zeta,\chi)$, with lines of constant $\zeta$ crossing out of $\C$. Then $f = \varepsilon_{12} r^\zeta = \Omega^{1/2} r^\zeta$. We have that $r_\zeta$ is the only nonvanishing component of $r_A$, and $1 = \Omega^{AB} r_A r_B$ implies $r_\zeta =(\Omega^{\zeta\zeta})^{-1/2}$. It follows that $r^\zeta = (\Omega^{\zeta\zeta})^{1/2}$, and we arrive at $f = (\Omega \Omega^{\zeta\zeta})^{1/2} =(\Omega_{\chi\chi})^{1/2}$. Finally, we have that $d\ell_A = r_A\, f\, d\chi = r_A (\Omega_{\chi\chi})^{1/2}\, d\chi = r_A\, ds$, in agreement with Eq.~(\ref{dell1}). 

\subsection{Grid velocity}
\label{subsec:grid} 

We follow the motion of a point on $\SSS(t)$ with a {\it fixed set} of intrinsic coordinates $\theta^A$. The description of this motion in the three-dimensional ambiant space is provided by
\begin{equation}
x^a = X^a(t, \theta^A=\mbox{constant}).
\end{equation}
The velocity of this grid point is then
\begin{equation}
W^a(t,\theta) := \partial_t X^a(t, \theta).
\label{grid_velocity} 
\end{equation}
We shall name this the {\it grid velocity}.

As can be gathered from the discussion surrounding Eq.~(\ref{partial_scalar}), the components of the grid velocity {\it do not} constitute a set of surface scalars (unlike, say, the components of the normal vector). Their transformation property can be inferred by taking the time derivative of Eqs.~(\ref{Y_vs_X}) and (\ref{X_vs_Y}), which gives
\begin{subequations}
\label{Wa_transf}
\begin{align}
W^a(t,\psi) &= W^a(t,\theta) + \partial_t \Theta^A\, e^a_A, \\ 
W^a(t,\theta) &= W^a(t,\psi) + \partial_t \Psi^M\, e^a_M.
\end{align}
\end{subequations}
These equations are compatible by virtue of Eq.~(\ref{diff_identities}). 

We decompose the grid velocity according to
\begin{equation}
W^a(t,\theta) = W_n(t,\theta)\, n^a + W^A(t,\theta)\, e^a_A,
\label{Wa_decomp}
\end{equation}
where $W_n := n_a W^a$ is the normal component, while $W^A := e^A_a W^a$ are the tangential components. Equation (\ref{Wa_transf}) reveals that $W_n$ is actually a surface scalar,
\begin{equation}
W_n(t,\psi) = W_n(t,\theta).
\label{Wn_transf}
\end{equation}
On the other hand, the tangential components transform according to
\begin{subequations}
\label{WA_transf}
\begin{align} 
W^M(t,\psi) &= \bigl[ W^A(t,\theta) + \partial_t \Theta^A \bigr] \partial_A \Psi^M, \\
W^A(t,\theta) &= \bigl[ W^M(t,\psi) + \partial_t \Psi^M \bigr] \partial_M \Theta^A,
\end{align}
\end{subequations}
and they are {\it not} surface vectors.

Because $W^A$ is not a surface vector, the action of the covariant derivative $D_B$ on this object is not defined {\it a priori}. We shall define it by the standard rule,
\begin{equation}
D_B W^A := \partial_B W^A + \Gamma^A_{BC} W^C,
\end{equation}
with the understanding that the outcome is {\it not} a surface tensor.

\subsection{Time derivative of the induced metric}
\label{subsec:partial_metric}

We calculate $\partial_t \Omega_{AB}$, in which the differentiation is carried out at fixed $\theta^A$. We simplify the computation by taking the ambiant coordinates $x^a$ to be Cartesian; because $\Omega_{AB}$ and its time derivative form sets of ambiant scalars, the outcome shall be independent of this choice, and there is no loss of generality.

We begin with Eq.~(\ref{Omega_def}), in which we substitute $g_{ab} = \delta_{ab}$, and we differentiate it with respect to $t$. Taking into account that
\begin{equation}
\partial_t e^a_A = \partial_{tA} X^a = \partial_A W^a,
\end{equation}
we have that
\begin{equation}
\partial_t \Omega_{AB} = \delta_{ab} \bigl( \partial_A W^a\, e^b_B + e^a_A\, \partial_B W^b \bigr).
\end{equation}
Next we differentiate the decomposition of Eq.~(\ref{Wa_decomp}) with respect to $\theta^A$, and find
\begin{equation}
\partial_A W^a = (\partial_A W_n) n^a + W_n \partial_A n^a
+ (D_A W^C) e^a_C + W^C D_A e^a_C. 
\label{partialA_Wa} 
\end{equation}
In this we insert Eq.~(\ref{GW2}) with $\Gamma^c_{ab} = 0$, and get
\begin{equation}
\partial_A W^a = \bigl( \partial_A W_n - K_{AC} W^C \bigr) n^a
+ W_n \partial_A n^a + (D_A W^C) e^a_C.
\end{equation}
We make the substitution within our previous expression for $\partial_t \Omega_{AB}$, and take into account the orthogonality of $n^a$ and $e^a_A$. We obtain
\begin{equation}
\partial_t \Omega_{AB} = W_n \delta_{ab} \bigl( \partial_A n^a\, e^b_B + e^a_A\, \partial_B n^b \bigr)
+ D_A W_B + D_B W_A.
\end{equation}
In the final step we return to Eq.~(\ref{extrinsic_def}), which we expand as
\begin{equation}
K_{AB} = e^a_A e^b_B\, \partial_a n_b = \frac{1}{2} e^a_A e^b_B (\partial_a n_b + \partial_b n_a)
= \frac{1}{2} \bigl( \partial_A n_b\, e^b_B + e^a_A\, \partial_B n_a \bigr)
= \frac{1}{2} \delta_{ab} \bigl( \partial_A n^a\, e^b_B + e^a_A\, \partial_B n^b \bigr),
\end{equation}
and which we recognize in the preceding expression for $\partial_t \Omega_{AB}$.

We have arrived at
\begin{equation}
\partial_t \Omega_{AB}(t,\theta) = 2 W_n K_{AB} + D_A W_B + D_B W_A.
\label{partial_Omega1}
\end{equation}
This, as expected, is {\it not} a surface tensor, because $W_A$ is not a surface vector. From Eq.~(\ref{partial_Omega1}) we immediately obtain
\begin{equation}
\frac{1}{\sqrt{\Omega}} \partial_t \sqrt{\Omega} = \frac{1}{2} \Omega^{AB} \partial_t \Omega_{AB}
= W_n K + D_A W^A,
\label{partial_Omega2}
\end{equation}
where $K := \Omega^{AB} K_{AB}$ is the trace of the extrinsic curvature.

\subsection{Other computations}
\label{subsec:computations}

In this section we compute derivatives of various geometrical quantities defined on $\SSS(t)$. In all cases $\partial_t$ indicates partial differentiation with respect to $t$ at fixed $\theta^A$, and $\partial_A$ indicates partial differentiation with respect to $\theta^A$. 

\subsubsection{Derivatives of the ambiant metric} 

The ambiant metric evaluated on $\SSS(t)$ is
\begin{equation} 
g_{ab}(t,\theta) := g_{ab}\bigl(x = X(t,\theta)\bigr). 
\end{equation}
Differentiation with respect to $t$ produces $\partial_t g_{ab} = \partial_t X^c\, \partial_c g_{ab}$, or
\begin{equation}
\partial_t g_{ab}(t,\theta) =W^c \bigl( \Gamma^d_{ac}\, g_{db} + \Gamma^d_{bc}\, g_{ad} \bigr).
\label{partial_g}
\end{equation}
Similarly,
\begin{equation}
\partial_t g^{ab}(t,\theta) =-W^c \bigl( \Gamma^a_{cd}\, g^{db} + \Gamma^b_{cd}\, g^{ad} \bigr), 
\label{partial_ginv}
\end{equation}
and almost identical computations produce
\begin{subequations}
\label{partialA_g}
\begin{align}
\partial_A g_{ab} &= e^c_A \bigl( \Gamma^d_{ac}\, g_{db} + \Gamma^d_{bc}\, g_{ad} \bigr), \\
\partial_A g^{ab} &= -e^c_A \bigl( \Gamma^a_{cd}\, g^{db} + \Gamma^b_{cd}\, g^{ad} \bigr).
\end{align}
\end{subequations}

\subsubsection{Angular derivatives of the normal vector} 

Next we relate $\partial_A n^a$ to the extrinsic curvature. We return to Eq.~(\ref{extrinsic_def}), which gives  
\begin{equation}
K_A^{\ B}\, e^b_B = \bigl( e^b_B e^B_d \bigr) e^a_A \nabla_a n^d 
= (\delta^b_d - n^b n_d) e^a_A \nabla_a n^d = e^a_A \nabla_a n^b,
\end{equation}
in which we used the completeness relation of Eq.~(\ref{completeness}), as well as the fact that $n^a$ is a unit vector, so that $n_d \nabla_a n^d = \frac{1}{2} \nabla_a (n_d n^d) = 0$. We then expand the covariant derivative on the right-hand side, and get
\begin{equation}
\partial_A n^a = K_A^{\ C}\, e^a_C - \Gamma^a_{bc} n^b e^c_A.
\label{partialA_na1}
\end{equation}
Lowering the $a$-index with the help of Eq.~(\ref{partialA_g}), we also have that
\begin{equation}
\partial_A n_a = K_{AC}\, e^C_a + \Gamma^b_{ac} n_b e^c_A.
\label{partialA_na2}
\end{equation}

\subsubsection{Time derivative of the tangential vectors} 

We now proceed with the computation of $\partial_t e^a_A = \partial_A W^a$. We insert Eqs.~(\ref{GW2}) and (\ref{partialA_na1}) within Eq.~(\ref{partialA_Wa}), and obtain
\begin{equation}
\partial_t e^a_A = \bigl( \partial_A W_n - K_{AB} W^B \bigr) n^a
+ \bigl( D_A W^C + W_n K_A^{\ C} \bigr) e^a_C
- \Gamma^a_{bc} W^b e^c_A
\label{partial_eaA}
\end{equation}
after some simplification. 

\subsubsection{Time derivative of the normal vector} 

Next we calculate $\partial_t n^a$, which we decompose according to
\begin{equation}
\partial_t n^a = C n^a + C^A e^a_A,
\label{partial_n_decomp} 
\end{equation} 
with $C := n_a \partial_t n^a$ and $C^A := e^A_a \partial_t n^a$. To get an expression for $C$ we differentiate the identity $g_{ab} n^a n^b = 1$ with respect to $t$, and insert Eq.~(\ref{partial_g}). We obtain
\begin{equation}
C = -\Gamma^a_{bc} n_a n^b W^c.
\label{C_result}
\end{equation}
For $C^A$ we differentiate $g_{ab} e^a_A n^b = 0$ and insert Eqs.~(\ref{partial_g}) and (\ref{partial_eaA}). This gives
\begin{equation}
C^A = -D^A W_n + K^A_{\ B} W^B - \Gamma^a_{bc} e^A_a n^b W^c.
\label{CA_result}
\end{equation}
Our final expression for $\partial_t n^a$ is
\begin{equation}
\partial_t n^a = -\bigl( D^A W_n - K^A_{\ B} W^B \bigr) e^a_A - \Gamma^a_{bc} n^b W^c.
\label{partial_n1}
\end{equation}
Lowering the $a$-index with the help of Eq.~(\ref{partial_g}), we also have that 
\begin{equation}
\partial_t n_a = -\bigl( \partial_A W_n - K_{AB} W^B \bigr) e^A_a + \Gamma^b_{ac} n_b W^c.
\label{partial_n2}
\end{equation}

The appearance of connection terms in Eqs.~(\ref{partial_n1}) and (\ref{partial_n2}) may seem surprising. To elucidate this, let us examine a vector field $p^a(t,x)$ defined in the ambiant space. Its restriction to $\SSS(t)$ is $p^a(t,\theta) :=  p^a(t,x=X(t,\theta))$, and its derivative with respect to $t$ at fixed $\theta^A$ is
\begin{equation}
\partial_t p^a(t,\theta) = \partial_t p^a(t,x) + W^b \partial_b p^a
= \partial_t p^a(t,x) + W^b \nabla_b p^a - \Gamma^a_{bc} p^b W^c.
\end{equation}
The connection term, therefore, arises when a partial derivative at fixed $\theta^A$ is written in terms of a partial derivative at fixed $x^a$. While $\partial_t p^a(t,x)$ is an ambiant vector, $\partial_t p^a(t,\theta)$ does not constitute a set of surface scalars.

\subsubsection{Time derivative of the extrinsic curvature} 

As a final exercise we compute $\partial_t K_{AB}$. To simplify the task we adopt the strategy of Sec.~\ref{subsec:partial_metric} and let $x^a$ be Cartesian coordinates. We then have
\begin{equation}
K_{AB} = e^a_A e^b_B \partial_a n_b = e^b_B \partial_A n_b,
\end{equation}
and differentiation with respect to $t$ produces
\begin{equation}
\partial_t K_{AB} = \bigl( \partial_t e^b_B \bigr) \partial_A n_b + e^b_B\, \partial_{tA} n_b.
\end{equation}
The time derivative of $e^b_B$ was obtained in Eq.~(\ref{partial_eaA}), and to compute $\partial_{tA} n_b$ we differentiate Eq.~(\ref{partial_n2}) with respect to $\theta^A$. Recalling that $\Gamma^a_{bc} = 0$ in the context of this calculation, we have that
\begin{equation}
\partial_{tA} n_b = \bigl( -D_{AB} W_n + D_A K_{BC}\, W^C + K_{BC} D_A W^C \bigr) e^B_b
+ \bigl( K_A^{\ B} \partial_B W_n - K_A^{\ B} K_{BC} W^C \bigr) n_b.
\end{equation}
We make the substitutions in $\partial_t K_{AB}$, and simplify the result with the Codazzi equation [{\it Toolkit} Eq.~(3.40)],
\begin{equation}
D_A K_{BC} = D_C K_{AB}.
\end{equation}
We arrive at
\begin{equation}
\partial_t K_{AB} = -D_{AB} W_n + W_n K_{AC} K^C_{\ B} + W^C D_C K_{AB}
+ K_{AC} D_B W^C + K_{BC} D_A W^C.
\label{partial_K}
\end{equation} 

\subsection{Hadamard time derivative}
\label{subsec:hadamard}

It is possible to introduce an alternative time-derivative operator that acts on a surface tensor and returns a surface tensor. The construction apparently originates with Hadamard in Ref.~\cite{hadamard:03}. In his text \cite{grinfeld:13}, Grinfeld introduces a variant of the Hadamard derivative, but we shall not consider it here.

\subsubsection{Scalar}

To begin we examine a scalar field $f(t,x)$ defined in the ambiant space. When acting on such a function, the Hadamard time derivative is defined to be
\begin{equation}
\D_t f := \partial_t f(t,x) + W_n n^a \partial_a f.
\label{Df_def1} 
\end{equation}
The meaning of the right-hand side goes as follows. Let $P$ be a point on $\SSS(t)$ identified by the intrinsic coordinates $\theta^A$. The point $Q$ on $\SSS(t+dt)$ with the same values of $\theta^A$ is displaced from $P$ by the vector $\bm{W}\, dt$, where $\bm{W}$ is the grid-velocity vector. We decompose $\bm{W}$ according to $W_n \bm{n} + W^A \bm{e}_A$, and instead of $Q$, we select the point $R$ on $\SSS(t+dt)$ that is displaced from $P$ by the normal vector $W_n \bm{n}\, dt$. Then $\D_t f\, dt = f(t+dt,R) - f(t,P)$.

We now wish to rewrite the right-hand side of Eq.~(\ref{Df_def1}) in terms of $\partial_t f(t,\theta)$, with $f(t,\theta) := f(t,x=X(t,\theta))$ representing the restriction of $f$ to $\SSS(t)$. Differentiating with respect to $t$ at fixed $\theta^A$, we have that
\begin{equation}
\partial_t f(t,\theta) = \partial_t f(t,x) + \partial_t X^a\, \partial_a f
= \partial_t f(t,x) + W^a \partial_a f
= \partial_t f(t,x) + \bigl( W_n n^a + W^A e^a_A \bigr) \partial_a f.
\end{equation}
Making the substitution in Eq.~(\ref{Df_def1}), we arrive at
\begin{equation}
\D_t f := \partial_t f(t,\theta) - W^A \partial_A f.
\label{Df_def2}
\end{equation}
We shall take this to be the official definition of the Hadamard time derivative on a surface scalar, when the scalar $f(t,\theta)$ is defined on $\SSS(t)$ only. The geometrical meaning of the operation remains unchanged. 

It is easy to show, using Eqs.~(\ref{partial_scalar}) and (\ref{WA_transf}), that $\D_t f$ transforms as a surface scalar:
\begin{equation}
\D_t f(t, \psi) = \D_t f(t, \theta).
\label{D_transf_scalar} 
\end{equation}
This property confers its utility to the Hadamard time derivative.

\subsubsection{Ambiant tensors}

Let $T^{a\cdots}_{b\cdots} (t,x)$ be a tensor defined in the ambiant space. Its Hadamard time derivative shall be the natural generalization of Eq.~(\ref{Df_def1}), 
\begin{equation}
\D_t T^{a\cdots}_{b\cdots} := \partial_t T^{a\cdots}_{b\cdots}(t,x)
+ W_n n^c \nabla_c T^{a\cdots}_{b\cdots}. 
\label{DT_def1}
\end{equation}
The operation produces another ambiant tensor. 

The restriction of the tensor to the surface is
\begin{equation}
T^{a\cdots}_{b\cdots}(t,\theta) := T^{a\cdots}_{b\cdots}\bigl(x = X(t,\theta)\bigr),
\end{equation}
and its partial time derivative at fixed $\theta^A$ is
\begin{equation}
\partial_t T^{a\cdots}_{b\cdots}(t,\theta)
= \partial_t T^{a\cdots}_{b\cdots}(t, x)
+ \bigl( W_n n^c + W^A e^c_A \bigr) \partial_c T^{a\cdots}_{b\cdots}.
\end{equation}
We convert the $\partial_c$ derivative to a covariant derivative and make the substitution within Eq.~(\ref{DT_def1}). We arrive at
\begin{equation}
\D_t T^{a\cdots}_{b\cdots} := \partial_t T^{a\cdots}_{b\cdots}(t,\theta)
- W^A \partial_A T^{a\cdots}_{b\cdots}
+ W_n n^c \bigl( \Gamma^a_{cd} T^{d\cdots}_{b\cdots} + \cdots
- \Gamma^d_{cb} T^{a\cdots}_{d\cdots} - \cdots \bigr), 
\label{DT_def2}
\end{equation}
and take this as the official definition of the Hadamard derivative when the tensor is defined on $\SSS(t)$ only.

A calculation analogous to the one leading to Eq.~(\ref{D_transf_scalar}) reveals that
\begin{equation}
\D_t T^{a\cdots}_{b\cdots}(t, \psi) = \D_t T^{a\cdots}_{b\cdots}(t, \theta). 
\label{D_transf_ambiant} 
\end{equation} 
The ambiant tensor $T^{a\cdots}_{b\cdots}$ constitutes a set of surface scalars --- they transform as such under a change of intrinsic coordinates. Equation (\ref{D_transf_ambiant}) therefore states that this scalar property is preserved by the Hadamard derivative. It may be useful to note that in Eq.~(\ref{DT_def2}), $W_n$, $n^c$, and $\Gamma^a_{bc}$ are all surface scalars.

\subsubsection{Surface tensors}

Noting that $W^A \partial_A f = \Lie_W f$ is the Lie derivative of the scalar $f$ in the direction of $W^A$, a natural tensorial generalization of Eq.~(\ref{Df_def2}) to surface tensors is
\begin{subequations}
\label{DS_def}
\begin{align} 
\D_t S^{A\cdots}_{B\cdots} &:= \partial_t S^{A\cdots}_{B\cdots}(t,\theta)
- \Lie_W S^{A\cdots}_{B\cdots}, \\
&= \partial_t S^{A\cdots}_{B\cdots}(t,\theta)
- W^C \partial_C S^{A\cdots}_{B\cdots}
+ S^{C\cdots}_{B\cdots}\, \partial_C W^A + \cdots  
- S^{A\cdots}_{C\cdots}\, \partial_B W^C - \cdots, \\
&= \partial_t S^{A\cdots}_{B\cdots}(t,\theta)
- W^C D_C S^{A\cdots}_{B\cdots}
+ S^{C\cdots}_{B\cdots}\, D_C W^A + \cdots  
- S^{A\cdots}_{C\cdots}\, D_B W^C - \cdots.
\end{align}
\end{subequations}
The operation takes a surface tensor and returns another surface tensor; the transformation rule is 
\begin{equation}
\D_t S^{M\cdots}_{N\cdots}(t, \psi)
= \partial_A \Psi^M \cdots \partial_N \Theta^B \cdots
\D_t S^{A\cdots}_{B\cdots}(t, \theta), 
\label{D_transf_surface} 
\end{equation}
as required of a surface tensor. 

The explicit demonstration of this property relies on a fairly laborious computation, which makes use of Eqs.~(\ref{diff_identities}) and (\ref{WA_transf}). Some of the differential identities of Eq.~(\ref{diff_identities}) appear in differentiated form. For example, differentiation of
\begin{equation}
\partial_t \Psi^M + \partial_t \Theta^B\, \partial_B \Psi^M = 0
\end{equation}
with respect to $\theta^A$ produces
\begin{equation}
\partial_{tA} \Psi^M + \partial_{tN} \Theta^B\, \partial_A \Psi^N\, \partial_B \Psi^M
+ \partial_t \Theta^B\, \partial_{AB} \Psi^M = 0.
\end{equation}
As another example, differentiation of
\begin{equation}
\partial_B \Psi^N\, \partial_N \Theta^A = \delta^A_B
\end{equation}
with respect to $\psi^M$ produces
\begin{equation}
\partial_{BC} \Psi^N\, \partial_M \Theta^C\, \partial_N \Theta^A
+ \partial_B \Psi^N\, \partial_{MN} \Theta^A = 0.
\end{equation}
Making these substitutions at various points in the computation returns Eq.~(\ref{D_transf_surface}) after simplification.

\subsubsection{Hybrid tensors}

The previous definitions of the Hadamard derivative can be adapted to handle hybrid tensors with a number of ambiant and surface indices. The idea is to follow Eq.~(\ref{DT_def2}) and include a connection term for each ambiant index, and to follow Eq.~(\ref{DS_def}) and exploit the Lie derivative to take care of the surface indices. A concrete example is
\begin{equation}
\D_t P^a_A := \partial_t P^a_A(t,\theta) - W^B \partial_B P^a_A - P^a_B\, \partial_A W^B
+ W_n n^c \Gamma^a_{bc} P^b_A.
\label{DP_def} 
\end{equation}
This is a vector when viewed from the ambiant space, and a covector when viewed from the surface.

\subsubsection{Examples}

The definition of Eq.~(\ref{DT_def1}) implies immediately that
\begin{equation}
\D_t g_{ab} = 0.
\label{D_g}
\end{equation}
This can also be recovered from Eq.~(\ref{DT_def2}), by making use of Eqs.~(\ref{partial_g}) and (\ref{partialA_g}).

The partial time derivative of the induced metric was given in Eq.~(\ref{partial_Omega1}). Combining this with Eq.~(\ref{D_transf_surface}) applied to $\Omega_{AB}$, we have that
\begin{equation}
\D_t \Omega_{AB} = 2 W_n K_{AB}.
\label{D_Omega}
\end{equation}
Both sides of the equation are proper surface tensors. 

When we insert Eq.~(\ref{partial_eaA}) within Eq.~(\ref{DP_def}) for $e^a_A$ and make use of Eq.~(\ref{GW2}), we obtain
\begin{equation} 
\D_t e^a_A = (\partial_A W_n) n^a + W_n K_A^{\ C} e^a_C.
\label{D_eaA} 
\end{equation}
Like the left-hand side, for each value of $a$ the right-hand side is a surface covector. 

To compute $\D_t n^a$ we make use of Eq.~(\ref{partial_n1}) for $\partial_t n^a$ and Eq.~(\ref{partialA_na1}) for $\partial_A n^a$. Making the substitutions in Eq.~(\ref{DT_def2}), we get
\begin{equation}
\D_t n^a = -\bigl( D^A W_n \bigr) e^a_A.
\label{D_n1} 
\end{equation}
Lowering the metric with Eq.~(\ref{D_g}) gives
\begin{equation}
\D_t n_a = -\bigl( \partial_A W_n \bigr) e^A_a.
\label{D_n2} 
\end{equation}

Finally, we compute $\D_t K_{AB}$ by combining Eq.~(\ref{partial_K}) with Eq.~(\ref{DS_def}). This yields
\begin{equation}
\D_t K_{AB} = -D_{AB} W_n + W_n K_{AC} K^C_{\ B}, 
\label{D_K}
\end{equation}
and we confirm once more that the right-hand side is a surface tensor. 

\subsection{Time derivative of an integrated quantity}
\label{subsec:integral}

Let $\UU(t)$ be a portion of the surface $\SSS(t)$, and let it correspond to a time-independent domain $D$ of some system $\theta^A$ of intrinsic coordinates. Let $F(t)$ be the integral of a surface scalar $f$ over the region $\UU(t)$,
\begin{equation}
F(t) := \int_{\UU(t)} f\, dS = \int_D f(t,\theta)\, \sqrt{\Omega}\, d^2\theta.
\end{equation}
Then the time derivative of this quantity is
\begin{equation}
\frac{dF}{dt} = \int_D \bigl[ \D_t f + f W_n K + D_A(f W^A) \bigr] \sqrt{\Omega}\, d^2\theta, 
\label{dFdt}
\end{equation}
where $K := \Omega^{AB} K_{AB}$ is the trace of the extrinsic curvature. The integral is formulated in the special coordinate system. While $\D_t f$ and $f W_n  K$ are surface scalars, $D_A (f W^A)$ is not because $W^A$ is not a surface vector. Notice that this term could be sent to a boundary integral by appealing to Gauss's theorem, and that the remaining integral over $\UU(t)$ would then involve surface scalars only. The failure of $dF/dt$ to be expressed solely in terms of surface scalars comes from the fact that the region $\UU(t)$ can vary arbitrarily as time marches on; it is mapped to a fixed domain $D$ in a very specific coordinate system.

The derivation of Eq.~(\ref{dFdt}) relies on the time-invariance of the domain $D$, so that
\begin{equation}
\frac{dF}{dt} = \int_D \frac{\partial}{\partial t} \bigl( f \sqrt{\Omega} \bigr)\, d^2\theta.
\end{equation}
We obtain
\begin{equation} 
\frac{dF}{dt} = \int_D \bigl[ \partial_t f + f(W_n K + D_A W^A) \bigr] \sqrt{\Omega}\, d^2\theta
\end{equation}
after making use of Eq.~(\ref{partial_Omega2}), and we arrive at Eq.~(\ref{dFdt}) after recalling the definition of Eq.~(\ref{Df_def2}) for the Hadamard time derivative.

\section{Gravitational Lagrangian of a surface fluid}
\label{sec:interface_gravity} 

In this Appendix we examine the gravitational piece of the Lagrangian of a surface fluid, given by
\begin{equation}
L_{\rm grav} := \int_{\SSS(t)} \sigma U\, dS - \frac{1}{8\pi G} \int \nabla_a U \nabla^a U\, dV,
\label{L_grav}
\end{equation}
with the second integral defined over all space. We are interested in the consequences of the fact that the potential $U$ is not differentiable on $\SSS(t)$.

We begin in Sec.~\ref{subsec:potential} with a characterization of the gravitational potential of a singular surface, and a demonstration that its normal derivative is discontinuous across the surface. In Sec.~\ref{subsec:regularized} we imagine that the surface actually possesses a nonvanishing thickness, so that it produces a smooth potential; this device allows us to resolve ambiguities that manifest themselves in some calculations. We introduce the Eulerian and Lagrangian variations of the gravitational potential in Sec.~\ref{subsec:variation1}, take the variation of the Lagrangian of Eq.~(\ref{L_grav}) with respect to the potential in Sec.~\ref{subsec:variation2}, and its variation with respect to the fluid configuration in Sec.~\ref{subsec:variation3}.

The computations presented here are used in Secs.~\ref{sec:interface} and \ref{sec:twophase} of the paper. 

\subsection{Potential of a singular surface}
\label{subsec:potential} 

To deduce the properties of $U$ in the vicinity of a surface with a finite area density $\sigma$ but a formally infinite volume density $\rho$, it is helpful to work in Gaussian coordinates $(\ell,\theta^A)$. Here $\ell$ is the distance away from $\SSS(t)$ as measured along geodesics (straight lines) that intersect the surface orthogonally, and $\theta^A$ are intrinsic coordinates on $\SSS(t)$, extended away from the surface by preserving their values on each orthogonal geodesic. We continue to take $\theta^A$ to be Lagrangian coordinates on the surface, so that fluid elements move with constant values of $\theta^A$. The construction allows us to express the volume density as the distribution
\begin{equation}
\rho = \sigma(\theta^A)\, \delta(\ell),
\end{equation}
where $\delta(\ell)$ is the Dirac delta function. The unit normal to the surface can be written as $n_a = \nabla_a \ell$. The tangential vectors are $e^a_A := \partial x^a/\partial \theta^A$. 

Anticipating that the variation of $L_{\rm grav}$ with respect to the gravitational potential will produce a distributional Poisson's equation, 
\begin{equation}
\nabla^2 U = -4\pi G \sigma\, \delta(\ell),
\label{poisson_interface} 
\end{equation} 
we recognize that the solution also will have a distributional nature. We write it as
\begin{equation}
U = U_-\, \Theta(-\ell) + U_+\, \Theta(\ell),
\end{equation}
where $U_-$ is the potential for $\ell < 0$, $U_+$ is the potential for $\ell > 0$, and $\Theta(\ell)$ denotes the Heaviside step function. Substitution within Eq.~(\ref{poisson_interface}) produces $\nabla^2 U_\pm = 0$ together with the junction conditions
\begin{equation}
[U] = 0, \qquad n^a [\nabla_a U] = -4\pi G \sigma,
\label{junction}
\end{equation}
where $[\psi] := \psi_+(\SSS(t)) - \psi_-(\SSS(t))$ is the jump of the quantity $\psi$ across the surface. Continuity of $U$ implies that the tangential derivatives are also continuous, $e^a_A [\nabla_a U] = 0$, and we conclude that
\begin{equation}
[\nabla_a U] = -4\pi G \sigma\, n_a.
\label{jump_gradU}
\end{equation}
In words, we have that the gradient of $U$ is discontinuous across the surface, and that the discontinuity is directed along the normal vector. 

\subsection{Regularized surface}
\label{subsec:regularized} 

The distributional methods utilized in the preceding subsection can be very convenient in many instances, but in others they can lead to ambiguities. In such cases it is helpful to step away from the idealization, and to think of the surface as a thin (but not infinitely thin) layer of fluid with a large (but not infinitely large) volume density. We imagine the layer to be extending from $\ell = -\ell_0$ to $\ell = +\ell_0$, with $\ell_0$ small, and we take its regularized volume density $\tilde{\rho}$ to be sharply peaked around $\ell = 0$. The layer's area density is then
\begin{equation}
\sigma = \int_{-\ell_0}^{\ell_0} \tilde{\rho}\, d\ell.
\end{equation}
The solution to the regularized Poisson equation is denoted $\tilde{U}$, so that $\nabla^2 \tilde{U} = -4\pi G \tilde{\rho}$. We let $\tilde{g}_a := \nabla_a \tilde{U}$ be the regularized gravitational field, and write Poisson's equation as
\begin{equation}
\nabla_a \tilde{g}^a = -4\pi G \tilde{\rho},
\end{equation}
in the form of Gauss's law.

To investigate the behavior of $\tilde{g}_a$ near the layer we perform a pillbox integration over a small volume that straddles it. The divergence theorem allows us to write $\int \nabla_a \tilde{g}^a\, dV = [\tilde{g}_a] n^a A$, where $[\tilde{g}_a]$ is now the difference of $\tilde{g}_a$ between the positive and negative sides of the pillbox, while $A$ is the area of each side. On the other hand we have that $\int \tilde{\rho}\, dV = \sigma A$, and Gauss's law produces
\begin{equation}
[\tilde{g}_a] n^a = -4\pi G \sigma,
\label{junction_regularized} 
\end{equation}
in agreement with Eq.~(\ref{junction}). We also perform an integration of $\tilde{g}_a$ over a closed curve that straddles the layer. Because $\tilde{g}_a\, dx^a = dU$ the integral necessarily returns zero, and we obtain
\begin{equation} 
[\tilde{g}_a] e_A^a = 0. 
\end{equation}
Continuity of the tangential derivatives of $\tilde{U}$ implies that
\begin{equation}
[\tilde{U} ] = 0,
\end{equation} 
and we again have agreement with Eq.~(\ref{junction}).

\subsection{Eulerian and Lagrangian variations of the potential}
\label{subsec:variation1} 

The foregoing results reveal that while $U$ is continuous on a singular surface, its normal derivative is discontinuous. Suppose now that $\SSS(t)$ is deformed by a variation of the fluid configuration described by the Lagrangian displacement vector $\xi^a$. The potential $U$ continues to be continuous across the deformed surface, and this implies that
\begin{equation}
\mbox{$\Delta U$ is well defined on $\SSS(t)$},
\end{equation}
where $\Delta U$ is the Lagrangian variation defined by $\Delta U := U(t,\theta;\epsilon) - U(t,\theta;0)$. But $\nabla_a U$ is discontinuous, and this implies that
\begin{equation}
\mbox{$\delta U$ is not defined on $\SSS(t)$},
\end{equation}
where $\delta U$ is the Eulerian variation, related to $\Delta U$ by $\delta U = \Delta U - \xi^a \nabla_a U$. 

\subsection{Variation of the Lagrangian with respect to the potential}
\label{subsec:variation2}

We shall return to the relation between $\Delta U$ and $\xi^a \nabla_a U$ below. For the time being we go back to Eq.~(\ref{L_grav}) and carry out a variation with respect to the gravitational potential, keeping the fluid configuration fixed. This means that we set $\xi^a = 0$ in the calculation, and that there is no distinction between the Lagrangian and Eulerian variations of the potential. In this specific context, $\delta U$ is well defined on $\SSS(t)$. The manipulations are identical to those of Sec.~\ref{subsec:var_U}, and we obtain
\begin{equation}
\delta L_{\rm grav} = -\frac{1}{4\pi G} \oint_\infty \delta U \nabla^a U\, dS_a
+ \frac{1}{4\pi G} \int \nabla^2 U\, \delta U\, dV + \int_{\SSS(t)} \sigma \delta U\, dS.
\end{equation}
We eliminate the surface integral at infinity by appealing to the usual variation rules, and we convert the integral over $\SSS(t)$ to a volume integral with the help of $\delta(\ell)$. We get
\begin{equation}
\delta L_{\rm grav} = \frac{1}{4\pi G} \int \bigl[ \nabla^2 U + 4\pi G \sigma \delta(\ell) \bigr] \delta U\, dV.
\end{equation}
The variation of the action is then $\delta S_{\rm grav} = \int \delta L_{\rm grav}\, dt$, and demanding that $\delta S_{\rm grav} = 0$ for an arbitrary $\delta U$ produces Eq.~(\ref{poisson_interface}), as expected. With the understanding that $U$ is distribution-valued, the discontinuity of $\nabla_a U$ on $\SSS(t)$ does not constitute an obstacle in these computations. 

\subsection{Variation of the Lagrangian with respect to the fluid configuration}
\label{subsec:variation3} 

We now carry out a variation of the gravitational action with respect to the fluid configuration. The variation is described by $\delta U = 0$ and $\xi^a \neq 0$, and an immediate difficulty is that the relation $\Delta U = \xi^a \nabla_a U$ is meaningless for a singular surface: while the left-hand side is well defined, the right-hand side is not, in view of the discontinuity of $\nabla_a U$. To make sense of the variation we must therefore step away from the idealization and re-introduce the regularized version of the surface layer, with its smooth density $\tilde{\rho}$ and potential $\tilde{U}$. The relevant part of the regularized Lagrangian is
\begin{equation}
\tilde{L}_{\rm grav} = \int \tilde{\rho} \tilde{U}\, dV,
\end{equation}
and its variation with respect to the fluid configuration is given by
\begin{equation}
\delta \tilde{L}_{\rm grav} = \int \tilde{\rho} \Delta\tilde{U}\, dV
= \int \tilde{\rho}\, \xi^a \nabla_a \tilde{U}\, dV.
\end{equation}
We wish to evaluate this, and eventually express the result in terms of surface quantities, in the limit $\ell_0 \to 0$.

First we decompose $\xi^a$ into normal and tangential components, according to $\xi^a = \xi_n\, n^a + \xi^A\, e^a_A$, and substitute within the integral. We obtain
\begin{equation}
\delta \tilde{L}_{\rm grav} = \int \tilde{\rho}\, \xi_n \tilde{g}\, dV
+ \int \tilde{\rho}\, \xi^A \nabla_A \tilde{U}\, dV,
\end{equation}
where $\tilde{g} := n^a \partial_a \tilde{U} = \partial_\ell \tilde{U}$ is the normal component of the gravitational field. Because $\xi^A \nabla_A \tilde{U}$ varies gently across the layer, we may approximate it by its value at $\ell = 0$ inside the second integral, and get
\begin{equation} 
\int \tilde{\rho}\, \xi^A \nabla_A \tilde{U}\, dV \simeq \int_{\SSS(t)} \sigma \xi^A \nabla_A \tilde{U}\, dS
\end{equation}
after integration with respect to $\ell$. In this equation, and in other approximate results obtained below, the approximation improves as the layer is made thinner and thinner.

The evaluation of the first integral requires more work. We use the regularized Poisson equation to express $\tilde{\rho}$ in terms of $\nabla^2 \tilde{U}$. Evaluation of the Laplacian in Gaussian coordinates, taking into account the fact that the normal derivative of $\tilde{U}$ is much larger than its tangential derivatives, reveals that it is well approximated by $\partial_\ell \tilde{g}$. We then have that
\begin{equation} 
\int \tilde{\rho}\, \xi_n \tilde{g}\, dV \simeq -\frac{1}{4\pi G} \int \xi_n\, \tilde{g} \partial_\ell \tilde{g}\, d\ell dS
= -\frac{1}{4\pi G} \int \xi_n \partial_\ell \bigl(\tfrac{1}{2} \tilde{g}^2\bigr)\, d\ell dS.
\end{equation}
Because $\xi_n$ varies gently across the layer, we approximate it by its value at $\ell = 0$, and we integrate with respect to $\ell$. We obtain
\begin{equation} 
\int \tilde{\rho}\, \xi_n \tilde{g}\, dV \simeq
-\frac{1}{4\pi G} \int_{\SSS(t)} \xi_n \bigl[\tfrac{1}{2} \tilde{g}^2 \bigr]\, dS. 
\end{equation}
We next write
\begin{equation}
\bigl[\tfrac{1}{2} \tilde{g}^2\bigr] = \frac{1}{2} \bigl( \tilde{g}_+^2 - \tilde{g}_-^2 \bigr)
= \frac{1}{2} ( \tilde{g}_+ + \tilde{g}_- ) ( \tilde{g}_+ - \tilde{g}_- )
= \langle \tilde{g} \rangle[\tilde{g}],
\end{equation}
where $\langle \tilde{g} \rangle := \frac{1}{2} (\tilde{g}_+ + \tilde{g}_-)$ is the arithmetic average of $\tilde{g}$ on the positive and negative sides of the layer. Invoking Eq.~(\ref{junction_regularized}), we finally arrive at
\begin{equation}
\int \tilde{\rho}\, \xi_n \tilde{g}\, dV \simeq
\int_{\SSS(t)} \sigma \xi_n \langle n^a \nabla_a \tilde{U} \rangle\, dS 
\end{equation}
for the first integral.

Collecting results, we have that the variation of the regularized Lagrangian is given by
\begin{equation}
\delta \tilde{L}_{\rm grav} \simeq \int_{\SSS(t)} \sigma \Bigl( \xi_n \langle n^a \partial_a \tilde{U} \rangle
+ \xi^A \nabla_A \tilde{U} \Bigr)\, dS.
\end{equation}
We recall that the approximation improves as the layer is made thinner and thinner. In the idealized limit of a singular surface, the preceding result becomes
\begin{equation}
\delta L_{\rm grav} = \int_{\SSS(t)} \sigma \Bigl( \xi_n \langle n^a \partial_a U \rangle
+ \xi^A \nabla_A U \Bigr)\, dS.
\end{equation}
At this stage the regularized layer has served its purpose and can be discarded.

The preceding manipulations reveal that
\begin{equation}
\Delta U = \xi_n \langle n^a \partial_a U \rangle + \xi^A \nabla_A U 
\label{DeltaU_interface}
\end{equation}
under a variation of the fluid configuration on a singular surface. In the singular limit we have that
\begin{equation}
\langle \psi \rangle := \frac{1}{2} \Bigl\{ \psi_+\bigl(\SSS(t)\bigr) + \psi_-\bigl(\SSS(t)\bigr) \Bigr\}  
\end{equation} 
is the average of the quantity $\psi$ on each side of the surface. 

\bibliography{aniso}
\end{document}